\documentclass[11pt,a4paper]{article}
\pdfoutput=1
\usepackage{jheppub}
\usepackage{lscape}
\usepackage{array}
\usepackage{hyperref}
\usepackage{amsfonts}
\usepackage{amssymb}
\usepackage{bbm} 
\usepackage{graphicx}
\usepackage{caption}
\usepackage{subcaption}

\usepackage[normalem]{ulem}

\renewcommand{\=}{\,= \,}

\renewcommand{\a}{\alpha}

\newcommand{\s}{\sigma}
\renewcommand{\t}{\tau}

\newcommand\G{\Gamma}

\newcommand{\CH}{\mathcal{H}}

\newcommand{\CI}{\mathcal I}

\newcommand{\IZ}{\mathbb Z}

\renewcommand{\i}{{\rm i}}

\newcommand{\Ge}{\Gamma_\text{e}}
\newcommand{\e}{{\bf e}}

\renewcommand{\th}{\theta}

\newcommand{\z}{\zeta}

\newcommand{\imt}{\tau_2}

\newcommand{\El}{E}

\newcommand\bi{\begin{itemize}}
\newcommand\ei{\end{itemize}}

\newcommand\bspl{\begin{split}}
\newcommand\espl{\end{split}}

\newcommand\mtn{m\t+n}

\newcommand{\be}{\begin{equation}}
\newcommand{\ee}{\end{equation}}
\newcommand{\bea}{\begin{eqnarray}}
\newcommand{\eea}{\end{eqnarray}}

\newcommand\ub{\underline{u}}
\newcommand\vb{\underline{v}}

\newcommand\Sell{S_{\text{ell}}(z(u))}
\newcommand\ReSell{\text{Re}S_{\text{ell}}}

\title{From multi-gravitons to Black holes: The role of complex saddles}

\author{Alejandro Cabo-Bizet}
\emailAdd{alejandro.cabo\_bizet@kcl.ac.uk}
\affiliation{Department of Mathematics, King's College London,\\
The Strand, London WC2R 2LS, U.K.}

\abstract{~By applying the Atiyah-Bott-Berline-Vergne equivariant integration formula upon double dimensional integrals, we find a way to compute the matrix integral representations of~$4d$~$\mathcal{N}=1$ superconformal indices.~The final formula allows us to easily extract analytic results in the large-rank expansion of certain theories.~As an example, we compute the leading one-loop corrections to the effective action of the known complex saddles in those theories.~For a superconformal index of~$SU(N)$~$\mathcal{N}=4$ SYM,  we use the equivariant integration formula and the Picard-Lefschetz method to show that at large enough values of~$N$, only two, among the known complex saddles, dominate the counting of~\emph{large operators} i.e.~of operators with charges of order~$N^2$.~Contributions from other known complex saddles are present, but we show they are exponentially suppressed in that range of charges; the smaller the charges the less suppressed they are, and eventually, to count~\emph{small operators} i.e.~operators with charges smaller than~$N^{\frac{2}{3}}$, like multi-gravitons, they can not be neglected. 

}

\usepackage{amsmath}

\begin{document}
 
\maketitle


\section{Introduction and summary \label{sec:Intro}}

Following the general principles of string theory/gauge theory conjecture~\cite{Maldacena:1997re2}, and inspired by previous seminal results in the context of string theory~\cite{Strominger:1996sh}; a fresh perspective has emerged in the last few years regarding the microscopic structure of supersymmetric~$AdS_{d+1}$ black holes in~$d=3$, starting with~\cite{Benini:2015eyy}, and~$d=4$, starting with~\cite{Cabo-Bizet:2018ehj,Choi:2018hmj,Benini:2018ywd}.

Regardless of which of the current approaches one chooses as tool to study the asymptotic growth of BPS operators at large rank, via the use of matrix integrals~\cite{Cabo-Bizet:2019eaf,Cabo-Bizet:2020nkr} \cite{Benini:2018ywd,Benini:2018mlo}, there seems to be convincing evidence that complex eigenvalue configurations play an important role in understanding the problem. Currently, there are two approaches that work at large-$N$ and generic values of BPS charges. They are consistent with each other, and in important ways, also with the predictions coming from the gravitational side of the duality~\cite{Cabo-Bizet:2018ehj}, they both include two complex eigenvalue configurations, whose effective actions are entropy functionals dual to the Bekenstein-Hawking entropy of the corresponding~$AdS_5$ black holes~\cite{Gutowski:2004yv,Cvetic:2004hs,Cvetic:2004ny,Kunduri:2005zg,Kunduri:2006ek, Silva:2006xv,Hosseini:2017mds}. These two approaches are~$a)$ the {Bethe Ansatz formula} of~\cite{Benini:2018ywd,Benini:2018mlo,Closset:2017bse}, and $b)$ the {saddle-point analysis} of~\cite{Cabo-Bizet:2019eaf,Cabo-Bizet:2020nkr}.~\footnote{A very interesting third perspective has been put forward recently~\cite{Goldstein:2020yvj}. } This paper completes the approach of~\cite{Cabo-Bizet:2019eaf,Cabo-Bizet:2020nkr} by promoting the saddle point approximation to an exact formula at finite values of~$N$. The problem that remains is that for the theories we have analysed so far, it is hard to evaluate the formula at finite values of~$N$. However, at large $N$ the formula gives a natural expansion that matches exactly the current expectations from the string theory side of the duality.~\footnote{An incomplete subset of relevant references in both gravity and field theory sides is~\cite{Cassani:2019mms,Kantor:2019lfo,Suh:2018qyv,Bobev:2019zmz}\cite{Benini:2020gjh,Lezcano:2019pae,Lanir:2019abx}\cite{ArabiArdehali:2019orz}\cite{Murthy:2020rbd,Agarwal:2020zwm}\cite{Choi:2018vbz,Honda:2019cio,ArabiArdehali:2019tdm,Kim:2019yrz,Cabo-Bizet:2019osg,Amariti:2019mgp}\cite{Crichigno:2020ouj}~\cite{Goldstein:2019gpz}~\cite{Copetti:2020dil,Agarwal:2020pol,Agarwal:2019crm},~\cite{Larsen:2019oll,Nian:2020qsk,David:2020ems,Melo:2020amq,David:2020jhp,Larsen:2020lhg}. A pedagogical introduction to the topic can be found in~\cite{Zaffaroni:2019dhb}. } 

There were two issues that~\cite{Cabo-Bizet:2019eaf,Cabo-Bizet:2020nkr} left open:

\begin{enumerate}
\item  A precise analysis of a deformation of the original contour passing through all the complex saddles there studied (This would have justified the assumption of considering all such configurations as competing ones in determining the original integral). 

\item An exact formula that could be used to evaluate~$1/N$ perturbative and non-perturbative corrections. 
\end{enumerate}

This paper solves these two issues. The natural completion of the large-$N$ saddle point approach initiated in~\cite{Cabo-Bizet:2019eaf,Cabo-Bizet:2020nkr} is the {Atiyah-Bott-Beligne-Vergne}~(ABBV) equivariant integration formula~\cite{Duistermaat:1982vw,1983InMat..72..153D, Witten:1982im,berline1983,Atiyah:1984px}.~\footnote{We have followed as much as possible the presentation given by Atiyah and Bott in~\cite{Atiyah:1984px}. But there are many other useful references. An absolutely incomplete selection that we have found necessary to follow at some points, being~\cite{Duistermaat:1982vw,1983InMat..72..153D, Witten:1982im,berline1983} and~\cite{audin2012torus,Cordes:1994fc,Niemi:1994ej,Blau:1995rs,Pestun:2016qko,JeffreyLectures,Alekseev:2000fe,Cremonesi:2014dva}.} The formula does not require us to assume large~$N$, but to evaluate effective actions and one-loop determinants we will use the large-$N$ expansion.
  
It is well known that the superconformal index can be written as a matrix integral~\cite{Romelsberger:2005eg,Kinney:2005ej}. We will see that the ABBV formula can be used to compute a generic family of matrix integrals that includes the one associated to the 4d superconformal index as a particular case. For simplicity we will focus on unitary matrix integrals but the idea can be explored in other cases as well. The answer is similar to the one of~$a)$ in that it can be expressed as a sum over configurations that solve a Bethe Ansatz equation.~\footnote{That they compute the same quantity does not imply that they are the same formula. }

To prove the formula, we will make use of the ambiguity in the choice of complex extension of the integrand of the original matrix integral, away from the real contour of integration.
In appendix~\ref{sec:RegularizationFlow}, we will find that these extensions can be understood as regularizations of the divergent super-determinants in the original 4d physical problem.~The elliptic extension of~\cite{Cabo-Bizet:2019eaf,Cabo-Bizet:2020nkr} is one of those. The (piecewise-) meromorphic extension corresponds to another. 

The second goal of this paper is to understand the competition of the many large-$N$ complex configurations that are known to compete, at the moment: Is there a finite/small number of saddles dominating the large~$N$ expansion of the microcanonical index?~\footnote{For~$SU(N)$~$\mathcal{N}=4$ SYM the number of known fixed points is finite and bounded by a polynomial function of~$N$~\cite{Hong:2018viz,ArabiArdehali:2019orz}\cite{Cabo-Bizet:2019eaf}.} The observable we will focus on, will be the logarithm of the number of operators -- counted with~$(-1)^F$ -- at a given value of a one-dimensional section in the space of conserved BPS charges in~$SU(N)$~$\mathcal{N}=4$ SYM. We will call it the \emph{microcanonical index}. The generating function of those integer numbers is the superconformal index~\cite{Romelsberger:2005eg,Kinney:2005ej}.~\footnote{Early relevant work concerning the estimation of growth in the number of~$1/16$-th BPS operators in~$\mathcal{N}=4$ SYM, can be found in~\cite{Janik:2007pm,Janik2008,Berkooz:2006wc,Chang:2013fba}.}

We will observe that for the counting of operators with charges of order~$N^2$ i.e.~large operators, the competition among known saddles is dominated by only two configurations, the~$(1,0)$ and~$(1,1)$. For large enough values of~$N$ the contributions from these two configurations become the same and the absolute value of the microcanonical index matches the exponential of the Bekenstein-Hawking entropy of the dual Gutowski-Reall black hole. Contributions from other~$(m,n)$ saddles are present but they are shown to be exponentially suppressed for such counting problem. As one tries to count operators of lesser charges, the contributions from the remaining~$(m,n)$ configurations become less suppressed. In principle the formula could be used even to count small operators of charges~$\mathcal{O}(1)$~\footnote{As shown in~\cite{Murthy:2020rbd,Chang:2013fba,Kinney:2005ej} these small operators are in one-to-one correspond to multi-gravitons. }, at leading order in large-$N$ approximation, but doing so with precision, would require the resummation of contributions from all saddle configurations.

Before contributions from other~$(m,n)$ saddles start being relevant, something interesting happens for operators whose charges are not large enough, but still finite in units of~$N^2$. For those values of charges, the ABBV formula predicts that the logarithm of the absolute value of the microcanonical index must oscillate around the exponential of the Bekenstein-Hawking entropy and that these oscillations are determined mainly by the superposition of contributions coming from~$(1,0)$ and~$(1,1)$ saddles. For smaller values of charges of order~$0$ in units of~$N^2$, more precisely of order~$N^{2/3}$ or smaller, the~\emph{two-saddles approximation} becomes invalid and other~$(m,n)$ configurations need to be considered.

The previous understanding is consistent with recent numerical results that show that for relatively large (but still finite) values of~$N$ ($N$ of order~ $10$)~\cite{Murthy:2020rbd,Agarwal:2020zwm} indeed oscillations are present. Moreover, the analytic interpolation of the numerical oscillations given in reference~\cite{Agarwal:2020zwm}, is consistent with the theoretical understanding summarised in the previous paragraph. Specifically,~\footnote{Up to a factor of~$2$ that lies within the margin of systematic error that one makes after discarding subleading contirbutions to the effective action at large~$N$.} such interpolation corresponds to the profile of the previously mentioned two-saddles approximation to the microcanonical index.

To summarise, we will focus on answering the following question: is there a finite number of configurations~$P$ dominating the large-$N$ counting problem? Our answer will be:
\begin{itemize}
\item For the counting of large enough operators, only two of them dominate the large-$N$ expansion.~Every other fixed point/saddle point/root known to us at the moment is either non-contributing, or exponentially suppressed in that expansion.
\item The presence of oscillations at relatively large (but still finite) values of~$N$ around the Bekenstein-Hawking entropy is a consequence of the sum over saddles/fixed points/roots. 
\item At leading order in large-$N$ expansion the oscillations become negligible and the counting of large operators matches the exponential of the Bekenstein-Hawking entropy.
\end{itemize}

Our goal with this paper is not to make precise numerical predictions at subleading order in the~$\frac{1}{N}$ expansion, but to understand the competition and the relevant hierarchies of charges where the nature of the competition among saddles changes. We will mainly focus on analyzing the leading order in the~$1/N$ expansion, but the ABBV formula does not require assuming large~$N$. To have analytic control at finite~$N$,~\footnote{Or large~$N$ and small enough scale of charges.} though,~a complete classification of the set of fixed points is needed. That classification must be completed in a case-by-case basis as the form of the conditions that determine the positions of the fixed points~$P$'s, depend on the representation content of the theory one wishes to analyze. Unfortunately, this is an open problem even in the simplest toy example that we study i.e.~a simple ensemble of~$1/16$-th BPS operators in~$SU(N)$~$\mathcal{N}=4$ SYM at large~$N$. Thus, given our current technical reach, we focus on analytically understanding the large-$N$ competition among known configurations, and use Picard-Lefschetz theory to identify which of them contribute, and finally, among those, which one dominates the large-$N$ counting.~\footnote{The claim that large-$N$ counting of large operators is dominated by two complex saddles, should not be extrapolated to a statement about the large-$N$ asymptotics of the \emph{grand canonical index} index~$\mathcal{I}$. These are two different quantities. It is possible that the canonical index could take relatively small values {at specific regions} of the chemical potential~$\tau$, meanwhile the underlying Hilbert space could still have an exponentially large number of states at large enough BPS charges, counted with the grading of~$(-1)^F$. }

In the remaining part of this introductory section we summarize the main two results of the paper. In subsection~\ref{subsec:IntroInd} we introduce the superconformal index and its relation to matrix integrals and elliptic hypergeometric integrals. In subsection~\ref{subsec:IndexFromABBV} we explain how the ABBV formula can be used to compute relevant families of matrix integrals. In subsection~\ref{subsec:ReloApp} we comment on the relation to the saddle point analysis of~\cite{Cabo-Bizet:2019eaf,Cabo-Bizet:2020nkr} and the Bethe Ansatz formula of~\cite{Benini:2018ywd,Benini:2018mlo,Closset:2017bse}. In subsection~\ref{sec:CompetitionIntro} we explain how the leading large-$N$ behaviour of a specific microcanonical index is controlled by the competition among complex saddles, and how such competition depends on how large the operators to be counted are. In subsection~\ref{subsec:ProspComm} we give interesting open questions that are left as future work. Towards the end we summarize the content of the main sections of the paper. The regularization of the 4d super-determinants is presented in appendix~\ref{sec:RegularizationFlow}.

\subsection{Superconformal indices, matrix integrals and Elliptic Hypergeometric integrals: a lightening review}\label{subsec:IntroInd}

The Hamiltonian definition of the~$\mathcal{N}=1$ superconformal index on~$S_3$~\cite{Romelsberger:2005eg,Kinney:2005ej}  is
\be \label{InI0rel}
\CI(p,q) 
\=  \, {\text{ Tr}}_{\CH_\text{phys}}\,  (-1)^F \text{e}^{-\beta \{\mathcal{Q},\bar{\mathcal{Q}}\} } 
p^{ J_1+\frac{Q}{2}}\,q^{ J_2+\frac{Q}{2}}\, t^{Q_{g}}\,.
\ee
The~$\mathcal{Q}$ and~$\overline{\mathcal{Q}}$ are complex conjugated supercharges.~$F$ is the Fermionic number.~$J_{1}$ and~$J_{2}$ are 
the left and right angular momenta on~$S_3$.~$Q$ is the~$U(1)$ R-charge. The~$J_{1,2}\,+\,\frac{Q}{2}$ are the two combinations out of~$J_{1,2}$ and~$Q$ that commute with $\mathcal{Q}$ and $\overline{\mathcal{Q}}$. The~$Q_g$ denotes a generic global charge. The~$p\=\e(\s)$,~$q\=\e(\t)$ and~$t\=\e(\Delta)$ are the rapidities associated to the charges~$J_{1,2}\,+\,\frac{Q}{2}$ and~$Q_g$, respectively.

The series coefficients of~\eqref{InI0rel} in an expansion around~$p=q=0$ (and for~$t=1$) count the difference between bosonic and fermionic gauge invariant states in the cohomology of~$\mathcal{Q}$, with fixed values of energy~$E$. In an abuse of terminology, we will call the energy~$E$ charge, as that is justified by the BPS relation that all relevant BPS states obey
\be\label{Energy}
E\= 2J\,+\,Q\, \quad \text{with}\quad 2J\,\equiv\,J_1\,+\,J_2\,.
\ee

For a theory carrying a representation~$R$ of a gauge group~$G$, which we assume to be a product of unitary groups
\be\label{IntegralSingleLetter} 
\mathcal{I}\=\int [dM] \, \exp\Bigl(\sum_{n=1}^{\infty} \,\frac{ \mathcal{I}_{\text{sl}}(p^n,q^n)}{n}\, \chi_{R(G)}(M^n) \Bigr)\,.
\ee
The~$\mathcal{I}_{sl}$ denotes the same index~\eqref{InI0rel} but computed after restricting the trace~$\text{tr}$ to run over gauge invariant operators that include only single fundamental fields~\footnote{For example, supposing the theory is Lagrangian these are the fundamental fields in the free limit of such Lagrangian. }, and any number of derivative insertions, instead of over the entire Hilbert space of the theory.~$[dM]$ is the invariant measure of integration over~$G$ which is the product of Haar measures of each of the simple factors in~$M$.~$\chi_{R(M)}$ is the character of the representation~$R$.~If~$\{\e{(u)}\}$ denotes the set of eigenvalues of~$M$ then~$\chi_R(U)\=\sum_{\rho}\e(\rho(u))$ where~$\{\rho\}$ is the set of weights of the representation~$R$. From~\eqref{IntegralSingleLetter} one can recover the index of a generic theory from the single-letter indices of chiral and vector multiplets
\be
\begin{split}
\mathcal{I}^\text{c}_{\text{sl}}(t,p,q)\= \frac{ t - pq/t }{(1-p)(1-q)}  \,,\quad
\mathcal{I}^\text{v}_{\text{sl}}(p,q)\=1\,-\, \frac{(1- pq)}{(1-p)(1-q)}\,.
\end{split}
\ee
For example, consider a~$U(N)$~$\mathcal{N}=1$ SCFT composed by one vector multiplet and three adjoint chiral multiplets. The vector and chiral multiplets are assumed to be charged under the~$U(1)_{R}$ superconformal~$R$-charge. The chiral multiplets are assumed to carry the fundamental representation of the~$SU(3)$ flavour symmetry group. In this case the superconformal index is a unitary matrix integral
\be\begin{split}\label{CharF}
\mathcal{I}&\= \int [dU]\, \exp\Bigl(\sum_{n=1}^{\infty} \,\frac{ \mathcal{I}_{sl}(t^n_1,t^n_2,t^n_3,p^n,q^n)}{n}\, \text{Tr}U^n\text{Tr}U^{\dagger n} \Bigr)\,, \\
\mathcal{I}_{sl}&\=1\,-\, \frac{(1\,-\, t_1)(1\,-\, t_2)(1\,-\, t_3)}{(1-p)(1-q)}\,,
\end{split}
\ee
where the variables~$t$,~$p$ and~$q$ are constrained by~$\frac{t_1 t_2 t_3}{p q}=1$ and initially one must assume~$0<|t_{1,2,3}|,\,|p|, |q|\,<1$ in order to guaranty this representation of~$\mathcal{I}$ to be well defined. The~$\mathcal{I}$ can be analytically continued in~$t$'s,~$p$ and~$q$ to larger regions. {Interestingly, it turns out that~$\mathcal{I}$ has singularities when~$p=q$ becomes a root of unit.~These singularities are important in the large-$N$ counting of BPS operators.~Their neighborhood encodes the essential information needed to extract the asymptotic form of such numbers. }

After integrating over angular variables, the integral~\eqref{CharF} can be recast in terms of integrals of products of elliptic functions \cite{Dolan:2008qi}\cite{Spiridonov:2010em}
\be\label{N4SYM}
\begin{split}
\mathcal{I}&\= \kappa \,\oint_{|\zeta|=1} \prod_{i=1}^{\text{rk}(G)}{\frac{d\zeta_i}{\zeta_i}}\,  \left(\prod_{i\,<\,j=1}^{N} \, {\theta_{\text{ell}}(\frac{\zeta_{i}}{\zeta_j},p)}\, {\theta_{\text{ell}}(\frac{\zeta_{j}}{\zeta_i},q)}\right)\,\prod_{i,j=1\atop i \neq j}^{N} \, \,{\prod_{a=1}^{3} \G_{\text{ell}}(\frac{\zeta_{i}}{\zeta_j}\,t_{a};\,p,\,q)}\,,\, \\
\kappa&\equiv \frac{(p;p)^{\text{rk}(G)}(q;q)^{\text{rk}(G)}}{N!}\,\prod_{a=1}^{3} \G_{\text{ell}}(t_{a};\,p,\,q)\,,
\end{split}
\ee
\footnote{The definition of elliptic Gamma function~$\Gamma_{\text{ell}}$ is given in equation~\eqref{GammaeDef}.}~where for~$G=U(N)$,~$\text{rk}(G)\,=\,N$, and for~$G=SU(N)$,~$\text{rk}(G)=N-1$. Moreover, for~$G=~SU(N)$ the~$\xi^{-1}_N=\prod_{i=1}^{N-1}\xi_{i}$.
~\footnote{For generic theories similar expressions exist. From this expression and after algebraic manipulations it follows that
\be\label{UNSUN}
\mathcal{I}_{SU(N)}\,=\, (p;p)^{-1}(q;q)^{-1}\, \mathcal{I}_{U(N)}\,.
\ee
In particular, this implies that the answer at leading order in large~$N$ expansion, is the same for both gauge groups.}
It is useful to note that the vector multiplet contribution can also be expressed in terms of elliptic Gamma functions by using the following identities
\be\label{IdentityThetaGamma}
\theta_{\text{ell}}(\zeta,p)\theta_{\text{ell}}(1/\zeta,q)\= \G_{\text{ell}}(\zeta\,p q;\,p,\,q) \G_{\text{ell}}(1/\zeta\,p q;\,p,\,q) \=  \frac{1}{\G_{\text{ell}}(\zeta;\,p,\,q)\G_{\text{ell}}(1/\zeta;\,p,\,q)}\,.
\ee
Sometimes it is convenient to express~\eqref{N4SYM} in terms of the potentials~$\underline{u}$,~$\s$,~$\t$ and~$\Delta$'s 
\be
\zeta_\rho\=\e(\rho(u)) \,,\,p\=\e(\s)\,,\, q\=\e(\t)\,,\, t_a\=\e(\Delta_{(a)}+\frac{\t+\s}{2})\,,
\ee
where~$\underline{u}=(u^1,u^2,\ldots)$ is an array that collects the~$N$ Cartan angles. For the moment, we can consider the vector~$\rho$ to be a weight of the adjoint of either~$U(N)$ or~$SU(N)$. For the case of~$SU(N)$ we can consider~$\rho$ to be a weight of~$U(N)$ upon the imposition of the constraint~$u^N\=-\sum^{N}_{i=1} u^i$.

The~$\Delta$'s encode the dependence on $R$-charges, and chemical potentials dual to global symmetries. They can be used to recover refined versions of the superconformal index.~\footnote{The point at which the flavour fugacities vanish $\Delta_g=0$ corresponds to~$\Delta_1=\Delta_2=\Delta_3=-\frac{\t\,+\,\s}{6}+ \mathbb{Z}$. As pointed out in~\cite{Cabo-Bizet:2019eaf} for this value of the parameters there is a pole in~$A\times B$ and thus in principle one would need to use the ABBV formula plus contributions coming from the non-vanishing residues. This correction seems to be essential at finite~$N$. However, at large-$N$ we argue that in principle one can compute the integral~\eqref{N4SYM} before imposing the constraint~\eqref{constraintSUSY} and then recover the constrained result after using meromorphic flow in the~$\Delta$'s. This is the approach we will follow in this paper. In forthcoming work, we will comeback to analyze the details of the contributions that come from residues.  } In these variables the constraint below~\eqref{CharF} takes the form
\be\label{constraintSUSY}
\Delta_{(1)}\,+\,\Delta_{(2)}\,+\,\Delta_{(3)}\,+\,\frac{\t+\s}{2} \,\in\, \mathbb{Z}\,.
\ee  

As mentioned before, for a generic theory with gauge group $G$ of rank $\text{rk}(G)$, the index can be written as elliptic hypergeometric integrals as well~\cite{Dolan:2008qi}\cite{Spiridonov:2010em}. The generic answer can be recast in the form
\be\label{IndexGeneral}
\begin{split}
\mathcal{I}&\= \int_0^1 d\underline{u}\,e^{-\,S(u)}\,\equiv\, \kappa_{G} \int^1_0 d\underline{u} \prod_{\a} \prod_{\rho_\a\,\neq\,0} \Ge(z_\a(u)+\frac{\t+\s}{2};\t,\s)\,, \\
\kappa_G&\equiv \frac{(p;p)^{\text{rk}(G)}(q;q)^{\text{rk}(G)}}{|\mathcal{W}|}\,\prod_{\a \atop \a\neq\text{vectors}}\prod_{\rho_\a=0} \G_{\text{e}}(z_{\a}(0)+\frac{\t+\s}{2};\,p,\,q)\,,
\end{split}
\ee
\footnote{The definition of elliptic Gamma functions~$\Gamma_e$ is given in equation~\eqref{GammaeDef}.}~where the integration measure is defined as~$d{\underline{u}}\equiv\prod_{i=1}^{\text{rk}(G)} du_{i}$.  The~$|\mathcal{W}|$ is the dimension of the corresponding Weyl group. The~$S(u)$ is a zero-dimensional analog of QFT quantum effective action, and thus we will call it effective action from now on. The label~$\alpha$ runs over all the multiplets in the theory. We define~$z_{\a}\,\equiv\,\rho_\a(u)\,+\,\Delta_\a\,$ where $\rho_\a$ is the vector of charges, i.e~a weight,  that the multiplet~$\a$ carries (under the Cartan elements of $G$). The complex number~$\Delta_\a$ encodes both the dependence on the R-charges of~$\alpha$ and the dependence on the set of potentials associated to global charges~$Q_g$, as defined in~\eqref{InI0rel}. This parameter can be used to recover more refined versions of the superconformal index that are obtained by turning on global symmetry holonomies.

We find convenient to analyze the generic integrals like~\eqref{IndexGeneral}, without imposing constraints like~\eqref{constraintSUSY}. Namely for each multiplet~$\a$ we associate a generic~$\Delta_\a$ (even for the vector multiplets) and only after the evaluation of the observable one wishes to study, the $\Delta_\a$ can be fixed to specific values. For example, in the case of \eqref{N4SYM} we can work by default with~$\Delta_\a\=\{\Delta_{(1)},\Delta_{(2)},\Delta_{(3)},\Delta_{(4)}\}$ and in the very end,~once a convergent result is obtained one can fix~$\Delta_{(3)}$ from condition~\eqref{constraintSUSY}.

\subsection{Recovering the index from double periodic extensions: a first incomplete but useful trial}  \label{subsec:IndexFromABBV0}

Let~$A\times B$ be a real torus with~$A$ and~$B$ cycles. Let~$e^{-S_\lambda}$ be a family of smooth complex functions in~$A\times B$ labelled by a real parameter~$\lambda\,$. Assume without loss of generality that $\lambda$ flows from zero to one.
\begin{figure}\centering
\includegraphics[width=11.8cm]{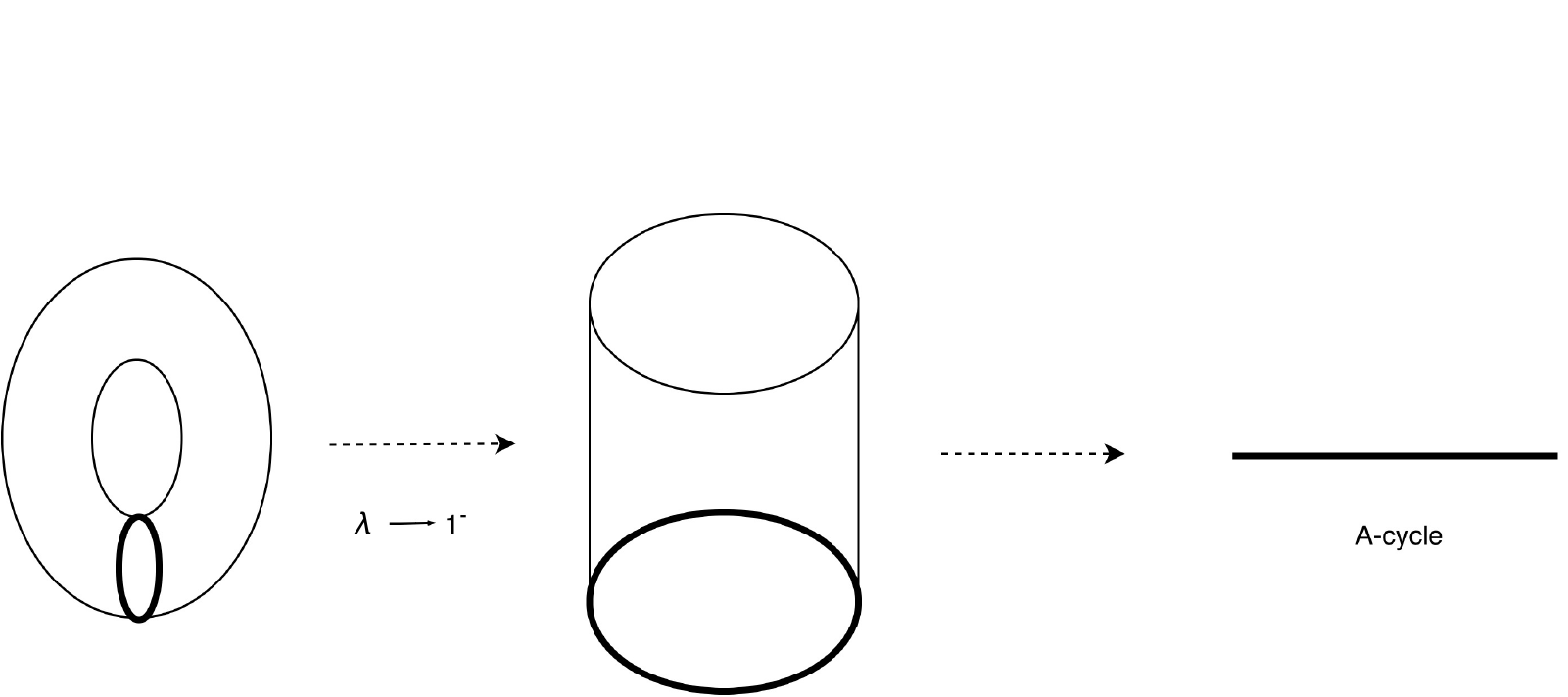} 
 \caption{A first trial to the problem: One starts with integrals of a family of exponentials of  moment maps which are smooth on the torus~$A\times B$. These integrals are parameterized by a real number~$\lambda$. As~$\lambda\,\to\,1^-$ the integrand develops a discontinuity at an~$A$-period and it is now well-defined and smooth in the cylinder (not in the torus). If the limit integrand is holomorphic and the height of cylinder is one, the integral over the cylinder collapses to an integral over an~$A$-period. Every~$A$-period gives the same answer. }
  \label{fig:Lambbda0}
\end{figure}
 Suppose that as~$\lambda\,\to\,1^-$ the~$e^{-S_\lambda}$ develops a discontinuity at an $A$-cycle of the torus. We can rephrase the previous statement by saying that in the limit~$\lambda\,\to\,1^-$ the domain on which~$S_\lambda$ was smooth, which used to be a torus~$A \times B$ before the limit, becomes a cylinder as sketched in Figure~\ref{fig:Lambbda0}. If the limit of~$e^{-S_\lambda}$ is holomorphic, its integral over the cylinder collapses to a middle-dimensional integral over any path homologous to an~$A$-period. Under such conditions, as it will be clear from the explanation we will give below, the limit of the integral over the original torus equals the integral of the limit function over any of the A-cycles.~\footnote{This last statement follows from the \emph{Lebesgue's dominated convergence theorem} (DCT)~\cite{Baxter1996}. We assume integration to be defined in the Lebesgue's sense. In summary, this theorem states that the limit of a Lebesgue integral coincides with the Lebesgue integral of the limit if the limit of the integrand is pointwise convergent, and the integral over the same domain of the absolute value of the limit is convergent. In particular, this implies that if the limit of the integrand is bound and the domain of integration is compact, the limit can always be commuted with the Lebesgue integral.} 
  
 Now, think of the~$A$-cycle in the previous paragraph as the Cartan torus of a gauge group. Let the torus~$A\times B$ be a double copy of~$A$. Suppose there exists a family of~$e^{-S_\lambda}$ on~$A\times B$ that flows to a holomorphic function living in the cylinder obtained by slicing~$A\times B$ across an~$A$-cycle. If the limit of~$e^{-S_\lambda}$ matches the integrand of the superconformal index (or the integrand of any other integrals that one wishes to compute) at one of these~$A$-cycles, any of them, we conclude that in such limit the integral over~$A\times B$ converges to the superconformal index. 
  
 A trivial example of how an integral over a slicing of~$A\times B$ collapses to a middle dimensional one over~$A$ (assuming~$\text{Im}(\Delta)>0$) is
 \be\label{DoubleIntegral}
 \int^1_0 du_1 \int^1_0 du_2 \, \Ge(u_1+u_2\t+\t+\Delta;\t,\t)\= \int^1_0 du_1 \, \Ge(u_1+\t+\Delta;\t,\t)\,.
 \ee
\footnote{This does not mean that there is no dependence on~$u_2$ in the integrand of the left-hand side. On the contrary, such integrand does depend on~$u_2$ but still the equality holds in virtue of the Cauchy theorem.} From inspection of the position of poles of~$\Ge$ that are given in appendix~\ref{App:Def}, one reaches the conclusion that there are no poles in the domain of integration of the left-hand side of~\eqref{DoubleIntegral}. That implies that the integral over~$u_2$ is spurious. The next obvious example is the~$n=2$ one in the following family
 \be
 \begin{split}
  \int^1_0  d\underline{u}_{1} \int^1_0\, d\underline{u}_{2} \, \prod_{i,j=1\atop i\neq j}^n\Ge(u_{1ij}+u_{2ij}\t+\t+\Delta;\t,\t)\\\= \int^1_0  d\underline{u}_{1}\, \prod_{i,j=1\atop i\neq j}^n\Ge(u_{1ij}+\t+\Delta;\t,\t)\,,
  \end{split}
 \ee
where~$u_{1,2 ij}=u_{1,2}^i\,-\,u_{1,2}^j\,$.
This can be generalized further to integrals of the following kind (assuming~$\text{Im}(\Delta_\a)>0$)
\be\begin{split}\label{InitialIntegral}
\int^1_0 d\underline{u}_1 \int^1_0 d\underline{u}_2 \prod_{\a}& \prod_{\rho_\a\,\neq\,0} \Ge(z_\a(u_1+\t u_2)+\t;\t,\t)\\ &\= \int^1_0 d\underline{u}_1 \prod_{\a} \prod_{\rho_\a\,\neq\,0} \Ge(z_\a(u_1)+\t;\t,\t)\,,
\end{split}
\ee
with~$\rho_\a$ being weights of a quiver gauge theory with matter in the fundamental, adjoint and/or bi-fundamental of say~$U(N)$ gauge nodes. The analysis for~$SU(N)$ nodes is more subtle, but one can always obtain it from~$U(N)$ by using the identity~\eqref{UNSUN}.

\subsection{The index from Atiyah-Bott-Berline-Vergne localization formula}

\label{subsec:IndexFromABBV}

\emph{What is the limitation of the previous approach?}

As we will explain in due time, integrals over~$A\times B$ of the smooth function~$e^{-S_\lambda}$ can be computed by using the Atiyah-Bott-Berline-Vergne formula. The expressions take the following form
\be
\int \,\frac{\omega^n}{n!}\, e^{-S_\lambda}\=\sum_P \, \frac{e^{-S_{\mathcal{\lambda}}(P)}}{ {E}(\nu_P)}\,.
\ee
Here~$n=\text{rk}(G)$ and the volume form~$\int \frac{1}{n!}\omega^n\=\int^1_0 d\underline{u}_1 \int^1_0 d\underline{u}_2\,$, is defined in terms of the two-form~$\omega\=\sum_{i=1}^n du^i_{1}\wedge du^i_{2}$. The remaining objects will be properly defined in due time. Here we just want to draw attention to a specific point.

The main claim of the previous subsection is summarized by the following equality
\be\label{threeEqualities}
\mathcal{I}\= \underset{\lambda\,\to\,1^-}{\lim}\,\int \,\frac{\omega^n}{n!}\, e^{-S_\lambda} \quad\Bigl( {\,\neq\,} \sum_P \,\underset{\lambda\,\to\,1^-}{\lim}\, \frac{e^{-S_{\mathcal{\lambda}}(P)}}{ {E}(\nu_P)}\Bigr)\,.
\ee
However, as stated by the inequality in parenthesis, the limit~$\lambda\,\to\,1^-$ of the previous subsection does not commute with the sum over~$P$'s. The reason being that
\be\label{HESSIANZEROMODE}
\underset{\lambda\,\to\, 1^-}{\lim}{E}(\nu_P)\=0\,.
\ee
This can be understood before doing computations. The quantity~$E(\nu_P)$, which will be properly defined below, is proportional to the square root of the determinant of the Hessian of~$S_\lambda$, and inversely proportional to the determinant of the matrix components of the two-form~$\omega$. As in that limit~$S_\lambda$ becomes holomorphic, the Hessian develops anti-holomorphic zero modes. Thus, as~$\det\omega$ is constant, equality~\eqref{HESSIANZEROMODE} follows through.

\paragraph{The integration measure}

The previous obstacle can be solved by using an appropriate integration measure, i.e by replacing the two form~$\omega$ by a~$\lambda$-dependent one that we will denote as~$\omega_{\lambda}$. The~$\omega_\lambda$ and~$S_{\lambda}$ will be defined in such a way that the corresponding integral will be independent of~$\lambda$. Thus, for such definitions the limit~$\lambda\,\to\,1^-$ is well defined. Finally, from our choice of~$S_\lambda$ it follows that for specific values of chemical potentials the integral along the real contour of integration is equal to the double dimensional integral for any~$0\leq\lambda<1$.

In particular, this implies that the ABBV equivariant integration formula can be used to compute 4d unitary matrix integrals of the form~\eqref{IntegralSingleLetter} or equivalently~\eqref{IndexGeneral}.

\begin{figure}\centering
\includegraphics[width=7.5cm]{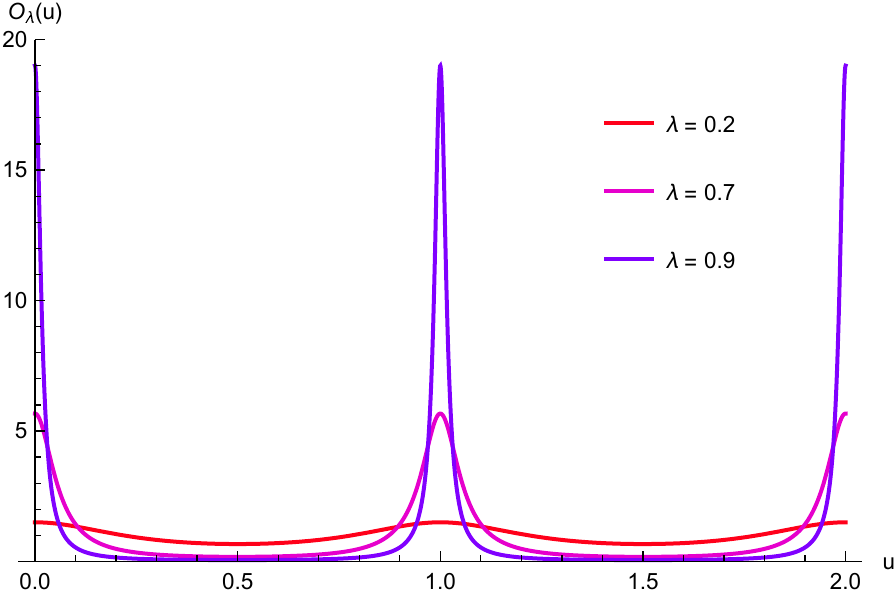} 
 \caption{ The profile of~$O_\lambda(u)$ that follows from the specific choice of function~$\{\cdot\}_\lambda$ that we have chosen to work with. The plot shows three different values of~$\lambda$.~$O_\lambda(u)$ is positive for any value of $0\leq\lambda<1$.~$O_{\lambda}(u)\to 1$ for small enough values of~$\lambda$.~$O_{\lambda}(u)$ tends to the periodic Dirac-delta function centered at integer values of~$u$ for~$\lambda\,\to\,1^-$. In fact, the specific details of the function~$\{\cdot\}_\lambda$, the one that defines the dependence on~$\lambda$ in both the action~$S_\lambda$, and the two-form~$\omega_\lambda$ (through~$O_{\lambda}$), are irrelevant as long as~$\{\cdot\}_\lambda$ is periodic, smooth and it preserves the conditions~\eqref{OLambdaCOnd},~\eqref{initialCondSForm} and~\eqref{measureLocalise}.  }
  \label{fig:Olambda}
\end{figure}

Let us go by steps. First we introduce the following measure of integration in~$A\times B$
\be\label{omegadef}
\int_{A\times B} \frac{\omega^n_\lambda}{n!}\,\equiv\, \int_{0}^1 d\underline{u}_{1}\int^1_0 d\underline{u}_{2} \prod_{i=1}^{n}\mathcal{O}_{i, \lambda}\,.
\ee
In this equation the closed two-form~$\omega_\lambda$ is defined as~
\be
\omega_\lambda\=\sum_{i=1}^n du^i_1\wedge du^i_2\, \mathcal{O}_{i,\lambda}\,.
\ee
For~$0\leq\lambda<1$
\be
\mathcal{O}_{i, \lambda}\= \mathcal{O}_{\lambda}(u_{2 i}) \,\equiv\, 1\,-\,\frac{\partial\{u_{2 i}\}_\lambda}{\partial u_{2i}}\,,
\ee
is a smooth and strictly positive function of a single real variable~$u_{2i}$ that is equal to $1$ at $\lambda\=0$ i.e.
\be\label{OLambdaCOnd}
\begin{split}
\mathcal{O}_\lambda(u_2)\,>\,0\,,\qquad
\mathcal{O}_\lambda(u_2+1)\=\mathcal{O}_\lambda(u_2)\,,\qquad\mathcal{O}_{i,\lambda=0}\=1\,.
\end{split}
\ee
These properties imply that~$\omega_\lambda$ is simply a smooth and globally well-defined passive diffeomorphism transformation of the two-form
\be\label{initialCondSForm}
\omega\=\omega_{\lambda=0}\=\sum_{i=1}^{n}du^i_1\,\wedge\, du^i_{2}\,.
\ee
The diffeomorphism is induced by the transformation~$u^i_2 \,\mapsto\, u^i_2 \,-\,\{u^i_2\}_\lambda$. The definition and properties of the function~$\{\cdot\}_\lambda$ will be given in the main body of the paper, in equation~\eqref{bracketMainText}. Here we just note that from that definition it follows that
\be\label{measureLocalise}
\mathcal{O}_\lambda(u_2)\,\underset{\lambda\,\to\,1^-}{\longrightarrow}\, \sum_{m\in\mathbb{Z}}\,\delta(u_2\,+\,m)\,,
\ee
as illustrated by the plots in figure~\ref{fig:Olambda}. In fact, the specific details of the function~$\{\cdot\}_\lambda$ are irrelevant as long as we use a~$\{\cdot\}_\lambda$ that is periodic, smooth and that preserves the conditions~\eqref{OLambdaCOnd},~\eqref{initialCondSForm} and~\eqref{measureLocalise}. This is consistent with the results presented in appendix~\ref{sec:RegularizationFlow}. There, it was shown that the choice of~$\{\cdot\}_\lambda$ corresponds to a choice of regularisation of the divergent one-loop determinants of the original 4d physical problem. Consequently different such choices should not affect the outcome of physical observables. We have chosen to work with the explicit form~\eqref{bracketMainText} for concreteness.

From property~\eqref{measureLocalise} and~\eqref{omegadef} it follows that {for any smooth action~$\mathcal{S}$ on $A\times B$}
\be\label{Limlambda1}
\int_{A\times B} \,\frac{\omega^n_\lambda}{n!}\, e^{-\mathcal{S}}\, \,\underset{\lambda\,\to\,1^-}{\longrightarrow}\,\int_{u^i_{2}=0} \,\prod_{i=1}^n du^i_{1} \, e^{-\mathcal{S}}\,.
\ee
If when restricted to the real contour of integration~$u^i_{2}=0$ the action~$\mathcal{S}$ {equals} the effective action~$S$ defined in equation~\eqref{IndexGeneral} i.e.
\be\label{PropScal}
\mathcal{S}(u)\= {S}(u) \qquad \text{at} \qquad u^i_{2}\=0\,,
\ee
then the desired integral e.g. the superconformal index, is recovered via the following limit
\be\label{c}
\int_{A\times B} \,\frac{\omega^n_\lambda}{n!}\, e^{-\mathcal{S}}\, \,\underset{\lambda\,\to\, 1^-}{\longrightarrow}\, \mathcal{I}\,.
\ee

\paragraph{The story is slightly more involved though} \label{par:FormulaHidden}

To guaranty smoothness of~$\mathcal{S}$ at intermediate stages we will need to introduce a cut-off~$\Lambda$~\cite{BrownLevin} that eventually must be fixed to~$\Lambda\,=\,1^-$. The limit~$\Lambda=1^-$ introduces discontinuities at~$u^i_{2}\=0$. Thus, the correct version of~\eqref{Limlambda1} in that case is a product of semi-sums of two single dimensional contour integrals i.e.
\be\label{Semisum}
\int_{A\times B} \,\frac{\omega^n_\lambda}{n!}\, e^{-\mathcal{S}}\, \,\underset{\lambda\,\to\,1^-}{\longrightarrow} \,\prod_{i=1}^n \frac{\Bigl(\int_{C^i_{+}}\,+\,\int_{C^i_{-}}\Bigr)}{2} \, {e^{-\,\mathcal{S}(u)}}\,,
\ee
~where
\be
\int_{C^i_{+}}\=\int^1_{0\,,\text{ at }u^i_{2}\,=\,0^+} d u^i_{1}\,,\qquad \int_{C^i_{-}}\=\int^1_{0\,,\text{ at }u^i_{2}\,=\,1^-} d u^i_{1}\,.
\ee
\eqref{Semisum} comes from the fact the the periodic delta function~\eqref{measureLocalise} picks up~$1/2$ times the value of the corresponding integrand at~$u^i_{2}=0^+$ plus~$1/2$ times the value of the corresponding integrand at~$u^i_2=1^-$.

At this point the global property (e.g. away from the $A$-period~$u^i_{2}\=0$) of our choice of function~$\mathcal{S}$ plays an essential role. Thus, we choose the action mentioned in subsection~\ref{subsec:IndexFromABBV},~$\mathcal{S}\= S_\lambda\,$ where~$S_\lambda$ will be defined in equation~\eqref{SLambdaDefinition}. In the limit~$\lambda,\,\Lambda\,\to\,1^-$ 
\be
\mathcal{S}\,\to\, S(u)\,,
\ee
where~$e^{-S(u)}$ is the meromorphic extension of the integrand of~\eqref{IndexGeneral}. In that case due to the periodicity property ~$e^{-S(u)}\,=\,e^{-S(u\,+\,2\pi)}$ and Cauchy theorem,
it follows that
\be\label{OperatorContour}
\int_{C^i_{-}}\,\blacksquare\=\int_{C^i_{+}}\,\blacksquare\,+ \,\widehat{\text{Res}}_{i}[\blacksquare]\,.
\ee
Both linear operators,~$\int_{C^i_{+}}$ and~$\widehat{\text{Re}}_{i}\,$, annihilate the dependence on a single complex variable~$u^i\,\equiv\,u^i_{1}\,+\,\t\, u^i_2$ of the function they act upon and give a new function of the remaining~$u^{j\,\neq\, i}$ complex variables. The integral fixes~$u^i_{2}\,=\,0$ and integrates~$u^i_{1}$ from~$0$ to~$1$. The operator~$\widehat{\text{Res}}_{i}$ when acting on a function~$f(u^i)$ of the variable~$u^i$'s, gives back a new function of the remaining variables ~$u^{j\,\neq\, i}$ which equates to the sum over residues 
\be\label{Residues}
\sum_{\xi_0}\text{Res}\Bigl[\frac{f(u(\xi))}{\xi^i},\,\xi^i\,=\,\xi_0\Bigr], \qquad \xi^i\,\=\,\e{(u^i)}\,.
\ee
$\xi_0$ runs over all the simple poles of the function~${f(u^i(\xi^i))}$ located in the annular domain~$|q|<|\xi^i|<1$. The corresponding residues can be computed systematically, but we will not do that in here. Plugging~\eqref{OperatorContour} into~\eqref{Semisum} and from the fact that~$\mathcal{I}\=\int_{C^i_{+}}\,e^{-S(u)}$ we obtain
\be\label{Variant}
\int_{A\times B} \,\frac{\omega^n_\lambda}{n!}\, e^{-{S}_\lambda}\, \,\underset{\lambda\,\to\, 1^-}{\longrightarrow}\, \mathcal{I}\,+\,\text{Residues}.
\ee
By Residues we mean contributions that involve the application of at least one of the operators~$\widehat{\text{Res}}_i$,~$i=1,\ldots, \text{rk}(G)$.  
These contributions vanish if~$e^{-S(u)}$ is a holomorphic function in the sliced~$A\times B$. We will assume there is always a chamber of~$\Delta$'s, where there is no contribution from simple poles.~\footnote{This very same assumption was used in reaching the Bethe Ansatz formula of~\cite{Benini:2018mlo}. The improved integration formula includes contributions that involve the action of at least one of the residue operators defined in equation~\eqref{OperatorContour}. We will analyze that point in forthcoming work.} From now on we focus on the simplest variant of~\eqref{Variant}
\be\label{c2}
\int_{A\times B} \,\frac{\omega^n_\lambda}{n!}\, e^{-\,S_\lambda(u)}\, \,\underset{\lambda\,\to\, 1^-}{\longrightarrow}\, \mathcal{I}\,.
\ee
A complete study of chambers of~$\Delta$'s where residues do contribute is left for future work.

\subsection{The final formula}\label{subsec:FormulaIntro}

Let us wrap up and move forward.~The relation~\eqref{c2} can be used to compute unitary matrix integrals like~\eqref{IndexGeneral} or~\eqref{IntegralSingleLetter} with the help of the Atiyah-Bott-Berline-Vergne integration formula. The first step is to define the following sequences of integrals
\be\label{ILambda1}
\mathcal{I}_\lambda \equiv \int_{A\times B} \,\frac{\omega^n_\lambda}{n!}\, e^{- S_\lambda}\,.
\ee 
As said earlier, the action~$S_\lambda$ will be defined in equation~\eqref{SLambdaDefinition}. The deformation parameter~$\lambda$ ranges from~$0$ to~$1$, including~$0$ but not~$1$! As we will explain in details in section~\ref{sec:FlowtoSCI} the sequence of integrals~\eqref{ILambda1} is computed exactly by the ABBV equivariant integration formula i.e.
\be\label{RepNewInd}
\mathcal{I}_{\lambda} \=  \sum_P \, \frac{e^{-S_{\mathcal{\lambda}}(P)}}{ {E}(\nu_P)}\,\,\equiv\,\,\sum_P\,\mathcal{I}_{\lambda}(P)\,,
\ee
where the~$P$'s are configurations of complex eigenvalues that are fixed points of the moment maps defined by the real and imaginary parts of the action~$S_{\lambda}$.~For $0\,\leq\,\lambda\,<\, 1$ they are also saddle points (or stationary phase points) of~$S_{\lambda}$.

The factor in the denominator is an equivariant Euler class that equates to
\be\label{EqEulerClass}
 {E}(\nu_P)\= \frac{1}{(2\pi \i)^{\text{rk}(G)}}\det{\frac{\,\partial B^i(P)}{\partial u_1^j}}\,.
\ee
For the examples we will analyze,~{half} of the fixed-point~(saddle-point) conditions can be written in terms of the objects~$B^i(u)$ (See definition~\eqref{EoMExplic}). They take the form of Bethe Ansatz equations~$B^i(P)=1$. This object does not seem to be the same as the Bethe operator of~\cite{Benini:2018mlo}, but it is closely related.~\footnote{A more detailed comparison will be left as future work.}

We will assume~$G=\otimes SU(N)$ and choose~$S_\lambda$ and~$\omega_\lambda$ in such a way that
\be\label{ConstantI}
\frac{\text{d} \mathcal{I}_\lambda(P)}{\text{d}\lambda}\=\Bigl(\frac{\partial}{\partial\lambda}\,+\,\frac{\partial P}{\partial \lambda}\cdot\nabla_{P}\Bigr)\, \mathcal{I}_\lambda(P)\=0\,.
\ee
Here~$\partial/\partial \lambda$ means derivative with respect to~$\lambda$ at fixed values of~$P$.

For fixed points~$P$ that are not continuously connected to other, the proof will go in two steps.~First, we will show that for any such~$P$
\be\label{EqsToProve}
\partial_\lambda S_{\lambda}(P)\=0\,,\qquad \partial_\lambda E(\nu_P)\=0\,.
\ee
and thus~$\frac{\partial\mathcal{I}_\lambda(P)}{\partial\lambda}\,=\,0$. The first equation follows from the fact that~$\partial_\lambda S_{\lambda}(u)$ is proportional to a linear combination of fixed point conditions and thus it vanishes at any~$P$ (not just for those with non-vanishing Hessian). The second condition follows trivially from the fact~$\partial_\lambda B^i(u)\=0$. Second, we will show that the position of that type of fixed points~$P$ can not depend on~$\lambda$ in virtue of ABBV theorem. That will complete the proof of~\eqref{ConstantI} for such cases. 

For a fixed point whose Hessian of the action vanishes,  and thus it is continuously connected to others, the variation in~$\lambda$ generates motion in the corresponding connected manifold of fixed points. As the ABBV formula includes integration over such manifolds, it follows that even in that case there is no dependence on~$\lambda$. 

The final conclusion is that in appropriate chambers of chemical potentials
~$\dfrac{d \mathcal{I}_\lambda}{d\lambda}\,=\,0$ and thus in virtue of~\eqref{c2}
\be\label{relIndexLambda}
\mathcal{I} \= \mathcal{I}_{\lambda}\quad\text{ for }\quad 0\,\leq\,\lambda\,<\,1\,,
\ee
where the superconformal index~$\mathcal{I}$, was defined in~\eqref{IndexGeneral}.
At last, plugging~\eqref{RepNewInd} into~\eqref{relIndexLambda} we obtain
\be\label{SumFixedPoints}
\mathcal{I}\= \sum_{P} \, \frac{e^{-S_{\mathcal{\lambda}}(P)}}{ E(\nu_{P})}\,.
\ee
As just mentioned, the~$P$'s can be part of connected manifolds and in that case one needs to integrate over such components with the measure induced by the inclusion map of the corresponding submanifold in~$A\times B$ upon the Liouville measure of~$A\times B$. We will elaborate on this later on.

\paragraph{Comment about complex fixed points} We focus on quiver theories with~$\nu$~$SU(N)$ gauge nodes and matter in the adjoint and bi-fundamental representations. As~$S_\lambda$ is smooth and double periodic, there will always exist complex fixed points~$P$ carrying finite Abelian group structure. That follows from the analysis of~\cite{Cabo-Bizet:2019eaf,Cabo-Bizet:2020nkr}. For instance for~$N$ prime, only~$\mathbb{Z}_N$ complex fixed points of such kind exist. They take the form
\be\label{ZNFP}
u^{i\beta}\=u^i\= \Bigl(\frac{i}{N}\,-\,\frac{N+1}{2 N}\Bigr)\,T\,, \qquad i\=1,\ldots, N-1\,.
\ee
$T\=m\t\,+\,n$ and~$m$,~$n$ are co-prime integers s.t.~$0 \leq m < N$ and~$0\leq n<N$.~$\beta=1,\ldots, \nu$ is the label of the~$SU(N)$ gauge nodes. For large-$N$ the~$(m,n)=(1,0)$ and~$(m,n)=(1,1)$ fixed points are the ones whose on-shell action corresponds to the dual AdS$_5$ BPS black holes.~For fixed $N$ the number of solutions are finite. Note that these fixed points lie outside the $A$-period~$\underline{u}_2=0$ i.e~they are not on the middle dimensional contour of integration that defines the superconformal index.

\subsection{Relation to other approaches and questions} \label{subsec:ReloApp}

\paragraph{Elliptic extension} Our initial motivation was to strengthen the basis of a large-$N$ saddle-point approach of~\cite{Cabo-Bizet:2019eaf,Cabo-Bizet:2020nkr}. We have found that the method put forward in those references is completed by the use of the ABBV formula at~$\lambda\=0$. The saddle-point or stationary phase approximation turns out to be exact in this case due to equivariant integration. In this way we have shown that complex saddles do contribute to the original integral.~On the way, we have fixed minor ambiguities regarding the computation of contributions coming from vector multiplets, and~$\tau$-independent constant phases in the effective action.

\paragraph{Bethe Ansatz formula} Possibly,
formula~\eqref{RepNewInd} and the Bethe Ansatz formula of~\cite{Benini:2018mlo,Closset:2017bse} are the same.~\footnote{As previously said, the fact fact they compute the same quantity does not mean that they are the same formula.} There are three observations we should make though:
\begin{itemize}
\item The Bethe operator~$B^i$ does not seem a priori to be the same one of~\cite{Benini:2018mlo,Closset:2017bse}. Finite~$N$ checks are perhaps a good way to start a more concrete comparison. We have computed the leading contribution of~$E(\nu_P)$ at large-$N$ for any~$(m,n)$ fixed point, and for the only case at disposal for comparison, the~$(1,0)$, the result matches an analogous semi-analytic result obtained via the Bethe Ansatz formula~\cite{GonzalezLezcano:2020yeb}. That is a large-$N$ check. Finite~$N$ checks from both perspectives will be addressed elsewhere.

\item The Bethe condition~$B^i=1$ {is only half of the saddle point conditions}. If all saddle point solutions carry Abelian group structure then the other half is automatically satisfied. However, should other solutions of the Bethe root conditions of~\cite{Benini:2018mlo,Closset:2017bse} not carrying Abelian group structure exist, we do not know if they will also be fixed points of our moment maps. This is an important issue that needs to be understood.

\item A complete classification of saddle points is missing. This is the main obstacle to make checks at finite and generic values of~$N$.~\footnote{ It would be interesting to explore if already-known integrability techniques could be of help to complete such classification~\cite{Nekrasov:2009uh,Nekrasov:2009ui}. }
\end{itemize}

\subsection{The large-$N$ expansion of the microcanonical index}\label{sec:CompetitionIntro}

In the second part of the paper, the ABBV formula will be combined with Picard-Lefschetz method to study how complex eigenvalue configurations compete at leading order in the large-$N$ expansion of the microcanonical index. So far, that is the only case where numerical results have been reported in the literature~\cite{Murthy:2020rbd,Agarwal:2020zwm}. The analysis will focus on the toy example of the case of~$\mathcal{N}=4$ SYM with~$\sigma=\tau$ and no flavour potentials. More general cases~$\sigma\neq\tau$ can be analyzed by using the results in appendix~\ref{sec:RegularizationFlow}.

Before entering in details let us explain in simple terms the conclusions of our study. As we will explain below, the perturbative parameter controlling the competition among saddles will be the exponential of minus the Bekenstein-Hawking entropy of the Gutowski-Reall black holes. To ease presentation let us denote this parameter as~$g\,\equiv\,e^{-\frac{A_{H}}{4}}$, where~$A_H$ is a function of the single independent charge carried by this BPS black hole, and which from now on we identify with the very same charge of the dual operators in the field theory side. For charges of order~$N^2$,~it turns out that~$g<<1$ for large enough values of~$N$, and thus only the two saddle contributions dominate the complete sum over~$P$. The contributions from other complex saddles to the microcanonical index are exponentially suppressed and then can be safely truncated when counting large operators.~For small enough charges, the~$g$ starts to approach one and indeed for charges of order~$N^{\frac{2}{3}}$,~$\log g$ becomes of order one. When the charges become of order one in the large-$N$ expansion ((.e~for multi-gravitons) the~$g\sim 1$ and then every known complex saddle contributes.

Next, we introduce some notation and explain our conclusions in more detail. Further details are postponed until section~\ref{sec:CompetitionFixedPoint}.

We will find that at large~$N$ the microcanonical index
\be
d(\mathfrak{q})\= \text{tr}_{\El} (-1)^F
\ee
defined as the trace over $\frac{1}{16}$-th BPS operators with charge~$\El=N^2 \,\mathfrak{q}$ for $\mathfrak{q}$ finite, is defined by the competition of only two fixed points~$(1,0)$ and $(1,1)$ out of the many $(m,n)$'s. We should mention that the associated grand canonical index is a function of a single chemical potential~$\tau$, namely
\be
\mathcal{I}(\tau)\= \,\sum_{\El\,=\,0}^\infty\, d(\mathfrak{q})\, q^{\El}.
\ee
This quantity can be obtained from the more general ensemble~$\mathcal{I}\=\mathcal{I}(\Delta_1,\Delta_2,\Delta_3,\sigma,\tau)$ (See~\eqref{N4SYM}) after imposing the following linear constraint among chemical potentials~$\Delta_1+\Delta_2+\Delta_3+\frac{\sigma+\tau}{2} \,=\,  p$, $p\,\in\, \mathbb{Z}$.~For instance~$p$ can be fixed to an integer that we denote as~$n_0=\pm 1$ e.g.~$p=n_0=\pm 1$.  The ensemble to be compared \emph{versus} the numerical results obtained in references~\cite{Murthy:2020rbd,Agarwal:2020zwm} corresponds to fixing~$\Delta_1\,=\,\Delta_2\,=\,\Delta_3$,~$\sigma\,=\,\tau$ and finally~$p$ to a specific integer value. We will fix~$p\=n_0=-1$ and proceed. 

The formula~\eqref{RepNewInd} predicts that the microcanonical index is a linear combination of contributions from each fixed point~$P$ i.e.
\be
d(\mathfrak{q})\= \sum_P d_P(\mathfrak{q})\,.
\ee
The single contributions~$d_P$ are extracted out of the grand canonical index by a contour integral in~$\tau$-plane. We denote such integral as~$C_\eta$. For each~$P$, this integral can be decomposed in an integer combination of $6$ Lefschetz-thimbles in~$\tau$-plane.~To understand whether a fixed point~$P$ contributes or not to~$d(\mathfrak{q})$, we need to compute the intersection numbers of~$C_\eta$ with its~$6$ thimbels.~ In that way one reaches a large-$N$ expansion of competing exponential terms the form
\be\label{ExpansionBlocks}\begin{split}
d(\mathfrak{q})&\= \sum_{ 0 \,\leq \,m,\,n \,< \,N\atop \text{gcd}(m,n)\,=\,1}\, \eta_{(m,n)} \,e^{ H_{(m,n)}(\mathfrak{q})\,+\,\dots}e^{\i \,(I_{(m,n)}(\mathfrak{q})\,+\,\ldots)}\,+\,\ldots \,,\end{split}
 \ee
where the magnitude of the~$(m,n)$-th exponential is controlled by the Morse function
\be\label{ExpoSuppr}
H_{(m,n)}(\mathfrak{q}) \= \frac{1}{m}\, H_{(1,0)}(\mathfrak{q})\=\frac{1}{4 m}\, A_H\,  \geq \,0\,,
\ee
and~$I_{(m,n)}$ are two real functions.~$H_{(m,n)}$ is the would-be entropy of the supposed-to-exist gravitational configuration dual to~$(m,n)$ fixed points.~\footnote{As computed in section 4.3 of~\cite{Cabo-Bizet:2019eaf}.} The~$\ldots$ denote sub-leading contributions in the large-$N$ expansion at finite~$\mathfrak{q}$. It denotes also potential contributions from other large-$N$ fixed points which are unknown to us at the moment. 

The integer intersection numbers~$\eta_{(m,n)}$ take three possible values
\be
\eta_{(m,n)}\,=\,\pm 1 \quad \text{or }\quad 0\,.
\ee 
For instance for~$(m,n)=(1,0)$ or~$(1,1),$~$\eta_{(m,n)}\,=\,1$.~\footnote{The $(0,1)$ does not contribute at leading order in large~$N$, so effectively we can take~$\eta_{(0,1)}=0$.}~The sufficient condition for~$(m,n)\neq(0,1)$ and~$\chi_1(m+n)\neq 0$ to contribute, i.e~for~$\eta_{m,n}\neq 0$, is that~$C_\eta$ must enclose its corresponding Cady-like point~$\tau=-\frac{n}{m}$~\cite{Cabo-Bizet:2019eaf}. For the chosen~$C_\eta$ that condition is equivalent to
\be\label{IntersectionDomainIntro}
-2<-\,\frac{n}{m}\,<\,1\,.
\ee
In that case the intersection numbers will be non vanishing i.e.~$|\eta_{(m,n)}|=1$. The~$(m,n)$ fixed points for which~\eqref{IntersectionDomainIntro} does not hold, have vanishing intersection number~$\eta_{(m,n)}=0$, and thus they do not contribute to the counting. That implies that~\emph{replica}~\footnote{The grand canonical index we study has periodicity~$\t\to \t+3$. This implies that the effective action of saddle $(m,n)$ is mapped to the effective action of $(m,n+3m)$ by such transformation. We say then that they are \emph{replica} of each other.  } fixed points of~$(1,0)$ e.g. like~$(1,3)$, do not contribute to the microcanonical index expansion defined by the contour~$C_\eta$, as expected.

Then, at large enough values of~$N$ with fixed and finite~$\mathfrak{q}\=\frac{\El}{N^2}\=\frac{\ell}{3 N^2}$, and up to exponentially suppressed contributions, the prediction coming from the ABBV formula to the microcanonical index~\eqref{ExpansionBlocks} is 
\be\label{TwoSaddlesApprox}
d(\mathfrak{q})\= e^{\, H_{(1,0)}(\mathfrak{q})\,+\,\ldots}\,\times \,2\, \cos{\bigl(I_{(1,0)}(\mathfrak{q})\,+\,\ldots\bigr)}\,+\,\ldots\qquad \,.
\ee
{Here the~$\ldots$ denote sub-leading corrections in large-$N$ expansion.} Note that in virtue of~\eqref{ExpoSuppr}, the~$(m,n)$ with~$m>1$ are exponentially suppressed. We will call~\eqref{TwoSaddlesApprox} the~\emph{two-saddles approximation}  to the microcanonical index.

The Morse function~$H_{(1,0)}$ coincides with the Bekenstein-Hawking entropy of the dual AdS$_5$ black hole~\cite{Gutowski:2004ez}. The cosine modulation 
\be\label{Interference}
\cos{\bigl(I_{(1,0)}(\mathfrak{q})\,+\,\ldots\bigr)}
\ee
explains the oscillations of~$d(\frac{\ell}{3N^2})$ as a function of~$\ell$ at relatively large but still finite values of~$N$ e.g.~$N\sim 10$.~These oscillations have been recently observed in the~$U(N)$ index~\footnote{At leading order in large~$N$ approximation the~$U(N)$ and~$SU(N)$ indices are equivalent. Differences come in at finite values of~$N$. The two saddle approximation can be used to approximate the microcanonical index of both,~$U(N)$ and~$SU(N)$, finite values of~$N$ e.g.~$N\sim 10$. The approximation should be even better for larger values of the charge~$\ell\,$, or equivalently for finite values of~$\mathfrak{q}=\frac{\ell}{3 N^2}$. This is consistent with the outcome of the fit made in~\cite{Agarwal:2020zwm} to their numerical results~\cite{Agarwal:2020zwm,Murthy:2020rbd}. The reason is that the small parameter that controls the order of the corrections coming from other fixed points is~$e^{-H_{(1,0)}(\mathfrak{q})}$; and this parameter increases when~$\mathfrak{q}$ decreases, at fixed~$N$. In particular it approaches~$e^{-\pi N^2\mathfrak{q}^{3/2}}\,\sim 1$ when~$\mathfrak{q}\to0$.  Roughly speaking, for values of charges~$\ell$ (resp.~$\mathfrak{q}$) smaller enough than~$N^{2/3}$ (resp.~$\frac{1}{N^{\frac{4}{3}}}$) we need to start considering contributions from other fixed points in order to increase precision. Implementing this procedure to refine the counting lies beyond the scope of this paper. Here we always assume that charges are large enough in such a way that the two-fixed point approximation to the oscillations is a good approximation (e.g.~$\mathfrak{q}$ finite at large~$N$). } via explicit evaluation of the Fourier coefficients up to rank~$N=10$~\cite{Murthy:2020rbd,Agarwal:2020zwm}.

\begin{figure}\centering
\includegraphics[width=7.5cm]{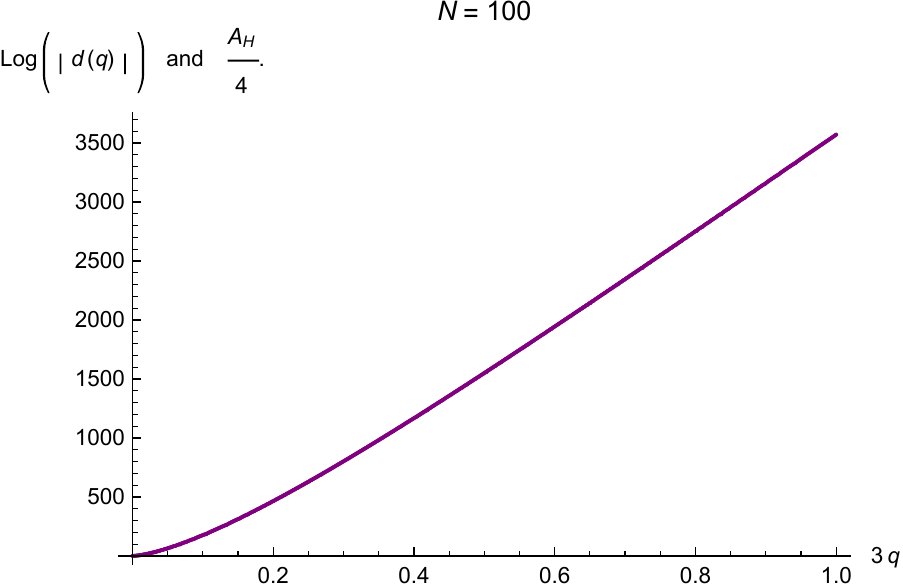}\qquad\\
 \caption{ In the figure in the top we have plotted values of the two-saddles approximation to the graded number of states~$d(\mathfrak{q})$ for~$N=100$, and the Bekenstein-Hawking entropy of the dual black holes. In this plot~$q\,=\,\mathfrak{q}$. In this scale, both plots are superposed and both are the same. In this scale of charges and entropy, the two saddles oscillations are negligible. For smaller values of~$N$ oscillation become noticeable as shown ``experimentally" in~\cite{Agarwal:2020zwm}(~\cite{Murthy:2020rbd}) and as analytically predicted by the two-saddles approximation.
 Something interesting happens at the point~$q=0$. This point encodes information about small operators, i.e~roughly speaking, operators with charges of order~$N^{\frac{2}{3}}$ or less. If we zoom in close enough into the point~$q=0$, oscillations start to enhance, first caused by the interference among contributions coming from~$(1,0)$ and~$(1,1)$, and at some point contributions from generic~$(m,n)$ saddles start to become important and can not be neglected. In principle, that flow evolution can be exactly probed (at least the one driven by the competition among~$(m,n)$ fixed points) with the help of the ABBV formula, but implementing that with numerical precision lies beyond the scope of the present paper and it is left for future work. }
  \label{fig:OscillationsVsN}
\end{figure}

Here we claim that oscillations around the Bekenstein-Hawking curve, like the example~\eqref{Interference} are an interference effect produced by the superposition of contributions of different complex fixed points of the matrix integral that include the~$(m,n)$ fixed points. However, for large enough value of~$N$, but not too large e.g.~$N\sim 10$, the main contribution comes from the couple of fixed points~$(1,0)$ and~$(1,1)$. This is, we claim that the oscillations are a consequence of 
\begin{itemize}
\item the grand canonical index~$\mathcal{I}(\tau)$ having an exact decomposition as a sum~\eqref{SumFixedPoints} over fixed points~$P$.~\footnote{Note that~$\mathcal{I}(\tau)$ depends only on a single parameter,~$\tau$.}
\end{itemize}
In the region of charges where the two-fixed point approximation to the oscillations can be trusted, we note that at large enough values of~$N$ the oscillatory pattern becomes negligible, and the curve~$d(\mathfrak{q})$ approaches the profile of the Bekenstein-Hawking entropy of the Gutowski-Reall solution~\cite{Gutowski:2004ez,Gutowski:2004yv} as shown in the plot in figure~\ref{fig:OscillationsVsN}.

\subsection{Final comments}\label{subsec:ProspComm} 

Before entering in technical details, let us share few questions/problems that we plan to study in the future:

\begin{itemize}

\item[1.] To check the ABBV formula for~$SU(N)$~$\mathcal{N}=4$ SYM against numerical data for the lower rank~$N=2,\,3,\ldots$. 

\item[2.] As mentioned in the caption of figure~\ref{fig:OscillationsVsN}, in principle, for~$SU(N)$~$\mathcal{N}=4$ SYM, the flow evolution from large to small charges~$E$, always at large $N$, can be exactly probed (at least the approximation to the exact flow driven by the competition among~$(m,n)$ fixed points) with the help of the ABBV formula. It would be extremely interesting to implement this with numerical precision and to compare against numerical data. 

\item[3.] To understand how this approach relates to the more orthodox large-$N$ methods~\cite{Brezin:1977sv,Marino:2015yie}.~\footnote{A fresh revision of many related topics have been recently given in~\cite{Anninos:2020ccj}.} Is the continuum limit of the fixed points~$P$ related to large-$N$ cuts or not? Can one understand the large-$N$ expansion of our approach from within the frame of loop-equations, WKB expansions and/or resurgence?  Answers to these questions could uncover interesting observations.~\footnote{Interesting approaches to the analysis of unitary matrix models of the kind addressed in this paper, have been recently put forward in~\cite{GonzalezLezcano:2020yeb,Copetti:2020dil}. They suggest points of contact between classical large-$N$ saddle point approaches and the one presented in this paper.}

\item[4.] Do these techniques give a hint on how to construct an explicit and tractable representation of the Hilbert space of~$\frac{1}{16}$-th BPS operators (in~$\mathcal{N}=4$ SYM language) in a large-$N$ expansion? The order zero problem would be to try to understand the form of the large operators (in terms of fundamental letters) that dominate the counting.

\item[5.] The exact solvability of this subsector of~$SU(N)$~$\mathcal{N}=4$ SYM,~\footnote{In relation with point $4.$, similar exact formulas, to the one here presented, can be analogously found for other BPS observables.} is indicative of the existence of an analogous subsector in the string theory side of the duality. It would be very interesting if such subsector could be explicitly identified.

\item[6.] It would be very interesting to explore how this sector of large BPS operators embeds and interact with non-BPS operators in the Hilbert space of weakly coupled~$\mathcal{N}=4$ SYM. This will allow us to probe how the complete physical system behaves when perturbed away from zero temperature and supersymmetry.

\end{itemize}

This paper is organized as follows. In section~\ref{sec:FlowtoSCI} we introduce the ABBV equivariant integration formula. In section~\ref{subsec:SLambda} contact is made with the superconformal index.~In section~\ref{sec:LargeNExpansion} we re-compute the effective actions of~$(m,n)$ fixed point and in the way fix previous ambiguities ~\cite{Cabo-Bizet:2019eaf,Cabo-Bizet:2020nkr} associated to vector multiplet contributions. In the second part of that section, in subsection~\ref{Enp}, we compute the equivariant Euler class of generic~$(m,n)$ fixed points at leading order in large-$N$ expansion and compare against a known result in the simplest ensemble of~$SU(N)$~$\mathcal{N}=4$ SYM for the of~$(1,0)$ fixed point. In section~\ref{sec:Counting} the ABBV formula and Picard-Lefschetz method are combined to analyze the competition among fixed points. The focus is on the simplest ensemble of~$SU(N)$~$\mathcal{N}=4$ SYM that captures the growth of states of interest. In appendix~\ref{sec:RegularizationFlow} we revisit the issue of regularization of the supersymmetric partition function~\cite{Cabo-Bizet:2018ehj} and show that there exists a scheme on which the latter matches the superconformal index. Although we mainly work in the case~$\sigma\,=\,\tau$ that condition can be relaxed. In appendix~\ref{ExtRegul} we show how the results obtained in this paper, can be easily generalized to infinitely many other cases~$\sigma\,\neq\, \tau$. In appendix~\ref{app:ReminderSmooth} the deformed action~$S_\lambda$ that will be used to apply the equivariant localization formula, is shown to correspond to a particular choice of regularization scheme. Other supporting material is relegated to the remaining appendices.

\section{Equivariant integration on a real torus}\label{sec:FlowtoSCI}
This section introduces the Atiyah-Bott-Berline-Vergne (ABBV) equivariant integration formula. This formula can be used, for instance, to compute integrals over symplectic manifolds. The relevant integrands can contain exponentials of a linear combination of the so-called moment maps~$f^a$. These are real analytic functions whose gradients generate isometries and preserve the symplectic structure of the manifold in question. The original integral localises to lower dimensional ones over sub-manifolds that are left invariant by the flows generated by the~$f^a$'s. As pointed out by Atiyah and Bott in~\cite{Atiyah:1984px}, one convenient feature of the formula is that the linear combination of $f^a$'s can be complex. That will be advantageous to our purposes.

\subsection{The moment maps and the equivariant Euler class}

Let~$A$ be the Cartan torus of the gauge group~$G$. Let~$\mathcal{M}\,=\,A\times B$ denote a real torus of rank~$2n\,\equiv\,  2\text{rk}(G)$. The original Cartan torus is a middle dimensional cycle~$A$ in~$A\times B$.~$A\times B$ can be thought as a product of~$n$ two-tori with labels~$i=1,\ldots,n$~i.e~a double copy of a period~$A$ with real angles
\be
u^i_{1} \,\sim\, {u}^i_{1}\,+\,1\,, \qquad {u}^i_{2}\, \sim\, {u}^i_{2}\,+\,{1}\,.
\ee 
We find convenient to define the following symplectic 2-form on~$A\times B$~
\be\label{Liouville}
\omega_\lambda\,\equiv\, \sum_{i=1}^{n} du_{1}^{i}\,\wedge\, du_{2}^{i}\,\mathcal{O}_{i, \lambda}(u_{2}^{i})\,=\, \sum_{i=1}^n du^i\,\wedge\, dv^i \,\mathcal{O}_{i, \lambda}(v^i)\,,
\ee
with complex twisted differentials
\be
du\,\equiv\, du_{1} \,+\,\t du_{2}\,, \qquad dv\,\equiv\, du_2\,.
\ee
This 2-form allows to define integrals via the corresponding Liouville measure~$\frac{1}{n!}\,\omega_\lambda^n\,$,
\be\label{PathIntegral}
\int_{A\times B} \,\frac{\omega_\lambda^n}{n!}   \, e^{-\,\mathcal{S}}\,\equiv\, \int_0^1\,d\underline{u}_1  \int^1_0  \,d\underline{u}_{2}\, \prod_{i=1}^{n}\,\mathcal{O}_{i, \lambda}\, e^{-\mathcal{S}}\,.
\ee 
For any value of~$\lambda$,~$\int^1_0 du^i_2 \,\mathcal{O}_{i\,\lambda}\=1$.

Let~$\mathcal{S}$ be an arbitrary linear combination of $d$ functions i.e.~$\mathcal{S}\equiv\sum_{a=1}^d \i t_a f^a\,$. The~$f^a:\mathcal{M}\mapsto \mathbb{R}$ with~$a\,=\,1,\ldots, d\,$, are smooth and real functions on~$\mathcal{M}$.~The functions~$f^a$ are called moment maps of a~$d$-torus action on~$\mathcal{M}$, if there exists a set of globally defined~$d$ vector fields~$V^a$ on~$\mathcal{M}$ that solve the following set of equations
\be\label{DefV}
{\imath}(V^a)\,\omega_\lambda \= df^a\,.
\ee
In that case, a particularization of the Atiyah-Bott-Beligne-Vergne (ABBV) integration formula~\cite{Atiyah:1984px} states that for~$t^a \in \mathbb{C}$
\be\label{ABFormula}
\begin{split}
\int_{\mathcal{M}}  \frac{\omega_\lambda^n}{n!}  \,e^{-\,\i \sum_{a} t_a\, f^a }\= \sum_{P}\,\int_{P}\, \imath_{P}^*\Bigl(\frac{\omega_\lambda^n}{n!}\Bigr)\, \frac{e^{-\,\i \sum_{a}t_a f^a(P)}}{E(\nu_P)}\,.
\end{split}
\ee
Usually, the parameters~$t^a$ are assumed to be pure imaginary numbers, and thus the action in the exponent is a real function, a Morse function. That it is possible to use complex~$t^a$'s, was noticed by Atiyah and Bott in their seminal work~\cite{Atiyah:1984px}.~Here we build upon their suggestion (See footnote~\ref{TComplex} below).

The~$P$'s are connected sub-manifolds of~$\mathcal{M}$ that solve the fixed-point conditions
\be\label{FixedPoints}
\imath(V^a)\,\omega_\lambda \=df^a \=0\,,\qquad a\=1,\ldots, d\,.
\ee 
The vector fields~$V^a$ generate isometries of~$\omega_\lambda$. That can be shown by using Cartan's formula upon the Lie derivative along~$V^a$ of~$\omega_\lambda$, and the defining equation~\eqref{DefV} i.e.
\be\begin{split}
\mathcal{L}_{V^a}\,\omega_\lambda&\= i(V^a)d\omega_\lambda\,+\, d(i(V^a)\omega_\lambda)  \\
&\= d(i(V^a)\,\omega_\lambda) \=d^2 f^a\=0\,.
\end{split}
\ee
The measure~$\imath_P^*(\frac{\omega_\lambda^n}{n!})$ is the pullback induced by the inclusion map~$\imath_P:P \hookrightarrow \mathcal{M}$ over the Liouville measure. Let us assume the~$P$'s to be flat submanifolds of~$\mathcal{M}$.
The~$E(\nu_P)$ is the equivariant Euler class of the normal bundle~$\nu_P$ to~$P$ in~$\mathcal{M}$.~Let the real co-dimension of~$P$ be~$2k$ then~\cite{Niemi:1994ej,Blau:1995rs}
\be\label{EnuP}
E(\nu_P)\= \frac{1}{(2\pi)^{k}}\,\frac{\sqrt{\det\Bigl( \mathcal{H}_{\mu \nu}(p_0) \Bigr)}}{\sqrt{\det\Bigl( (\imath_P^*\omega_\lambda)_{\mu\nu}\Bigr)}}\,, \qquad p_0\in P\,,
\ee
where the greek indices~$\mu$ and~$\nu\=1,\ldots, 2k$ label the  coordinates normal to~$P$ on~$\mathcal{M}$ with~$a=1,\ldots,k$. We note that for a complex choice of equivariant parameters~$t_a$'s the Hessian needs not to be real.~\footnote{We obtained this expression by following the analysis in section~3 of~\cite{Niemi:1994ej} but implementing the new feature that we are working with complex equivariant parameters and thus complex actions.}

~$(\imath_P^*\omega)_{\mu\nu}$ stands for the components of the two-form~$\imath_P^*\omega$.~The Hessian matrix~$\mathcal{H}$ is defined as 
\be\label{HDef}
\mathcal{H}\=
\left(
    \begin{array}{r@{}c|c@{}l}
  &  \mbox{$\frac{\partial^2{ \mathcal{S}}}{\partial u^a \partial{u}^b}$} & \mbox{$\frac{\partial^2{ \mathcal{S}}}{\partial u^a \partial{v}^b}$} \\\hline
  &     \mbox{$\frac{\partial^2{ \mathcal{S}}}{\partial v^a \partial{u}^b}$} &  
     \mbox{$\frac{\partial^2{ \mathcal{S}}}{\partial v^a \partial{v}^b}$}
    \end{array}  
\right) \,,\qquad a, b=1\,,\ldots, k\,,
\ee
where the twisted derivatives are
\be\label{UVChange}
\partial_{\ub}\, \equiv\, \partial_{\ub_{1}}\,,\qquad \partial_{\vb}\,\equiv\, \partial_{\ub_{2}} -\t\,\partial_{\ub_{1}}\,.
\ee
This twisted derivatives need not to be interpreted as coming from a complex change of variables in~$A\times B$. The coordinate will always be assumed to be~$u^i_1$ and~$u^i_2$. 

It is not rare to find continuous fixed sub-manifolds and not just isolated fixed points~$P$. For example, for~$U(N)$~$\mathcal{N}=4$~SYM the fixed manifolds are not points but two dimensional real surfaces. To avoid this extra technical involvement only~$SU(N)$ gauge groups will be analyzed.~\footnote{The~$U(N)$ case requires a minor redefinition of the closed~$2$-from ~$\omega_\lambda$ in~\eqref{Liouville} that we will not try to obtain in this paper.}

\subsection{The fixed point conditions }

Let us focus on the case of a two-torus action i.e~fix~$d=2$. Fix the moment maps~$f^a$ to be the real and imaginary parts of the action~$\mathcal{S}$ and the equivariant parameters accordingly i.e.
\be\label{ActionIdent}
f^1\,= \text{Re}\,\mathcal{S}\,,\qquad f^2\= \text{Im}\,\mathcal{S}\,,\qquad \i t_1 = t_2=1\,.
\ee
As announced before our choice of equivariant parameters is complex.~\footnote{\label{TComplex}As pointed out in~\cite{Atiyah:1984px} in page 15 in the paragraph below the comment numbered as 2). For the case of a 2-torus action, the component~$u_a\in \mathbb{C}$ in the~$l$-array mentioned in that paragraph (in this case~$l=2$), corresponds to the~$\i t_a$ in here.}

From this choice of moment maps and equivariant parameters~$t_a$ in~\eqref{ActionIdent}, it is immediate to see that {the fixed point conditions~\eqref{FixedPoints} are the saddle-point condition of} the corresponding {action~$\mathcal{S}$} i.e.~$\partial_{\underline{u}_{1}}\mathcal{S}=\partial_{\underline{u}_{2}}\mathcal{S} =0$. Or equivalently 
\be\label{fixedPointA}
\partial_{\ub}\,\mathcal{S} \=0\,, \qquad \partial_{\vb}\,\mathcal{S} \=0\,,
\ee
when written in terms of the twisted derivatives~$\partial_{\ub}$ and~$\partial_{\vb}$. This form of the equations will simplify the analysis later on.

\subsection{Examples of fixed points: quiver gauge theories} \label{subsec:QuiverSaddels}

Let~$\mathcal{S}$ be a smooth and double periodic extension to the double torus~$A\times B$, of the effective action~$S(u)$ in~\eqref{IndexGeneral}.

Any such~$\mathcal{S}$, with the spectrum of gauge charges~$\rho$ of the adjoint and/or bi-fundamental of~$SU(N)$ nodes, contains a simple set of stationary configurations that carries a finite Abelian group structure of rank~$N$.~That was shown in~\cite{Cabo-Bizet:2019eaf,Cabo-Bizet:2020nkr}. See section~$6$ of that reference for a proof.~\footnote{For finite $N$ there are a finite number of these fixed points. For instance for solutions with~$\mathbb{Z}_N$ structure, which are those labelled by a couple of integers~$(m,n)$ with~gcd$(m,n)=1$, and rank~$N=2$ there are only three solutions~$(1,0)$,~$(1,1)$ and~$(0,1)$. For simplicity in this paper we will always assume~$N$ to be prime. In that case, among solutions carrying finite Abelian group representations, only those carrying~$\mathbb{Z}_N$ representation exist.~It is possible that there are other kind of solutions that do not carry these previously mentioned finite Abelian group structure. }

Let us assume the action~$\mathcal{S}$ to be of the form 
\be\label{action}
\mathcal{S}(u)\= \sum_{\a}\sum_{\rho_\a\,\neq\,0} \mathcal{S}_{\a}(\rho(u))\,-\,\log \kappa_G\,,
\ee
where for~$z\,=\,z_1\,+\,\tau z_2$ and the single-particle action~$\mathcal{S}_\alpha(\rho(u))$ is a smooth and double periodic complex function of the real components~$z_1$ and~$z_2$ of~$z_\a(u)=\rho_\a(u)+\Delta_\a$, and~$\rho_\a$ are weights with respect to the total gauge group~$G={\oplus}_\beta\,SU_\beta(N)$. The index~${\beta}=1,\ldots, \nu$ labels the gauges nodes.~$\rho_\a$ is the array of~$\nu (N-1)$ charges under the action Cartan generators of~$G$. We will always assume multiplets in the the adjoint, or the bi-fundamental representations. The variable~$u$ stands for the array of~$\nu (N-1)$ complexified Cartan angles i.e~the complexified Cartan angles of all gauge nodes are collected in a single array of dimension~$\nu \times (N-1)$ and denoted by the letter~$u$ in this equation.

For any smooth and double periodic single particle action~$\mathcal{S}_\alpha$ the solutions to~\eqref{fixedPointA} include configurations
\be\label{Psaddles}
u^{i\beta}\=u^i\= \Bigl(\frac{i}{N}\,-\,\frac{N+1}{2 N}\Bigr)\,T\,, \qquad i\=1,\ldots, N-1\,.
\ee
To simplify presentation we can insert an extra auxiliary~$N$-th gauge variable in some intermediate equations
\be\label{largeNAuxiliary}
u^N\= \Bigl( 1\,-\,\frac{N+1}{2 N}\Bigr)\,T\,.
\ee
In the previous equations~
\be
T\=m \t+n\,,\qquad~0\leq m\leq N-1\,, \qquad~0\leq n\leq N-1\,,\qquad gcd(m,n)=1\,.
\ee
As the ansatz~\eqref{Psaddles} is the same for every node~$\beta$, effectively, we can consider all weights~$\rho_\a$ to be weights of the Adjoint of the diagonal~$SU(N)$ subgroup of~$G$. Thus, for any such quiver gauge theory, the effective action of ansatz~\eqref{Psaddles} can be written in the convenient form
\be
\nu\,\sum_{\a}\sum_{\rho\, \in\, \text{Adj}(SU(N))\atop \rho\,\neq\,0} \mathcal{S}_\alpha\Bigl(\rho(u)\Bigr)\,\=\,\nu\,\sum_{\a}\sum_{i,\,j\,=\,1 \atop i\,\neq\, j}^{N} \mathcal{S}_\alpha\Bigl(\frac{i-j}{N}\, T\Bigr)\,,
\ee
where all the effective weights are in the adjoint of the diagonal~$SU(N)\subset G$ and the index~$\a$ runs over every multiplet in the quiver theory.

In the large-$N$ limit the set of fixed points~\eqref{Psaddles} becomes a continuum i.e.~$\frac{i}{N}\to x\in [0,1)$ and~$\sum_{i}\to N \int^1_0 dx$. The effective action in that limit is computed by the Fourier averages along the period~$T=m\t+n$
\be\label{EffActionLambdaMN}
\begin{split}
\mathcal{S}_{\text{eff}}(m,n) &\=  \, \sum_\a \mathcal{S}_{\text{eff}\,\a}(m,n) \,  \\ &\,\equiv\,\nu N^2\,\sum_{\a} \int_{0}^{1}\int_0^1 dx dy\, \mathcal{S}_\a\bigl((x-y) T\bigr)\\ &\= \nu N^2\,\sum_{\a} \int_{0}^{1} dx\, \mathcal{S}_\a(x T)\,.
\end{split}
\ee
The index~$\a$ runs over every multiplet in the theory.
Note that we have not specified the form of the action~$\mathcal{S}_\a$ yet.

\section{The superconformal index} \label{subsec:SLambda}

\newcommand{\cS}{\mathcal{S}}
\newcommand{\cV}{\mathcal{V}}
As mentioned in section~\ref{par:FormulaHidden} the first step to obtain a formula for the superconformal index~$\mathcal{I}$~is to construct the action~$S_\lambda$ that together with the measure~$\omega_\lambda$ defines the~$\lambda$-independent quantity~$\mathcal{I}_\lambda$ introduced in equation~\eqref{ILambda1}. The real and imaginary parts of the action~$S_\lambda$ must be smooth moment maps in~$A\times B$ and thus, we can use the ABBV equivariant integration formula to compute~$\mathcal{I}_\lambda$ at any value of~$\lambda$ in between zero and one, including~$0$ and excluding~$1$. If by construction~$\mathcal{I}_{\lambda}$ reduces to~$\mathcal{I}$ in the limit~$\lambda\,\to\,1^-$, and if the former happens to be independent of~$\lambda$, then the ABBV formula for~$\mathcal{I}_{\lambda}$ is guarantied to compute~$\mathcal{I}$.~\footnote{ In specific regions of chemical potentials for which there are no poles emerging in the limit~$\lambda\to 1^-$(and~$\Lambda=1^-$) of the integrand of~$\mathcal{I}_\lambda$ inside the fundamental domain~$A\times B$. In case there are poles, the formula needs to be ammended as explained in section~\ref{par:FormulaHidden}.}

To simplify the presentation, at first we will focus on a single multiplet~$\alpha$ with gauge charge vector~$\rho_\alpha$. Later on a theory will be assembled.

~For~$0\,\leq\,\lambda\,<\,1$ we define
\be\label{VLambda}
e^{-S_{\lambda\a}}\=  \Gamma_\lambda(z_\a)\,\equiv\,  \frac{{R}_\Lambda^{-1}(z_\a)}{Q_{0\Lambda}(z_\a)}\,\Bigl(P_{0\Lambda}(z_\alpha)\Bigr)^{\,z_{2\a}(\{u_2\}_\lambda)}\,.
\ee
~Here~$z_{\a}\,=\,z_{\a}(u)\,=\,\rho_\a(u)\,+\,\Delta_\a\,$, and~$\Delta\,=\,\Delta_{\a1}\,+\,\t \Delta_{\a2}$ with real~$\Delta_{\a1}$ and~$\Delta_{\a2}$. From now on we assume~$-1\,<\,\Delta_{2\a}\,<\,1\,$.

The double periodic functions~$P_{0\Lambda}$ and~$Q_{0\Lambda}$ are defined via the logarithms
\be\label{logPnewText}
\log P_{0\Lambda}(z)\=\log P_{0\Lambda}(z_1,z_2)\,\equiv -\frac{\i}{2\pi}  \!
{\underset{m,n \in \IZ \atop m \neq 0}{\sum}} \; \Lambda^{|m|\,+\,|n|}\, \frac{\e(n z_2 - m z_1)}{m(m\t+n)} \,,
\ee
and
\be\label{logQnewText2}
\log Q_{0\Lambda}(z)\=\log Q_{0\Lambda}(z_1,z_2) \,\equiv\,  - \frac{1}{4\pi^2} \!
{\underset{m,n \in \IZ \atop m \neq 0}{\sum}} \; \Lambda^{|m|\,+\,|n|}\,\frac{\e(n z_2 - m z_1)}{m(m\t+n)^{2}} \,.
\ee
Here~$0\,\leq\, \Lambda\,<\,1$. The role of~$\Lambda$ is explained in appendix~\ref{app:P0Q0Props}. 
 
The pre-factor~$R_\Lambda^{-1}$ has two contributions. One is a power of $q=\e(\t)$ and the other a pure~$\tau$-independent phase
\be\label{FuncRepMB}
R_\Lambda^{-1}(z)\= R_\Lambda^{-1}(z_1,z_2)\=
q^{-\El_\Lambda}\,\,\e\bigl(-\frac{\Phi_\Lambda}{2}\bigr)\,.
\ee
Explicit definitions of~$\El_{\Lambda}$ and~$\Phi_\Lambda$ are given in appendix~\ref{app:ReminderSmooth} equation~\eqref{PhaseSmooth}. But for concrete purposes we will only be interested in their form at~$\Lambda\=1^-$.

From~\eqref{VLambda} we obtain the following expression for the action of a generic chiral multiplet
\be\label{PotentialChiral}
\begin{split}
S_{\lambda\a}&\=\,\log Q_{0\Lambda}(z_\a(u))\, -\, z_{2\a}(\{u_2\}_\lambda)\,\log P_{0\Lambda}(z_\a(u)) \\ & \qquad\qquad+\, 2 \pi\i \t\El_{\Lambda}(z_{2\a}(u))\,+\,\pi\i\Phi_{\Lambda}(z_\a(u))\,.
\end{split}
\ee
We define the bracket function of a real variable~$x$ as
\be\label{bracketMainText}
\{x\}_\lambda \,\equiv\,B_{1\lambda}(\{x\})+\,\frac{1}{2}\,.
\ee
It is a smooth version of the usual~$\{x\}\,\equiv\, x\,-\,\lfloor x \rfloor$.~$\{x\}$ is almost equal to $\lim_{\lambda\,\to\,1^-} \{x\}_\lambda\,$. The difference is located at integer values of~$x$. For such~$x$'s
\be\label{ProBrack}
\lim_{\lambda\,\to\,1^-} \{x\}_\lambda\, \= \frac{\{x^+\}\,+\,\{x^-\}}{2}\,.
\ee
We define the bracket of the vector~$\underline{x}$ as the vector of the brackets of the components~$\{\underline{x}^i\}$. The function~$B_{1\lambda}$ is defined in~\eqref{BperiodLambda}.

The phase function~$\Phi_{\Lambda}(z(u))$ can be any smooth function that in the limit~$\lambda\,\to\,1^-$ equates to a specific discrete periodization of the phase~$\Phi$ in identity~\eqref{Prefactor}. As mentioned before, one such an example is given in the second line of equation~\eqref{PhaseSmooth}. However, for concrete purposes there is no unique such choice i.e~one can choose any real~$\Phi_{\Lambda}(z(u))$ s.t. in the limit~$\Lambda\to 1^-$ the exponential~$\e\Bigl(\frac{\Phi_{\Lambda}}{2}\Bigr)$ goes to
\be\label{pragmatic}
\begin{split}
\frac{\e\Bigl( \frac{1}{2}\,B_1(\rho(u_{1})+\Delta_1- K) \Bigl(B_2(\lfloor \rho(\{u_{2}\})+\Delta_2\rfloor+1)\Bigr)\Bigr)}{ \,\e\Bigl( \frac{1}{2}\,B_1(\rho(u_{1})+\Delta_1- K) \Bigl(B_2(\lfloor \Delta_2\rfloor+1)\Bigr)\Bigr)}\,.
\end{split}\,
\ee
For large-$N$ evaluation, we will take the weights~$\rho$ to be the ones of the adjoint representation of~$U(N)$ which are labelled by two integers~$i,j=1,\ldots N$ and are s.t.~$\rho(u)\,\equiv\, u_{ij}\=u^i-u^j$. Although we are interested in~$SU(N)$, in order to obtain closed expressions for the contribution coming from the phase function~$\Phi_{\Lambda}$, it will be technically convenient to effectively work with~$U(N)$ instead of~$SU(N)$. In virtue of identity~\eqref{UNSUN}, the final contribution is bound to be the same at leading order in large~$N$ expansion.~Thus, to compute the large-$N$ contribution to the effective action of an~$(m,n)$ fixed point out of equation~\eqref{pragmatic} we will consider
\be
\rho(u_1)\,=\,u^i_{1}\,-\,u^j_{1}\,, \qquad \rho(\{u_2\})\,=\,u_{2\{i\}\{j\}}\,\equiv\,\{u^i_{2}\}-\{u^j_{2}\}\,.
\ee

The object~$K\in\mathbb{Z}$ is a choice of branch. It is an arbitrary piece-wise constant and integer function, which guaranty the exponent to be double periodic. When computing large-$N$ objects by using the continuum limit, different definitions of~$K$ could~{seem to lead} to different contributions to the phase of the effective actions of~$(m,n)$ fixed points. However, coming from the discrete finite~$N$ case, one can check that the difference between any two such choices is always an integer multiple of~$2\pi\i$, as it is bound to be the case.~ This implies that in order to compute values of the on-shell actions, we can even choose a branch~$K$ that breaks double periodicity, and that could even depend on specific saddles. This is because the value of the imaginary part of the effective action is only determined up to an integer multiple of~$2\pi\i$, thus we can always compute the double periodic value of the exponential by relaxing double periodicity of the exponent up to~$2\pi\i$ identifications. In due time~\footnote{This is a trick and it will only be used in the very end, to evaluate effective actions of specific fixed points.} we will use this feature for technical convenience.

Before moving on, let us summarize three relevant properties of~$e^{-S_\lambda(u)}$ that follow from definition~\eqref{VLambda}
\begin{itemize}
\item[a.] At finite~$0\leq \lambda<1$ and~$\Lambda$, it is smooth and double periodic, but~{non-meromorphic} in~$\underline{u}$. 
\item[b.] In the limits~$\lambda= 1^-$ and~$\Lambda=1^-$, it becomes~{locally meromorphic} i.e. 
\be\label{LimitingFunction}
\G_{\lambda}(z_\a(u))\= \Ge\bigl(z_{1\a}\,+\,\t z_{2\a}(\{u_2\})\,+\,\t;\,\t,\,\t\bigr) \,.
\ee  
\item[c.] For any~$0\leq\lambda<1$ and~$\Lambda\=1^-$, {it matches the elliptic gamma function in the limit} {to the~$A$-cycle~$u^i_{2}\=0^+$} i.e~at~$\Lambda=1^-$
\be
\G_{\lambda}(z_{1\a}(u)\,+\,\t z_{2\a}(0^+)) \= \Ge\bigl(z_{1\a}(u)\,+\,\t \Delta_{2\a}\,+\,\t;\,\t,\,\t\bigr) \,.
\ee
\end{itemize}
Property a. b. and c. were the initial assumptions needed in order to reach~\eqref{c2}.

\subsection{Half of the fixed point conditions: A Bethe Ansatz form} \label{subsec:SLambdaSubs}
In this subsection we show that for the choice of action~$S_\lambda$ to be defined below, half of the fixed point conditions can be written as a Bethe Ansatz condition.

Now that we have defined~$S_{\lambda\a}$ for a multiplet~$\alpha$, we can further refine the definition of~$\mathcal{I}_\lambda$ initially given in equation~\eqref{ILambda1} for the case of a generic quiver theory
\be\label{IndexGeneralDeformed}
\begin{split}
\mathcal{I}_\lambda&\= \int_{A\times B}\, \frac{\omega^n_\lambda}{n!}\,e^{-\,S_{\lambda}(u)}\,\\ &\equiv\, \kappa_{G} \int_0^1 d\underline{u}_1\int_0^1 d\underline{u}_2\, \prod_{i=1}^{n}\,\mathcal{O}_{i, \lambda} \, \prod_{\a} \prod_{\rho_\a\,\neq\,0} \G_{\lambda}(z_{1\a}(u)\,+\,\t\, z_{2\a}(u))\,.
\end{split}
\ee
The action of the complete theory is
\be\label{SLambdaDefinition}
S_{\lambda}(u)\,\equiv\,\sum_{\a}\sum_{\rho_\a} \,S_{\lambda\a}(u)\,,
\ee
where~$z_{1\a}\=\rho_\a(u_1)\,+\,\Delta_{1\a}$ and~$z_{2\a}\=\rho_\a(u_2)\,+\,\Delta_{2\a}$.
The function $\Gamma_{\lambda}(z_\alpha)=e^{-S_{\lambda\alpha}}$ was defined in~\eqref{VLambda}. It is a combination of $P_{0\Lambda}$, $Q_{0\Lambda}$ and Polylogs. 

The main goal of this subsection is to complement the explanations given in subsection~\ref{subsec:FormulaIntro}. To do so let us highlight the explicit form of half of the fixed point conditions. Once~$\Lambda=1^-$, such half takes the form
\be\label{EOMBA}
\begin{split}
\partial_{v^i} \,\mathcal{S_\lambda} &\= \sum_{\a}\sum_{\rho_\a}\rho^i\, 
\mathcal{O}_{i,\lambda} \log P_{0\Lambda=1^-}(z_\a(u)) \,+\,\text{c.t.}\, \\
&\=\mathcal{O}_{i,\lambda} \, \log B^{i}\,+\,\text{c.t.}\=0\,.
\end{split}
\ee
Recall that~$\partial_{v^i}\,\equiv\,\partial_{u^{i}_2}\,-\,\tau \partial_{u^i_{1}}\,$. In reaching~\eqref{EOMBA} we have used the definition of~$S_{\lambda\alpha}$~\eqref{VLambda}, the equations in appendix~\ref{Cutofff}, and also~\eqref{PhaseQLambdaRelation}.
The c.t. stands for contact terms contributions.
The~$B^i$ stands for
\be\label{EoMExplic}\begin{split}
B^i&\,\equiv\, \prod_{\a}\,\prod_{\rho_\a}\,P^{\rho_\a^i}_{0\Lambda=1^-}(z_\a(u))\,.
\end{split}
\ee
The vector multiplets do not contribute to the definition of~$B^i$ and can be ignored. That can be seen even before taking the limit~$\Lambda=1^-$. It follows from property~$\log P_{0\Lambda}(z)\=\log P_{0\Lambda}(-z)$, and from the fact that vector multiplets carry the weights of the adjoint representation which is a real representation. For the same reason, the same happens for the Bethe Ansatz equations of~\cite{Closset:2017bse,Benini:2018mlo}. 

\noindent From the positivity property~\eqref{OLambdaCOnd} i.e.~$\mathcal{O}_{i,\lambda}>0$ and the condition~$\partial_{v^i} S_\lambda=0$, it follows that the fixed points~$P$ must necessarily solve a Bethe Ansatz like condition
\be\label{BACOND}
\log B^i(P)\=0\,\implies\, B^i(P)\=1\,.
\ee
Moreover, they must not intersect the positions of the contact terms c.t..~\footnote{In case the latter do not vanish identically after summing over the corresponding matter content, which is the case for the examples we have studied.} Note that~\eqref{BACOND} is only half of the fixed point conditions.

The explicit form of these c.t.~contributions can be obtained from identity~\eqref{IdentityLogPalpba}. From the sufficient condition
\be
\forall_{i\,,\,j\,,\,\a}\, u^{i }\, \,-u^{j}\, \neq\, \Delta_\a\,\implies\, \text{c.t.}\=0\,, \,\qquad u^N\=-\sum^{N-1}_{i=1} \,u^i\,.
\ee
where~$i\,,\,j\,=\,1\,,\,\ldots\, N\,$, it follows that c.t.$\=0$ is solved by demanding the constraint to the left of the implication symbol upon any solution of~\eqref{BACOND}. For the theories we analyze, and the solutions of~\eqref{BACOND} that we are aware of (See subsection~\ref{subsec:QuiverSaddels}), that condition can be always satisfied.~\footnote{For instance by restricting to certain domains of the~$\Delta_\a$'s and/or by assuming~$N$ to be prime.} From now on we assume such choices and thus ignore the contact terms.

Before moving on, let us complete the proof of the first equation~\eqref{EqsToProve}, $\partial_\lambda S_{\lambda}(P)\=0$. Recall that~$\partial_\lambda$ in that equation means at fixed~$P$. This equation follows trivially from~\eqref{BACOND} as the only contribution of~$S_{\lambda}(P)$ that depends explicitly on~$\lambda$ is a linear combination of the logarithm of~\eqref{EoMExplic} which by definition vanishes at any fixed point~$P$.

\subsection{The final form of the equivariant Euler class}
In this subsection we refine the definition of the equivariant Euler class associated to the action~$S_\lambda$ and the symplectic form~$\omega_\lambda$.

For simplicity we focus on the generic case in which~$P$ is a point. That means that~$k=n$. From the definition of~$B^i$ given in~\eqref{EoMExplic} and identity~\eqref{IdentityLogPalpba}, it follows that for~$\Lambda=1^-$ and after avoiding contact terms
\be\label{DiagonalVanish}
\frac{\partial^2 S_\lambda}{\partial{v^i}\partial{v^j}}\=0\,.
\ee
Recall that~$\partial_{\ub}\, \equiv\, \partial_{\ub_{1}}\,,\, \partial_{\vb}\,\equiv\, \partial_{\ub_{2}} -\t\,\partial_{\ub_{1}}\,$.
Thus from the definition of the Hessian~$\mathcal{H}_{\mu\nu }$ of the action~$S_\lambda$~\eqref{HDef} and~\eqref{DiagonalVanish} it follows that only the off-diagonal blocks contribute to the corresponding determinant and thus
\be
\sqrt{\det\Bigl( \mathcal{H}_{\mu \nu}(p_0) \Bigr)}\= \pm \,\i^n\, \det\Bigl(\frac{\partial^2 B^i(u)}{\partial u^j}\Bigr) \,\times\, \prod_{i=1}^n \mathcal{O}_{i,\lambda}\,.
\ee

\be
\sqrt{\det\Bigl( (\imath_P^*\omega_\lambda)_{\mu\nu}\Bigr)}\=\sqrt{\det\Bigl( (\omega_\lambda)_{\mu\nu}\Bigr)}\=\pm \, \prod_{i=1}^n \mathcal{O}_{i,\lambda}\,.
\ee
From these equations and the definition of $E(\nu_P)$ in~\eqref{EnuP}, it follows that the equivariant Euler class equates to
 \be\label{TheFormula}
 {E}(\nu_P)\= \frac{1}{(2\pi \i)^{n}}\det{\frac{\,\partial B^i(P)}{\partial u_1^j}}\,.
\ee
In this definition we have fixed an overall constant ambiguity by comparing against other results in the literature. Note that this expression does not depend on~$\lambda$, thus it implies the second equation in~\eqref{EqsToProve}.

\subsection{The final integration formula: independence of~$\lambda$  }
In this subsection we show that the integral~$\mathcal{I}_\lambda$ is independent of~$\lambda$.

Again focus on~$\Lambda=1^-$. Assume that~$p_\lambda$ is a fixed point whose position depends on~$\lambda$, i.e.
\be
p_\lambda\= (\underline{u}_\lambda,\underline{v}_\lambda)\,,\qquad p_{\lambda+\delta\lambda}\=(u_\lambda+\delta\lambda \,{u}^{(1)},v_\lambda+\delta\lambda \,{v}^{(1)})\,.
\ee
It then follows that
\be\label{BiFlow}
B^i(p_\lambda)\= B^i(p_{\lambda+\delta\lambda})\=1\,.
\ee
The linear variation of this equation gives
\be\label{SecondEquality}
\sum_{j=1}^n\, u^{(1)j}\,\frac{\partial B^i(p_\lambda)}{\partial{u^j}}+\,v^{(1)j}\,\frac{\partial B^i(p_\lambda)}{\partial{v^j}}\=\sum_{j=1}^n\, u^{(1)j}\,\frac{\partial B^i(p_\lambda)}{\partial{u^j}} \=0\,,
\ee
The first equality follows from~\eqref{IdentityLogPalpba} and from the fact that only solutions of~\eqref{BiFlow} which do not collide with contact terms are fixed points. The second equality is telling us that the matrix~$\frac{\partial B^i(P_\lambda)}{\partial u_1^j}$ has a zero eigenvalue and thus
\be
\det{\frac{\,\partial B^i(p_\lambda)}{\partial u_1^j}}\=0\= {E}(\nu_{p_\lambda})\,.
\ee
In virtue of ABBV theorem, this can not be true for a fixed point~$p_\lambda$ that is not smoothly connected with others. Thus we conclude that for such class,~$p_\lambda$ can not depend on~$\lambda$ i.e.
\be
\frac{d p_\lambda}{d\lambda}\=0\,.
\ee
When~$p_{\lambda}$ is part of a connected set of fixed points~$P$, Atiyah-Bott localization theorem tells us that the matrix~$\frac{\,\partial B^i(p_\lambda)}{\partial u_1^j}$ can indeed have zero eigenvalues and the corresponding eigenvectors should be tangent vectors to the connected manifold~$P$ at~$p_\lambda$. The second equality in~\eqref{SecondEquality} is telling us that the variation induced by~$\lambda$ is constrained to move along the connected manifold of fixed points~$P$. But as the Atiyah-Bott prescription includes a covariant integration over such manifold, it then follows that even when~$p_\lambda$ belongs to a continuum set of fixed points~$P$, the 
associated contribution to the equivariant integral is independent of~$\lambda$. Of course it would be great to check this in a concrete fixed point, but we do not have an example of this type of solution yet.

In summary, for~$\Lambda\=1^-$ the integral~\eqref{IndexGeneralDeformed} does not depend on~$\lambda$, and in specific chambers in the space of potentials~$\Delta's$, it is by definition equal to the limit~$\underset{\lambda\,\to\, 1^-}{\lim} \mathcal{I}_\lambda$ (at~$\Lambda=1^-$). If we restrict attention to those specific regions in the domain of potentials~$\Delta$'s then
\be\label{IndexFormula}
\mathcal{I}\, \= \mathcal{I}_\lambda\= \sum_{P}  \, \frac{e^{-S_\lambda(P)}}{\,E(\nu_{P})}\,.
\ee
$E{(\nu_P)}$ was defined in~\eqref{TheFormula}. We should point out that generic fixed points~$P$ need to obey the saddle conditions~of $S_\lambda$.~The Bethe Ansatz condition~\eqref{BACOND} is only half of the latter.~$(m,n)$ fixed points satisfy both conditions trivially, but there could be other kind of solutions not carrying Abelian group structure.

\section{The leading contributions in large-$N$ expansion} \label{sec:LargeNExpansion}
In this subsection we revisit the evaluation of the large-$N$ effective action
\be\label{Action}
S_{\text{eff}}(p)\= S_{\lambda}(p)\,,
\ee
when~$p$ is an~$(m,n)$ fixed point.~In~\cite{Cabo-Bizet:2019eaf,Cabo-Bizet:2020nkr} the answer for~\eqref{Action} was fixed up to a~$\tau$ independent phase. Here we constraint such ambiguity.

In the large-$N$ limit, we can introduce the~$N$-th auxiliary variable~$u^N$ that at~$(m,n)$ saddles is defined as~\eqref{largeNAuxiliary} and work with the weights of~$U(N)$ instead of with the ones of~$SU(N)$. After substituting~$z$ by~$z_\a(u)= u_{i j}\,+\,\Delta_\a$ with the ansatz
\be
u_{ij}= \frac{i-j}{N}\, (m \t\,+\,n)\,,
\ee 
the discrete variables become continuum and the sums become integrals i.e.
\be
\frac{i}{N} \rightarrow x \in (0,1]\,, \qquad \sum_{i=1}^{N}\rightarrow N \int_{0}^1 dx\,.
\ee
In the large~$N$ for~$m\neq0$ one obtains
\be\label{OnshellEqs}\begin{split}
\int_0^1 \int_0^1 dx dy \,\log Q_{0\Lambda}\bigl((x-y) (m\t+n) +\Delta_\a\bigr)&\=  \frac{\pi\i\, B_{3\Lambda}(\{m \Delta_{\a1}\,-\, n\Delta_{\a2}\})}{3 m (m \t\,+\,n)^2}\,, \\
\int_0^1 \int_0^1 dx dy \,\log P_{0\Lambda}\bigl((x-y) (m\t+n) +\Delta_\a\bigr)&\=- \frac{\pi\i\, B_{2\Lambda}(\{m \Delta_{\a1}\,-\, n\Delta_{\a2}\})}{m (m \t\,+\,n)}\,,\\
\int_0^1 \int_0^1 dx dy  \,2\pi\i \t\,\El_{\Lambda}\bigl((x-y) (m\t+n) +\Delta_\a\bigr)&\=
\frac{\pi\i \t}{6}\,\bigl(2\Delta_{\a2}^3\,-\,\Delta_{\a2}\bigr)\,+\,\ldots\,.
\end{split}
\ee
The~Polylogs in~\eqref{OnshellEqs} become periodic Bernoulli polynomials at~$\Lambda\=1^-$.
The~$\ldots$ stands for corrections that go to zero at~$\Lambda=1^-$. 

From the limit~\eqref{pragmatic} of~$\Phi_\Lambda$ and after making a specific choice of branch~$K$ that we will detail below, we obtain a definite answer for the phase
\be
\lim_{\Lambda\to 1^-}\sum_{i,j\=0}^{N-1} \,\pi\i\Phi_{\Lambda}\Bigl(\frac{i-j}{N}\, (m\t+n) +\Delta_\a\Bigr) \,=\,\pi\i N^2\Phi(\Delta_a)\,.
\ee
Let us go step by step. For convenience we work in continuum variables~$x$ and~$y$ instead of discrete ones~$i$ and~$j$. To ease notation, let us drop the index~$\a$ for a moment. We recall that~$\Delta\,=\,\Delta_1\,+\,\tau\, \Delta_2\,$. For~$0\,<\,\Delta_{2}\,<\,1$ we can choose~(for $m>0$)
\be
K\Bigl((x\,-\,y)\, ( m \tau\,+\,n)\,+\,\Delta\Bigr)
\=
2m \lfloor m(\Delta_1\,-\,1)\,-\, n (\Delta_2\,-\,1)\rfloor\,+\,m(m\,+\,1)\,
\ee
for~$1\,-\,\Delta_2\,<\,(m x\,-\,m y)\,<\,{1-\sqrt{1\,-\,\frac{1}{2m}}\,\Delta_2}$ and zero otherwise. Below it will become clear why we have made this choice of~$K$.
A similar~$K$ can be defined for the case~$-1<\Delta_2<0$. 

{For~$-1\leq\Delta_2<1$} we can take the definition of~$\Phi_\Lambda$ in the exponent in the second line of~\eqref{pragmatic} and obtain
\be\label{PhiChiral}
\begin{split}
\pi\i N^2\Phi(\Delta)&\=\pi\i N^2\Phi(\Delta_1,\Delta_2)\,\\ &\equiv\,\pi\i N^2\,\int_0^1 \int_0^1 dx dy \, \bigl(B_1(nx\,-\,ny\,+\,\Delta_1)\,-\,K \bigr) \times \\ &\qquad \qquad\hfill\Bigl(B_2(\lfloor \{mx\}-\{my\}\,+\,\Delta_2\rfloor+1)\,-\,B_2(\lfloor \Delta_2\rfloor+1)\Bigr) \\
&\=  \pi\i N^2\,\Bigl(\frac{n}{3 m}\, \Delta^3_{2}\,+\,\frac{1}{m}\,\Delta^2_{2}\, B_{1}(\{m\Delta_{1}\,-\,n \Delta_{2}\})\Bigr)\,.
\end{split}
\ee
The choice of~$K$ mentioned before was chosen in order to reach the result in the third line. That will be convenient later on.

Next, we show an useful identity. Introducing the~$N$-th auxiliary variable~$u^{N}\,\equiv\,u^N_1\,+\,\tau u^N_{2}\=-\sum_{i=1}^{N-1} u^{i}$, 
and {defining its bracket as} $\{u^N_2\}_\lambda\,\equiv\, -\sum_{i\,=\,1}^{N-1}\,\{u^i_2\}_\lambda\,$ it follows that
\be\label{LambdaDependence}
\begin{split}
\sum_{\rho\in \text{Adj}(SU(N))} \rho(\{u_2\}_\lambda) \log P_{0\Lambda}(z_a(u)) &\=\sum_{i,j=1}^N\, (\{u^i_2\}_\lambda-\{u^j_2\}_\lambda) \log P_{0\Lambda}(z_a(u)) \\
&\= \sum_{i,j=1}^N\, (\{\frac{i}{N} m\}_\lambda-\{\frac{j}{N} m\}_\lambda) \log P_{0\Lambda}(\frac{(i-j)}{N} T+\Delta_\a)\\
&\=\sum_{i,j=1}^N\, \{\frac{i}{N} m\}_\lambda\, \Bigl(\log P_{0\Lambda}(\frac{(i-j)}{N} T+\Delta_\a)\,-\,i\leftrightarrow j\Bigr)\,\\
&=\sum_{i=1}^N\, \{\frac{i}{N} m\}_\lambda\, \sum_{j=1}^{N} \Bigl(\log P_{0\Lambda}(\frac{(i-j)}{N} T+\Delta_\a)\,-\,i\leftrightarrow j\Bigr)\, \\
&\=\sum_{i=1}^N\, \{\frac{i}{N} m\}_\lambda\,\times 0\=0\,.
\end{split}
\ee
In the last step we have used that
\be
 \sum_{j=1}^{N} \Bigl(\log P_{0\Lambda}(\frac{(i-j)}{N} T+\Delta_\a)\,-\,i\leftrightarrow j\Bigr)\, 
\ee
is a combination of the following vanishing quantities
\be\label{SumT}
\sum_{j=1}^{N} \sin{2\pi p \frac{(i-j)}{N} }\,.
\ee
The integer numbers $p$'s depend on~$m$ and $n$. One way to prove this analytically, is just to split the~$sin$ in the difference of exponentials, and then use the geometric sum~$\sum_{i=1}^{N} X^i\= \frac{X(1-X^N)}{1-X}$ for $X=\e(\frac{p}{N})\neq1$, then as~$X^N=1$ the sum cancels. When~$X=1$ every term in the sum~\eqref{SumT} vanishes. Note that identity~\eqref{LambdaDependence} holds for any finite~$N$,~$\lambda$ and~$\Lambda$. 

For every chiral multiplet, after using~\eqref{EffActionLambdaMN} we can obtain a closed expression for the deformed effective action of a chiral multiplet in terms of Polylogs:
\be
\begin{split}
\frac{S_{\lambda\a}(p)}{N^2}\,\equiv\,\frac{{S}_{\text{eff} \a}(m,n)}{N^2}&\=    \frac{\pi\i\, B_{3\Lambda}(\{m \Delta_{\a1}\,-\, n\Delta_{\a2}\})}{3 m T^2} \\ &\,+\, \Delta_{\a2}\,\frac{\pi\i\, B_{2\Lambda}(\{m \Delta_{\a1}\,-\, n\Delta_{\a2}\})}{m T}\,\\&
 \,+\,\frac{\pi\i \t}{6}\,\,\bigl(2\Delta_{\a2}^3\,-\,\Delta_{\a2}\bigr)\,+\,\pi\i \Phi_\Lambda\,.
\end{split}
\ee
Essentially, this effective action reduces to the one computed in~\cite{Cabo-Bizet:2020nkr} at~$\Lambda= 1^-$. This time the phase~$\Phi$ is fixed up to the previously mentioned ambiguity in the choice of central term~$K$. See~\eqref{PhiChiral}.

After taking the limits, we can use property~\cite{Cabo-Bizet:2020nkr}
\be
B_3(x+y)\= B_3(x)\,+\,3 B_2(x) y\,+\,3 B_1(x) y^2\,+\,y^3\,,\qquad x,\, y \,\in\,\mathbb{C}\,,
\ee
to obtain the following form of the effective action~\cite{Cabo-Bizet:2020nkr}
\begin{equation*}
\begin{split}
{S}_{\text{eff}\lambda \a}(m,n)&\=  \frac{ \pi \i N^2}{3m T^2}\,  [\Delta_{\a} ]^m_{T}  \,  \big([\Delta_{\a} ]^m_{T}  + \frac{1}{2}\big)   \big([\Delta_{\a} ]^m_{T}   + 1 \big)  \\
& 
 \, -\,\frac{\pi\i \tau}{6}\,N^2\,\Delta_{\a 2}\,+\, \pi \i  N^2 (C+\Phi )\,+\,\ldots.
\end{split}
\end{equation*}
For~$m\Delta_{\a1}\,-\,n \Delta_{\a2}\notin \mathbb{Z}$, the symbol bracket is the same as the one defined in~\cite{Cabo-Bizet:2020nkr} i.e.
\be\label{bracketDef}
[\Delta_\a]^{T}_m\,\equiv\, 
   T \Delta_{\a2}\,+\,\{m\Delta_{\a1}\,-\,n \Delta_{\a2}\}\,-\,1\,,
\ee 
for~$T=m\t+n\,$.~\footnote{To simplify the presentation, we have changed the definition of~$\Delta$ relative to the one we used in~\cite{Cabo-Bizet:2020nkr}. The translation is~$\Delta_{there}= \Delta_{here}+\tau$.}
However, in virtue of property~\eqref{ProBrack},~{for~$\gamma \equiv m \Delta_{\a1}\,-\,n \Delta_{\a2} \,\in\, \mathbb{Z}$} that definition needs to be slightly modified i.e~for~$\gamma\in\mathbb{Z}$
\be \label{Caveat}
\{\gamma\}\,\equiv\,\frac{\{\gamma^+\}+\{\gamma^-\}}{2}\,.
\ee
This follows from the fact that the corresponding discontinuities arise from limits of smooth and double-periodic Fourier expansions where uniform convergence is lost, and thus Dirichlet theorem applies (See around~\ref{DirichletTheo} below equation~\eqref{P0Theta} for more information on this theorem). The constant
\be
C\,\equiv\,
-\,\frac{n}{3 m}\, \Delta^3_{\a_2}\,\,-\, \frac{1}{m}\,\Delta^2_{\a2}\, B_{1}(\{\gamma\})\= -\Phi\,,
\ee
where the~$\Phi$ was defined in~\eqref{PhiChiral}.
Thus we reach
\be\label{EffActionFlavour}
{S}_{\text{eff}\lambda\a}(m,n)\=\frac{ \pi \i N^2}{3m T^2}\,  [\Delta_{\a} ]^m_{T}  \,  \big([\Delta_{\a} ]^m_{T}  + \frac{1}{2}\big)   \big([\Delta_{\a} ]^m_{T}   + 1 \big)
 \, -\,\frac{\pi\i \tau}{6}\,N^2\,\Delta_{\a 2}\,
\ee
but with the modification~\eqref{Caveat}.

\paragraph{Contribution from a vector multiplet}
The result for a vector multiplet can be recovered from the limit~$\Delta_{2\a}\to1^-$ and~$\Delta_{1\a}\to 0^{\pm}$ of~\eqref{EffActionFlavour}, by using the modified definition of~\eqref{bracketDef}.
In this case the effective action is
\be\label{VectorMultiplet}
{S}_{\text{eff}\lambda\text{v}}(m,n)\=\frac{\pi \i N^2}{6 m T}\,+\, \frac{\pi\i \t N^2}{3}\,+\, \frac{ \pi\i N^2 n}{3 m}\,.
\ee
Thus, having the previous detail in mind, we can write, for a complete gauge-anomaly free theory
\be
\begin{split}
{S}_{\text{eff}}(m,n)\,\equiv\,{S}_{\text{eff}
lambda}(m,n)&\=\sum_{\a}\,\frac{ \pi \i N^2}{3m T^2}\,  [\Delta_{\a} ]^m_{T}  \,  \big([\Delta_{\a} ]^m_{T}  + \frac{1}{2}\big)   \big([\Delta_{\a} ]^m_{T}   + 1 \big)
 \, -\,\frac{\pi\i \tau}{6}\,N^2\,\Delta_{\a 2}\, \\
 &\=\sum_{\a}\,\frac{ \pi \i N^2}{3m T^2}\,  [\Delta_{\a} ]^m_{T}  \,  \big([\Delta_{\a} ]^m_{T}  + \frac{1}{2}\big)   \big([\Delta_{\a} ]^m_{T}   + 1 \big)
 \,.
 \end{split}
\ee
In going to the second line we have used that
\be
\sum_\a \Delta_{2\a}\=\text{Tr} {R}\=0 \,+\,\mathcal{O}\Bigl(\frac{1}{N}\Bigr),
\ee
where~Tr${R}$ is the sum of superconformal R-charges of the fundamental fermionic fields of the corresponding theory. As noticed in~\cite{Cabo-Bizet:2020nkr}, this quantity vanishes at large~$N$ due to ABJ anomaly cancellation.  Please refer to subsection~4.1 of~\cite{Cabo-Bizet:2020nkr} around equations~(4.12) and~(4.21) for a detailed explanation.

\paragraph{The simplest ensemble of~$\mathcal{N}=4$ SYM}\label{par:N4SYM}
In this case we have one vector and three chiral multiplets with~$\Delta_{\a}\to\frac{1}{3}(-\t-n_0)$ where we can choose~$n_0= 1$ or~$n_0=-1$.~\footnote{ As it follows from setting up~$N=4$ SYM on the conformal boundary conditions dictated by the dual BPS black hole solution~\cite{Cabo-Bizet:2018ehj}.} We then proceed to complete the effective action computation of~\cite{Cabo-Bizet:2019eaf} 
\be\label{SeffN4SYM}
S_{\text{eff}}(m,n;\t)\= \frac{1}{m}\, \frac{\pi\i}{27}\, N^2 \, \frac{\bigl(2T+\chi_1(-n_0 m\,+\,n)\bigr)^3}{T^2}
    + N^2 \pi\i \,\varphi(m,n)\,.
\ee 
For our choice of branch~$K$
\be
\varphi(m,n)\= -\frac{\chi_1(-n_0 m \,+\, n)}{2 m}\,.
\ee
This answer is unique up to an addition of $\tau$-independent real numbers that after multiplication by~$N^2$ become an even integer.
From now on we will fix~$n_0=-1$.

 \subsection{The equivariant Euler class at leading order in the large-$N$ expansion} \label{Enp}
 
 In this subsection we study the equivariant Euler class~$E(\nu_P)$. The goal is to compute its leading behaviour in large-$N$ expansion.
 
For the~$S_{\lambda}(u)$ of a theory with a single~$SU(N)$ node and fields in the adjoint, let us say for~$\mathcal{N}=4$ SYM (if~$P$ is isolated),
\be\label{DetEqFinal}
\begin{split}
E(\nu_P)&\=\frac{1}{(2 \pi \i)^{N-1}}\, \underset{i,j}{\det} \,\frac{\partial}{\partial u^{j}} \,B^i(P)\,  \\
&\=\frac{1}{(2 \pi \i)^{N-1}}\, \underset{i,j}{\det} \,\frac{\partial}{\partial u_1^{j}}\, B^i(P)\,\\
&\=\frac{1}{(2 \pi \i)^{N-1}}\, \underset{i,j}{\det} \,\frac{\partial}{\partial u_1^{j}}\, \log B^i(P)\,.
\end{split}
\ee
In going to the second line we have used~\eqref{UVChange}. In going to the third line we have used~$B^i(P)=1$\,.

Basic algebraic manipulations show that the matrix elements can be arranged as follows
\be\label{DerL}
\partial_{u^j_{1}} \log B^{i}(u) \= \sum_\a \,\Bigl(\sum_{k\,=\,1\atop k\neq i}^{N-1}\,\mathcal{B}_{\a}(u_{j k})\, \delta_{ij} \,+\,\sum_{k\,=\,1\,\atop\, k\,\neq\, i \,\wedge\, k\, \neq\, j}^{N-1}\mathcal{B}_{ \a}(u_{N k})\Bigr)\,+\,\ldots\,.
\ee
For~$x=x_1+\t x_2$, with~$x_1$ and~$x_2$ being real, one obtains 
\be\label{DefLMath}
\begin{split}
\mathcal{B}_{\a}(x)&\,\equiv\,\frac{\partial }{\partial x_1} \Bigl(\log P_{0\Lambda}(x+\Delta_\a)\,-\,\log P_{0\Lambda}(-x+\Delta_\a)\Bigr)\, \\
&\=  \!-\,2 \i  
{\underset{m,n \in \IZ \atop m \neq 0}{\sum}} \; \Lambda^{|m|\,+\,|n|}\, \frac{\sin\Bigl(2\pi (n \Delta_{\alpha 2} - m \Delta_{\alpha 1})\Bigr)}{(m\t+n)}\, \cos\Bigl(2\pi (n x_2-m x_1)\Bigr)\,. 
\end{split}
\ee
The~$\ldots$ in~\eqref{DerL} stand for terms with single elements, namely without sum over~$N-1$. These terms are suppressed in the large-$N$ expansion. We write them for completeness
\be\label{ExtraTerms}
\ldots\=\, -\,\mathcal{B}_{\a}(u_{ij})\,|\epsilon_{ij}|\,+\,2\,\bigl(\mathcal{B}_{\a}(u_{N i})+\mathcal{B}_{\a}(u_{N j})\bigr) |\epsilon_{ij}|\,+\, 4\,\mathcal{B}_{\a}(u_{N i})\,\delta_{ij}\,,
\ee 
where~$1\leq i,\,j\leq N-1$.
At finite~$N$ these terms are relevant.~For instance at~$N=2$~the only non-vanishing contribution comes from the last term in~\eqref{ExtraTerms}.

At large~$N$ the~$(N-1)\times (N-1)$ matrix~\eqref{DerL}  takes the form
\be\label{Form}
D^{N-1} \times \begin{pmatrix} 
    2& 1 & \dots &1  \\
    \vdots &  & 2 & \vdots\\
    1 &     \ldots   & 1& 2 
    \end{pmatrix}\,,
\ee
with~$2$'s as diagonal elements and~$1$'s otherwise.  The matrix that multiplies~$D^{N-1}$ has determinant~$N-1$ and so the final form of~$E(\nu_P)$ has the following form at leading order in large-$N$ expansion
\be
(N-1) \, D^{N-1} \,\sim\, N \, D^{N-1}\,.
\ee
Let us show these statements and determine~$D$. The first contribution in~\eqref{DerL} is
\begin{equation*}
\label{EqPerS2}
\sum_{k\,=\,1\atop k\neq j}^{N-1}\,\mathcal{B}_{\a}(u_{j k})\,.
\end{equation*}
At large~$N$,~$\frac{j}{N}\to w\in [0,1)$ and~$\frac{k}{N}\to x\in[0,1)$ and the sums become integrals. In that limit~\eqref{EqPerS2} becomes
\be\label{EqSLambdaDeltas}
\begin{split}
N\,\int^1_0 dx  \mathcal{B}_{\a}((w-x) T)\=N\,\int^1_0 dx  \mathcal{B}_{\a}((x-w) T)\=N \,\int^1_0 dx \mathcal{B}_{\a}(x T)\,.
\end{split}
\ee
After fixing~$\Lambda\=1^-$, the last term in~\eqref{EqSLambdaDeltas} becomes
\be
N \,\int^1_0 dx \mathcal{B}_{\a}(x T)\= \frac{4\pi\i N}{ T}\,\Bigl(-T\,\Delta_{\a2}\,+\, [\Delta_\a]^T_m\,+\,\frac{1}{2} \Bigr)\,.
\ee
There is still a second contribution to analyze from~\eqref{DerL}. Evaluating it at fixed points, the sum becomes an integral at leading order in large-$N$ expansion
\be
\sum_{k\,=\,1\,\atop\, k\neq i \,\wedge\, k \neq j}^{N-1}\,\mathcal{B}_{\a}(u_{N k}) \longrightarrow N \,\int^1_0 dx \mathcal{B}_{\a}(x T)\,.
\ee
This integral equals the contribution from the first term in~\eqref{DerL}. Thus, we conclude that at leading order in large-$N$ expansion,~\eqref{DerL} takes the form~\eqref{Form}, with
\be
D\= \frac{4\pi\i N}{ T}\,\sum_{\a}\Bigl(-T\,\Delta_{\a2}\,+\, [\Delta_\a]^T_m\,+\,\frac{1}{2} \Bigr)\,.
\ee
The final result for~$(m,n)$ solutions is
\be\label{Determinant}
\underset{i,j}{\det}\,{\partial_{u_{j}} B^{i}}(P)\= N^N\,\Bigl(\frac{2}{ T} \,\sum_{\a}\Bigl(-T\,\Delta_{\a2}\,+\, [\Delta_\a]^T_m\,+\,\frac{1}{2} \Bigr)\Bigr)^{N-1}\,+\, \text{sub-leading}\,.
\ee
In the limit~$\Lambda\to 1^-$ the effective action~$S_\lambda$~\eqref{EffActionFlavour} develops discontinuities at walls defined by the condition~$m\Delta_1\,-\,n \Delta_2\in \mathbb{Z}\,$.  Dirichlet theorem fixes the value of $S_{\text{eff}\a}(m,n)$ and its derivatives~$\partial^\ell \,S_{\text{eff}\a}(m,n)$, at such discontinuities to
\be\label{DirichletTheoEq}
\frac{(\partial^\ell\,S_{\text{eff}\a}(m,n))^{L}\,+\,(\partial^\ell\, S_{\text{eff}\a}(m,n))^{R}}{2}\,,
\ee 
where~$L$ and~$R$ denote the limit values from the left and right sides of the corresponding discontinuity of~$\partial^\ell\,S_{\text{eff}\a}(m,n)\,$. 

To compare with other literature, let us analyze the previously introduced case of~$\mathcal{N}=4$ SYM. Focus on the saddle/fixed point~$(1,0)$ i.e~on~$T=\t$, then
\be\label{EqDelta10}
\sum_{\a}\,\Bigl(-T\,\Delta_{\a2}\,+\, [\Delta_\a]^\t_1\,+\,\frac{1}{2} \Bigr) \= -\,\frac{1}{2}\,.
\ee
We recall that the function bracket was defined in~\eqref{bracketDef}. For a generic gauge anomaly-free theory~$\sum_{\a}\Delta_{\a2}\=\text{Tr}R$ where~${R}$ denotes the superconformal R-charge of the fermionic fields in a given multiplet and~Tr denotes sum over all multiplets. This quantity vanishes at large~$N$ in virtue of ABJ anomaly cancellation~\cite{Cabo-Bizet:2020nkr}. For~$\mathcal{N}=4$ SYM this quantity vanishes at any value of~$N$.

Finally, we use~\eqref{EqDelta10} in the logarithm of~\eqref{Determinant} and obtain at leading order
\be
\log E(\nu_{(1,0)})\= N \,\log N\,-\,(N-1)\log(\t) \,+\,\text{sub-leading}\,.
\ee
This answer matches a recent result obtained for the fixed point~$(1,0)$ with a combination of a small~$\t$ expansion and numerical extrapolation, via the use of the Bethe Ansatz formula~\cite{GonzalezLezcano:2020yeb}.

\section{Towards an analytic approach to large-$N$ counting} \label{sec:Counting}

In this section we combine the ABBV formula with Picard-Lefschetz method to study how complex eigenvalue configurations compete at leading order in the large-$N$ expansion of the microcanonical index. We will use as toy example the case of~$\mathcal{N}=4$ SYM with~$\sigma=\tau$ and no flavour potentials. The generalization to the cases~$\sigma\neq\tau$ is explained in appendix~\ref{ExtRegul}. The outline of this section was given in the introduction section~\ref{sec:Intro}. Here we will focus on the details.

\subsection{The contour of integration}\label{sec:Contour}

We focus on the simplest ensemble of~$SU(N)$~$\mathcal{N}=4$~SYM that captures the graded counting of~$\frac{1}{16}$-th~BPS~operators. That is the family of integrals~\eqref{N4SYM} with~$t_{(1)}=t_{(2)}=t_{(3)}=t q$ and~$p=q$ and the constraint~\eqref{constraintSUSY}~$t^3= q^{-1}\,$. This is a function of a single chemical potential.

\begin{figure}\centering
\includegraphics[width=10.cm]{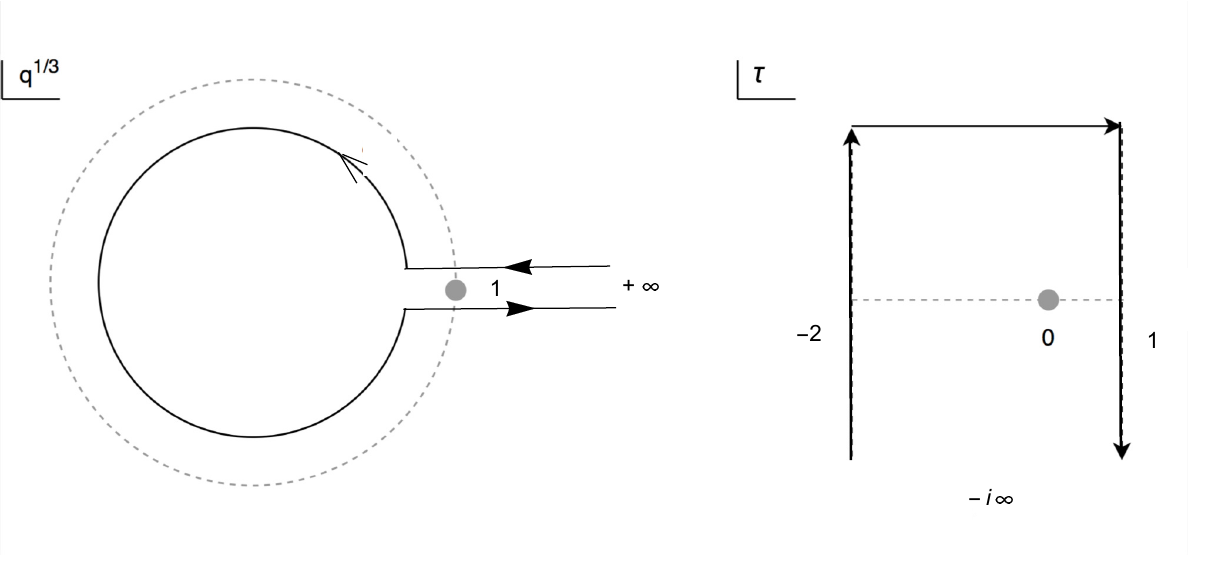} \qquad\qquad
 \caption{ The contour~$C_\eta$ in~$q^{\frac{1}{3}}$ and~$\tau$-planes. The original contour in~$q$-plane is a circle inside the unit disk~$|q|<1$. The grey dashed circle (segment) represents~$|q|=1$~($\text{Im}(\t)=0$). The integration along the two vertical contours in~$\tau$-plane cancel each other due to the periodicity properties of the index.  Due to reasons explained in~\ref{par:HolomorphicExtension} they must cross the real axis across irrational values. Thus, the integral along~$C_\eta$ equals the integral along the original contour~$C$.  }
  \label{fig:Countour}
\end{figure}

 For finite~$N$, the integral~$\mathcal{I}$~\eqref{N4SYM} can be computed order by order in a Taylor expansion around~$p=q=0$. The coefficients~$a_n$ in such expansion can be computed as a contour integral over the Cartan torus~\cite{Murthy:2020rbd,Agarwal:2020zwm} or via decomposition in characters of~$SU(N)$~\cite{Murthy:2020rbd,Dolan:2007rq}. As the R-charges are quantized in multiples of~$\frac{1}{3}$, the variable with the correct monodromy properties is~$q^{1/3}$. The Fourier coefficients can then be extracted from the~$q^{\frac{1}{3}}$-series, via the Laurent integral
\be\label{FourierCoeff}
a(\ell)\= \oint_C \,{d q^{\frac{1}{3}}} \,\frac{\mathcal{I}(q^{\frac{1}{3}})\,}{\,q^{\frac{\ell+1}{3}}\,}\,,\qquad j\in \mathbb{Z}\,.
\ee
The contour~$C$ is a closed contour surrounded by~$|q|=1$. In terms of~$\t$ the contour~$C$ is a segment that for convenience we take to run in between the vertical lines~$\tau_1=-2$ and~$\tau_1=1$. As~$\mathcal{I}(\t)\=\mathcal{I}(\t+3)$ we can add to~$C$ two vertical contours running in opposite sense from (resp. to)~$\t_2=-\eta$ to (resp. from) the left (resp. right) extremum of~$C$. The positive number~$\eta$ is assumed to be as large as wished, see Figure~\ref{fig:Countour}. On the contrary to~$C$,~$C_{\eta}$ has a natural decomposition in terms of Lefschetz thimbles. We will explain this later on. As explained in appendix~\ref{par:HolomorphicExtension} the vertical pieces of~$C_{\eta}$ must be slightly deformed (while preserving their mutual cancellation) in such a way they cross the real axis across two irrational values of~$\t$ at distance~$3$ of each other.

Plugging the ABBV formula~\eqref{RepNewInd} into~\eqref{FourierCoeff}, changing integration variables from~$q^{\frac{1}{3}}\mapsto \t$, using the fact that the integral of~$\mathcal{I}(\t)$ over~$C$ is equal to the integral over~$C_{\eta}$, and finally commuting the sum over~$P$'s with the integral over~$C_{\eta}$~\footnote{ If we consider only~$(m,n)$ saddles, as we are doing, then the number of terms in the sum over fixed points, is less than~$N^2$. Thus for large but finite~$N$, if the integrals of the summands is finite, then the integral can be commuted with the sum.} one obtains the microcanonical index as a sum over fixed points~$P$
\be\label{degeneracy}
\begin{split}
d(\mathfrak{q})\=\sum_{P}\,d_P\,,
\end{split}
\ee
where~$d(\mathfrak{q}) \equiv a(\ell+1)$ and~$\mathfrak{q}\= \frac{{\El}}{N^2} \equiv \frac{\ell+1}{3 N^2} >0$ and the contributions of single fixed point~$P$,~$d_P$,
are defined as
\be\label{IP}
{d}_P\,\equiv\, \int_{C_{\eta}} d\t\, \frac{e^{-S_\lambda(P)\,-\, 2\pi \i \t \El}}{E(\nu_P)}\=\int_{C_{\eta}} d\t\, e^{-S_\lambda(P)\,-\, 2\pi \i \t {\El}\,+\,\ldots}\,.
\ee
The~$\ldots$ denote sub-leading contributions in the large-$N$ expansion. 
The goal of this section is to study the leading in~$N$ asymptotics of the full sum by comparing the individual contributions of the~$d_P$. 

\subsection{Picard-Lefschetz method}

In this section we use Picard-Lefschetz method to analyze the integrals at large~$N\,$. Our approach follows the presentation of~\cite{Witten:2010cx}. The large-$N$ limit of integrals~$d_P$ is determined by the critical points of
\be\label{EqEntropyFunc}
\mathcal{E}_{P}(\t)\,\equiv\,-\,S(P) \,-\,2\pi \i \t {\El} \,.
\ee
This function is analytic in the region
\be\label{Ci}
\mathcal{X}_P\,\equiv\,\{\t\,:\, \t\in \mathbb{C}-\t_P\}\,,
\ee
and the point~$\tau_P$ is an essential singularity. The behaviour of~$\mathcal{E}_P$ in a limit to~$\t_P$ depends on how the limit is taken. For the example that we are considering,~$\t_P$ is always a rational number.

We need to study real one-dimensional curves in the space~$\mathcal{X}_P$, say~$\t\=\t(s)\in \mathcal{X}_P\,$, with worldline parameter~$s\in(0,\infty)$.
They must end or start at a non-degenerate critical~$\tau_c$ of~$\mathcal{E}_{P}$ i.e. they must satisfy the asymptotic initial condition
\be\label{InitialCond}
\tau_c^+\,\equiv\,\tau(0^+)\=\tau_c\quad \text{or} \quad\tau^{-}_c\=\tau(\infty)\=\tau_c\,.
\ee
Moreover, they should preserve the imaginary part of~$\mathcal{E}_P$ along the trajectory, and thus are determined by the differential equation
\be\label{ImConserv}
\partial_s \, I_P(\t(s))\=0\,,\qquad I_P(\t)\,\equiv\, \text{Im}\mathcal{E}_{P}(\t)\,,
\ee
subject to one of the initial conditions in~\eqref{InitialCond}. If the Morse function
\be
H_{P}(\t) \=\text{Re}\mathcal{E}_{P}(\t)
\ee 
is strictly decreasing along the flow i.e. 
\be\label{decrMorse}
\partial_s H_{P}(\t(s))<0\,,
\ee
then such flow~$\mathcal{J}$ is called a Lefschetz thimble and it satisfies the first of the conditions in equation~\eqref{InitialCond}. A more refined way of imposing condition~\eqref{decrMorse} upon solutions to~\eqref{ImConserv} is instead, to demand the flow to be a solution of the so-called~\emph{downward flow equations}~\cite{Witten:2010cx} with the initial condition~\eqref{InitialCond}.  In the end, every thimble~$\mathcal{J}$ obtained directly from~\emph{downward flow equations} can be obtained from restricting the space of  solutions of~\eqref{ImConserv} to the ones satisfying~\eqref{decrMorse}, and~\emph{viceversa}.

Instead of solving the flow equations, it is convenient to solve the algebraic relations in between imaginary and real part of the complex curves~$\tau(s)$ that arise from demanding the imaginary part of the entropy function, $I_P(\tau(s))$, to coincide with the imaginary part of $I_P(\tau_c))$ of the fixed point~$\tau_c$ of interest. The problem then reduces to that of solving algebraic equations. Out of those solutions one then needs to understand which ones are ascent or descent paths. Below we explain how that is done in this particular example.

The next step is to decompose the contour~$C_{\eta}$ in Figure~\ref{fig:Countour} into an integral combination of Lefschetz thimbles~$\mathcal{J}_P$
\be\label{decomposition}
C_{\eta}\= \sum_{\mathcal{J}_{P}} \,n_{\langle C_{\eta},\mathcal{J}_{P} \rangle}\, \mathcal{J}_P\,, \qquad n_{\mathcal{J}_P}\in \mathbb{Z}\,.
\ee
That this decomposition exists can be understood in two ways. One homological, and the other uses meromorphy of~$\mathcal{E}_P$. In the example we study, the latter turns out to be convenient to obtain the decomposition in thimbles~\eqref{decomposition}. Let us briefly comment on why the former turns out to be more involved in this case. This requires to recall the concept of relative homology. Again, we follow the presentation of~\cite{Witten:2010cx}.  

The thimbles~$\mathcal{J}_P$ are a basis of the relative homology group~$H_1(\mathcal{X}_P,\mathcal{X}_{P_{-T}},\mathbb{Z})$. This is the group of one-real dimensional paths that originate and end at the given connected piece of ``boundary" of~$\mathcal{X}_P$ (the space defined in~\eqref{Ci}). One-cycles that can be contracted to a point without crossing any obstruction are identified as trivial. We will see an instance of this in the next subsection. By one-cycles one means the following. Select a very large positive real number $T$, and excise from~$\mathcal{X}_P$ all the regions at which the Morse function~$H_P<-T$. Then, identify each connected subpart of such region as a ``point" and define the new punctured domain as~$\mathcal{X}_{P_{-T}}$. The one-cycles are the one-curves in $\mathcal{X}_P$ that start and end at the ``same" puncture of~$\mathcal{X}_{P_{-T}}$. Those cycles can be expanded as an integral combination of Lefschetz-thimbles~$\mathcal{J}_P$. Usually the integers are computed as intersection numbers between the corresponding one-cycle and the dual ascent paths~$\mathcal{K}_P$.~\footnote{Ascent paths are one-dimensional curves that solve the flow equation~\eqref{ImConserv}. They are also asymptotically close to a critical point, but this time they obey the opposite condition to~\eqref{decrMorse} i.e~they flow towards the critical point. Equivalently, they solve the so-called~\emph{upward flow equations}~\cite{Witten:2010cx}. } However, that method needs to be refined in the presence of Stokes' lines i.e~when there exists a flow running among two different critical points. In that case a thimble can be an ascent path as well, and one needs to slightly deform~$\mathcal{E}_P$ to solve this issue (See~\cite{Witten:2010cx} for a detailed discussion on this, and many other related issues).

The present case ``sits at" a Stokes line i.e.~there is a thimble~$\mathcal{J}_P$ that it is also ascent path~$\mathcal{K}_P$. That happens for every~$P$. The reason is that for every~$P$ there are two critical points of~$\mathcal{E}_P$ with the same imaginary part~$I_{P}$. Using meromorphy this will turn out not to be an issue to compute the intersection numbers.

Using the contour decomposition~\eqref{decomposition} upon the integrals~$d_P$ defined in~\eqref{IP} one obtains
\be
d_P \= \sum_{\mathcal{J}_P} n_{\langle C_{\eta},\mathcal{J}_{P} \rangle} \,\int_{\mathcal{J}_P}\,d\t\, e^{\mathcal{E}_P(\t)\,+\,\ldots\,}  \,.
\ee
We will see that for many~$P$'s the intersection numbers~$n_{\langle \ldots\rangle}$ vanish. But many others could contribute. The~$\ldots$ denote sub-leading contributions in the large-$N$ expansion.

\subsection{The competition among~$(m,n)$ fixed points} \label{sec:CompetitionFixedPoint}

Let~$N$ be a large integer, that we assume to be prime.~\footnote{This implies the~$(m,n)$ configurations here-analyzed to be the only fixed points carrying finite Abelian group structure of rank~$N$~\cite{Cabo-Bizet:2019eaf,Cabo-Bizet:2020nkr}. } The fixed points~$P$ are the~$(m,n)$'s configurations~\eqref{Psaddles}. The on-shell action of the~$(m,n)$ fixed point with~$0\leq m, n< N$ integer co-primes and~$(m,n)\neq(0,0)$, was computed in~\cite{Cabo-Bizet:2019eaf}
\be\label{SeffFFPaper}
S(m,n)\equiv S_{\text{eff}}(m,n;\t)\= \frac{1}{m}\, \frac{\pi\i}{27}\, N^2 \, \frac{\bigl(2T+\chi_1(m+n)\bigr)^3}{T^2}
    + N^2 \pi\i \,\varphi(m,n)\,.
\ee
In this formula~$T=m\tau+n$ and $\chi_1(\ell)=\{-1,0,1\}$ if~$\ell=\{-1,0,1\}$ mod $3$ respectively. From now on we focus on the cases~$\chi_1 \,\neq\, 0$ i.e.~$m\,+\,n \,\neq\, 0$ mod~$3$. Incorporating the cases~$\chi_1\=0$ will not affect the conclusions that will be presented below.
It will be useful to note the following conjugation relation
\be\label{ReflProperty}
S_{\text{eff}}(1,0; \t_{R})\=S_{\text{eff}}(1,1;\t)^*\,
\ee
where~$\t_R\=-\,1\,-\,\t_1\,+\,\i\, \t_2\,$.
For a given $(m,n)$ we must extremise the functional \eqref{EqEntropyFunc}
\be\label{Epsilon}
\mathcal{E}_{(m,n)}(\tau)\=- S_{\text{eff}}(m,n;\tau)-2 \pi \i \,\tau {\El}\,.
\ee
The extremization of this functional was analyzed in some detail in~\cite{Cabo-Bizet:2019eaf}. Here we will complete the analysis by following the Picard-Lefschetz method.

For any~$(m,n)$ there are three critical points in~$\mathcal{X}_{(m,n)}$. Let us denote them as
\be\label{criticalPoints}
\tau^*_+\,, \qquad  \tau^*_-\,,\qquad \tau^*_0\,.
\ee
We have not written down their explicit dependence on~$(m,n)$, but they do depend on~$m$ and~$n$~\cite{Cabo-Bizet:2019eaf}.
For the critical points~$\tau^*_{\pm}$ the Morse function~$H_{(m,n)}$ has the same absolute value and opposite sign: say plus for~$\t^*_+$ and minus for~$\t^*_-$. For the critical point~$\tau^*_0$
\be
H_{(m,n)}(\t^*_0)\=0\,.
\ee
The critical points~$\tau^*_{\pm}$ have equal phase function~$I_{(m,n)}$. Indeed for all~$(m,n)$ there will always be a descent/ascent flow from~$\t^*_+$ to~$\t^*_-\,$.

Let us start by studying the thimbles associated to the fixed point~$(1,0)$. For the critical point~$\t^*_+$ there are two thimbles, one flows to~$\t_2=-\infty$ and the other flows to~$\t\=0$. The point~$\tau\=0$ is the essential singularity of~$\mathcal{E}_{(1,0)}$, namely the difference between~$\mathcal{X}_{(1,0)}$ and~$\mathbb{C}$, as defined previously in equation~\eqref{Ci}. For a limit that approaches~$\tau=0$ from the first and third quadrants,~$H_{(1,0)}$ goes to~$-\infty$. For a limit that approaches~$\tau=0$ from the second and fourth quadrants,~$H_{(1,0)}$ goes to~$+\infty$. This means that Lefschetz thimbles of~$(1,0)$ that flow to~$\tau=0$, must approach it from the first and third quadrants, meanwhile ascent paths emanating from~$\tau=0$ must do it through the second and/or fourth quadrants. Indeed this is what happens.

 For~$(1,1)$ the pattern is analogous, but this time the essential singularity is located at~$\t\=-1$ and now the first (second) and third (fourth) quadrants take the roles of the second (first) and the fourth (third) had in the previous case, respectively. For a generic~$(m,n)$ fixed point, the story is also the same. Now the essential singularity is located at~$\tau=-\frac{n}{m}$ and the two opposite quadrants of that singularity through which the thimbles can enter, are determined by the sign of~$\chi_1(m+n)$. If~$\chi_1(m+n)=+1$ then thimbles should approach it through the first and third quadrants, otherwise through the second and fourth. The other region, apart from the essential singularity~$\tau=-\frac{n}{m}$, to which the thimbles can flow towards is the asymptotic region~$\t\=-\i\infty$. Namely in this region~$H_{(m,n)}$ goes to~$-\infty$. This is the same region where our integration cycle starts and ends at. 
 
 \begin{figure}\centering
\includegraphics[width=5.8cm]{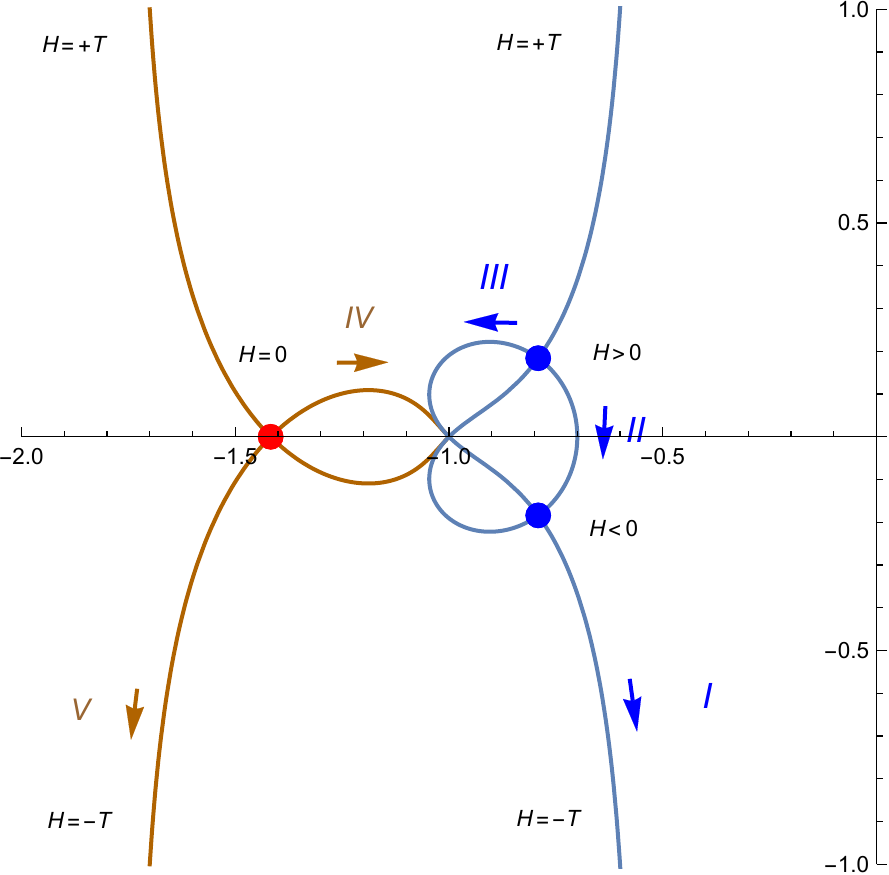}\qquad\qquad
\includegraphics[width=5.8cm]{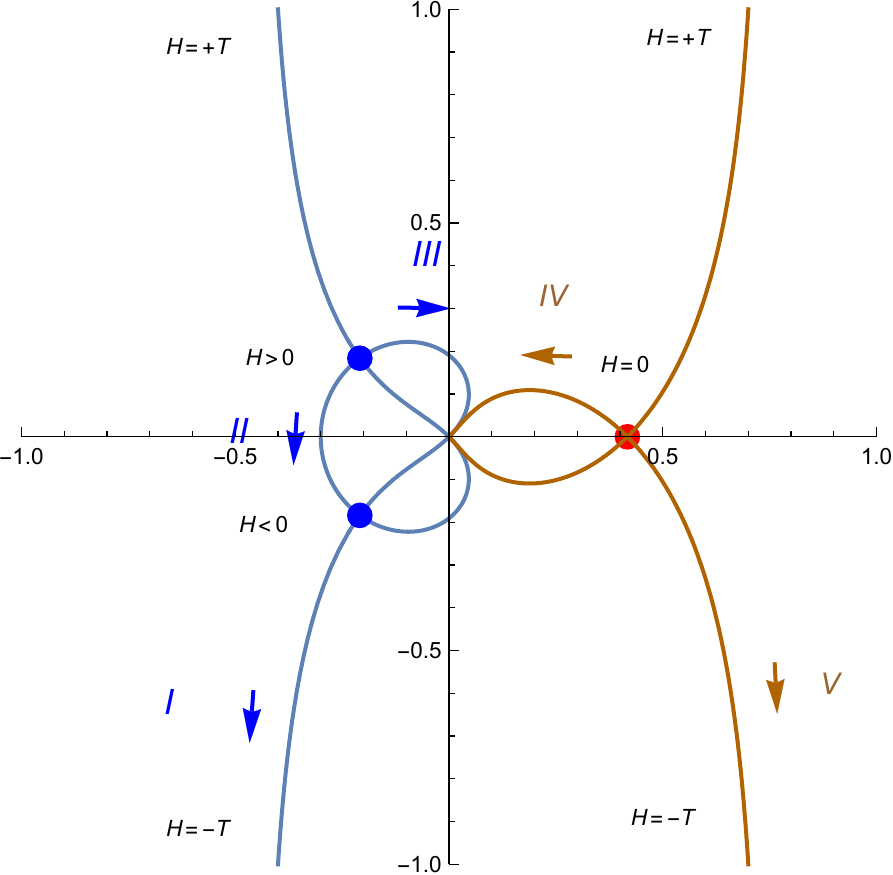} 
 \caption{ The plots of the analytic solutions (of the relevant algebraic equations) obtained for the Lefschetz thimbles and ascent paths associated to the~$(1,1)$ (left)  and~$(1,0)$ (right) fixed points. The solution for generic~$(m,n)$ has the same form as one of the two plots above. Each critical point of the entropy functional~$\mathcal{E}_{(m,n)}$ in~\eqref{Epsilon}, has two thimbles and two ascent paths associated. In the figure we have only indicated the direction of the flows corresponding to the thimbles that appear (have non-zero intersection number) in the decomposition of the original integration contour~$C_{\eta}$ which we have denoted as~$I$,~$II$,~$III$,~$IV$,~$V$.  The different~$H$'s in the plots denote the values of the Morse function at the closest critical point and also at~$\tau_2\approx\pm \infty$ where~$H=\pm T$, respectively, with~$T$ a very large positive number.   }
  \label{fig:Thimbles}
\end{figure}

To know whether the intersection numbers~$n_{\langle \ldots \rangle}$ that determine the contribution of fixed points~$(m,n)$ to the integral along~$C_{\eta}$ vanish or not, we need to know whether the cycle~$C_{\eta}$ is trivial in the relative homology~$H_{1}(\mathcal{X}_{(m,n)},\mathcal{X}_{(m,n)_{-T}};\mathbb{Z})\,$. That means to check whether the~$C_{\eta}$ encloses the essential singularity~$\tau=-\frac{n}{m}$. If the cycle~$C_{\eta}$ does not enclose~$\tau=-\frac{n}{m}$ the integral~$\mathcal{I}_{(m,n)}$ vanishes. It vanishes because the integrand is meromorphic and the contour can be contracted to a point at~$\t=-\i\infty$ (where the integrand vanishes).  If on the contrary the cycle~$C_{\eta}$ does enclose~$\tau=-\frac{n}{m}$ then the fixed point~$(m,n)$ does contribute and the precise intersection numbers can be computed easily.

Given our choice of contour~$C_{\eta}$, only fixed points for which
\be\label{IntersectionDomain}
-2<-\,\frac{n}{m}\,<\,1\,,
\ee
contribute to the entropy counting formula~\eqref{degeneracy}. Take an~$(m,n)$ that obeys this condition. The symmetry transformation~$\tau\mapsto \tau+3$ maps~$(m,n)\mapsto(m,n+3m)$. Naively, these two saddles should contribute with the same exponential weight, namely~$\mathcal{E}_{m,n}\=\mathcal{E}_{m,n+3m}$, however~$(m,n+3m)$ does not satisfy~\eqref{IntersectionDomain} and thus its intersection numbers vanish. Thus, these~\emph{replica} saddles do not contribute to the large-$N$ microcanonical index defined by the contour integral $C_{\eta}$.

In figure~\ref{fig:Countour} we have plotted the analytic solutions of all the three thimbles and three ascent paths, together with the corresponding critical points, for fixed points~$(1,0)$ and~$(1,1)$. In both cases the contour~$C_{\eta}$ can be explicitly checked to have intersection number equal to either~$1$,~$-1$ or~$0$ associated to thimbles~$\mathcal{J}_{(1,0)}$ and~$\mathcal{J}_{(1,1)}$ i.e.
\be
\begin{split}
C_{\eta}&\,\= \, -\,\mathcal{J}_{(1,0){I}}\,-\,\mathcal{J}_{(1,0){II}}\,+\,\mathcal{J}_{(1,0){III}}\,-\, \mathcal{J}_{(1,0){IV}} \,+\, \mathcal{J}_{(1,0){V}}\,,
\\
 &\,\=\, \quad \,\mathcal{J}_{(1,1){I}}\,+\,\mathcal{J}_{(1,1){II}}\,-\,\mathcal{J}_{(1,1){III}}\,+\, \mathcal{J}_{(1,1){IV}} \,-\, \mathcal{J}_{(1,1){V}}\,.
\end{split}
\ee
This decomposition can be obtained by simply comparing the contour~$C_\eta$ with the thimbles in figure~\ref{fig:Thimbles}.
 \begin{figure}\centering
\includegraphics[width=6.cm]{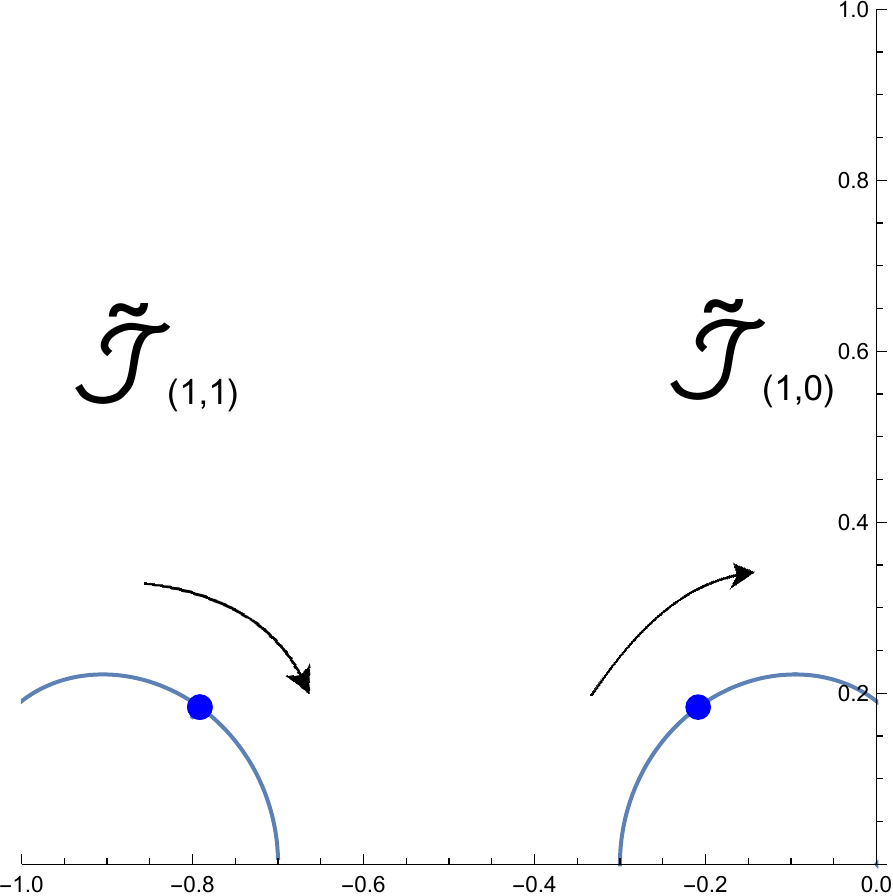} 
 \caption{ For large enough~$N$ we can focus on combinations of subregions of the dominating thimbles that include~$\t^*_+$. This is the plot of two such combinations, one of them associated to~$(1,1)$ (to the left), the other to~$(1,0)$ (to the right). These two combinations transform into each other under the~$\mathbb{Z}_2$ operation~$\tau\mapsto \tau^R$. The arrows denote the sense in which the original contour~$C_{\eta}$ flows. For convenience we define~$\widetilde{\mathcal{J}}_{(1,0)}$ and~$\widetilde{\mathcal{J}}_{(1,1)}$ to flow in the same sense~$C_{\eta}$ does.  }
  \label{fig:CountourJtildes}
\end{figure}

Integrals along these contours can be computed exactly with all the data we have. However, in this paper we are interested in the leading approximation at large~$N$ so it will be enough to identify the contribution of the leading critical point. That is always the~$\t^*_+$, thus, for instance, to analyze the contribution of the fixed points~$(1,0)$ and~$(1,1)$ we can focus on the following two portions of thimble
\be
\begin{split}
C_{\eta}\,\sim \widetilde{\mathcal{J}}_{(1,0)}\,+\, \ldots \,,\\
C_{\eta}\,\sim \widetilde{\mathcal{J}}_{(1,1)}\,+\, \ldots \,.
\end{split}
\ee
A computation shows that~$\widetilde{\mathcal{J}}_{(1,0)}$ and~$\widetilde{\mathcal{J}}_{(1,1)}$ transform into each other under the~$\mathbb{Z}_2$ operation
\be
(\tau_1,\tau_2(\tau_1))  \mapsto (\tau^R_1,\tau_2(\tau^R_1))\,.
\ee
Moreover, from property~\eqref{ReflProperty}, definition~\eqref{Epsilon} and the fact~${\El}\in\mathbb{R}$, it follows that
\be
\int_{\widetilde{\mathcal{J}}_{(1,1)}} d\tau_1 \, e^{\mathcal{E}_{(1,1)}(\t(\t_1))}\= \int_{\widetilde{\mathcal{J}}_{(1,0)}} d\tau_1 \, e^{\bigl(\mathcal{E}_{(1,0)}(\t(\t_1))\bigr)^*}\,.
\ee
This relation implies that contributions to the large-$N$ microcanonical index that come from~$(1,0)$ and~$(1,1)$, are complex conjugated to each other. We expect that this property will be preserved by sub-leading corrections in the large-$N$ expansion as well. That is because by direct evaluation one can check that the Fourier coefficients~$a_n$ are always real integers.

As explained before, generic~$(m,n)$ fixed points have analogous thimble structure. Thus, we must compare the contribution coming from thimbles associated to generic~$(m,n)$ fixed points that obey the condition~\eqref{IntersectionDomain}. For those~$(m,n)$, the intersection number~$n_{\langle C_{\eta},\ldots \rangle}\,$ with the~\emph{leading} thimble will always have absolute value one. Thus, the contribution of those fixed points is determined by the value of the Morse function at~$\t^*_+$ which is 
\be\label{MorseFMN}
\begin{split}
H_{(m)}(\mathfrak{q})\,=\,H_{(m,n)}(\t^*_+)&\,\equiv\,\frac{N^2}{m}\,\frac{\pi  \sqrt{(2 T^{*}+\,1)^3\, (1\,-\,6 T^*)}}{24 \sqrt{3}  T^{*2}}\,, \\
T^*&\,\equiv\,-\frac{54\, \mathfrak{q}\,+\,\sqrt[3]{2}\, D^2\,+\,8}{2\, 2^{2/3}\, (27\, \mathfrak{q}\,+\,4)\, D}\,,\\ D&\,\equiv\, \sqrt[3]{(27 \mathfrak{q}+4)^2+3 \sqrt{3} \sqrt{\mathfrak{q} (27 \mathfrak{q}+4)^3}}\,.
\end{split}
\ee
The Morse function is inversely proportional to~$m$ and it does not depend on~$n$. At a given level~$m$ there are only a finite number of~$n$'s with~$|\chi(m+n)|=1$ in the intersection domain~\eqref{IntersectionDomain}. For~$m=1$ there are two,~$n=0$ and~$n=1$. These are the two dominating contributions at large~$N$, and for charges of order~$N^2$. As explained above, both have the same leading exponential contribution in absolute value, but their phases are complex conjugated to each other and are given by the imaginary part
\be
I_{(1,0)}\= -\,I_{(1,1)}\=-\,\frac{\pi\,N^2\,  (2 T^*+1) (10 T^*-1)}{72 T^{*2}}\,+\,\frac{\pi N^2}{2}.
\ee
~\footnote{For large~$\mathfrak{q}$ after large-$N$
\be
\frac{H_{(1,0)}}{N^2}\sim\frac{1}{2} \sqrt{3} \pi  \mathfrak{q}^{2/3}-\frac{\pi  \sqrt[3]{\mathfrak{q}}}{\sqrt{3}}+O\left(\sqrt[3]{\frac{1}{\mathfrak{q}}}\right)\,,\qquad \frac{I_{(1,0)}}{N^2}\sim \frac{1}{2} \pi  \mathfrak{q}^{2/3}+\frac{\pi  \sqrt[3]{\mathfrak{q}}}{3}+\frac{\pi
   }{6}+O\left(\sqrt[3]{\frac{1}{\mathfrak{q}}}\right)\,.
\ee 
We should note that in terms of the charge~$\El$ of CFT operators,~$H_{(1,0)}\sim N^{\frac{2}{3}}\El^{\frac{2}{3}}$.}~If we discard sub-leading corrections coming from~$E(\nu_P)$,~\footnote{These were analyzed in subsection~\ref{Enp}.} and exponentially suppressed contributions coming from other fixed points, we obtain
\be\label{DegeneracyQ}
d(\mathfrak{q})\= e^{H_{(1)}(\mathfrak{q})\,+\,\ldots}\,\times 2 \cos{\bigl(I_{(1,0)}(\mathfrak{q})\,+\,\ldots\bigr)}\,+\,\ldots\,.
\ee
where we should stress that other fixed point configurations~$(m,n)$ also interfere to produce extra oscillations on top of~\eqref{DegeneracyQ}. However, at finite but relatively large values of~$N$ and at finite~$\mathfrak{q}$, these are exponentially suppressed with respect to the contributions coming from the pair~$(1,0)$ and~$(1,1)$, as one can note immediately from equation~\eqref{MorseFMN}. Moreover, at very first leading order in large-$N$ expansion all oscillations attenuate and the microcanonical index matches the exponential of the Bekenstein-Hawking entropy of the dual black holes (See figure~\ref{fig:OscillationsVsN})
\be
d(\mathfrak{q})\,\sim \,e^{H_{(1)}(\mathfrak{q})}\= e^{\frac{A_{H}}{4}}\,.
\ee
Here~$A_H$ is the area of the horizon of the Gutowski-Reall black hole.~\footnote{We have absorbed the~$5d$ Newton constant~$G_{5D}$, which is to be identified as proportional to~$\frac{1}{N^2}$, in the definition of~$A_H$.}

For small enough values of~$\mathfrak{q}$ after the large-$N$ limit is taken, the physical picture changes. In that case the relevant expansion is
\be
\frac{H_{(1,0)}}{N^2}\sim\pi  \mathfrak{q}^{3/2}\,-\,\frac{21}{8} \pi  \mathfrak{q}^{5/2}+O\left(\mathfrak{q}^{7/2}\right)\,,\qquad \frac{I_{(1,0)}}{N^2}\,\sim\, \frac{\pi }{2}\,+\,\pi  \mathfrak{q}\,-\,\frac{3 \pi  \mathfrak{q}^2}{2}\,+\,5 \pi  \mathfrak{q}^3\,+\,O\left(\mathfrak{q}^{7/2}\right)\,.
\ee
In terms of the charge~$\El$ of the CFT operators,~$H_{(1,0)}\sim N^{-1}\El^{\frac{3}{2}}$. Roughly speaking, this implies that to count operators with charges of the order~$E\sim N^{\frac{2}{3}}$ or smaller, the competition of other~$(m,n)$ saddles can not be neglected. Moreover, other critical points of the entropy functional~$\mathcal{E}_{(m,n)}$ in~\eqref{Epsilon}, mainly the real one (out of the triad of them), should also be analyzed in the regime of small charges~(at large $N$).~\footnote{Interestingly, note that for small enough values of~$\mathfrak{q}$, at large~$N$, the interference among~$(1,0)$ and~$(1,1)$ is destructive i.e~they tend to cancel each other.}

Let us summarize in words what we have learnt. The parameter that controls the magnitude of the exponential corrections coming from fixed points is~$\exp[-A_H/4]$ i.e. the inverse of the exponential of the entropy of the Gutowski-Reall black hole.  Generically, this parameter is small for large enough~$N$, meaning by this, values of~$N$ such as the discrete sums over eigenvalues are well approximated by the integrals.~However, for the two-fixed point approximation to work, at relatively large but still finite values of~$N$, operator charges need to be large enough in units of the specific values~$N$, in such a way that the perturbation parameter~$\exp[-A_H/4]$ remains small. That means that for small enough operators, the two-fixed point approximation breaks down and one needs to start considering many more configurations to approximate the correct micro-canonical index.~This means that when one approaches the transition from the black hole phase to the multigraviton phase, the ``two- fixed-point" approximation breaks down and competition from all~$(m,n)$ saddles must be considered. This observation deserves deeper analysis and exploration. The ABBV and the Bethe ansatz~\emph{formulae} are tools that could be useful to make this analytic picture precise at the numerical level (for instance at large~$N$ but for small operators).

\section*{Acknowledgements}
We thank Cyril Closset and Heeyeon Kim for a useful discussion regarding regularization procedures.~We are grateful to Davide Cassani and Dario Martelli for useful comments and discussions.~We are especially grateful to Sameer Murthy for his careful reading of the manuscript, as well as for many useful comments and discussions.~This work is supported by the ERC Consolidator Grant N. 681908, “Quantum black holes: A microscopic window into the microstructure of gravity”.

\appendix

\section{Definitions and identities} \label{App:Def}

The quasi-elliptic function~$\th_0(z)=\th_0(z;\t)$ has the following product representation
\be\label{productTheta}
\th_{\text{ell}}(\z;q)\=\theta_0(z;\t) \,=\,  (1-\z) \prod_{j=1}^\infty (1-q^j \z) \, (1-q^j \z^{-1}) \,.
\ee
The zeroes of~$\theta_0$ are located at~$z = j \t+k $ where~$j$ and~$k$ run over the integers. Representation~\eqref{productTheta} is convergent for~$\t_2>0$. 
It has also the following quasi-periodicity properties
\be\label{periodth0}
\th_0(z) \= \th_0(z+1) \= -\e(z) \, \th_0(z+\t) \,\quad\,\theta_0(\tau-z)\=\theta_0(z)\,.
\ee
The elliptic Gamma functions are defined out of the following product representation
\be\label{GammaeDef}
\G_{\text{ell}}(\z;p,q)\=\Ge(z;\s,\t) \=  \prod_{j,\,k=0}^{\infty}
\frac{1\,-\, \zeta^{-1} p^{j+1} q^{j+1}}{1\,-\,\zeta \,p^j\, q^k} \,.
\ee 
The poles and zeroes of~$\Ge$ are located at~$z= - j \s  - k \t + \ell$ and~$z= (j+1) \s  -(k+1) \t + \ell$ respectively  where the indices~$j$,~$k$ run over the non-negative integers and $\ell$ runs over the integers~\cite{Felder2000}. It has also the following quasi-periodicity properties
\be
\begin{split}
\Ge(z+\s;\s,\t)&\= \theta_0(z;\t)\,\Ge(z;\s,\t)\,,\\
\Ge(z+\t;\s,\t)&\= \theta_0(z;\s)\,\Ge(z;\s,\t)\,.
\end{split}
\ee
The first three Bernoulli polynomials are
 \bea
 B_1(z) &\,\equiv\,& z-\frac{1}{2}\,,\\
 B_2(z) &\,\equiv\,& z^2-z+\frac{1}{6}\,,\\
 B_3(z) &\,\equiv\,& z^3-\frac{3 \,z^2}{2}+\frac{z}{2}\,.
 \eea
Their Fourier series decomposition, for~$k> 1$ and~$0 \le z<1$ is 
\be\label{fourierexp}
B_k(z) \= - \frac{k!}{(2\pi \i)^k} \, \sum_{j \neq 0} \frac{\e(j z)}{j^k} \,.
\ee
The~$k=1$ is peculiar in the sense that~\eqref{fourierexp} is almost fine, but fails to hold at~$z=0$. The equality~\eqref{fourierexp} doesn't hold for~$k=1$ because at~$z=0$ the Fourier expansion vanishes meanwhile the Bernoulli polynomial~$B_1(\{0\})=-\frac{1}{2}$. Thus,~\eqref{fourierexp} fails to hold at~$z=0$. This will have zero impact in generic computations, and thus, generically, we will only highlight the difference when it will matter .

Clearly this also equals~$B_k(\{ x\})$ where~$\{x\}\,\equiv\, x-\lfloor x \rfloor$ is the fractional part of~$x$.
In particular we have for~$x \in \mathbb{R}$
\bea\label{Bperiod}
B_0(\{x\}) &\,\equiv\,& 1 - \sum_{j}\delta(x+j)\,, \\
B_1(\{x\}) &\,\equiv\,& -\frac{1}{2\pi \i} \sum_{ j\neq 0} \frac{\e(j x)}{j}\,\quad\text{for} \,\quad x\notin \mathbb{Z}\,, \label{B1Four}\\
B_2(\{x\}) & \,\equiv\,& \frac{1}{2 \pi^2} \, \sum_{j \neq 0} \frac{\e(jx)}{j^2} \,, \label{B2Four} \\
B_3(\{x\}) & \,\equiv\,& - \frac{3\,\i}{4\pi^3} \, \sum_{j \neq 0} \frac{\e(jx)}{j^3} \,. \label{B3Four} 
\eea
Equality~\eqref{B1Four} does not hold at~$x\in \mathbb{Z}$. These are the positions of the discontinuities of both he left and the right-hand objects. The value of the function in the left-hand side at~$x\in\mathbb{Z}$ coincides with the limit from the right. The value of the function in the right-hand side, matches the semisum of the left and right limits.

The generic polynomial~$B_k$ for~$k\,=\,1\,,\,2\,,\,\ldots $ obeys the following three properties (the first and second also apply to~$p=0$)
\bea
\label{IntegralBPols}
\int_0^1 dx\, B_k(\{x\})&\=&0\,, \\
\label{DerivativesBperiod}
B_k(\{x\})&\=& (k+1)\,\frac{d}{d x} B_{k+1}(\{x\}) \,,
\\ 
\label{CancellingContactTerms}
B_{k}(z+1)\,-\,B_{k}(z)&\=& k z^{k-1}\,.
\eea

At some points we will need to use the following smooth versions of the periodic Bernoulli polynomials
\bea\label{BperiodLambda}
B_{1\Lambda}(\{x\}) &\,\equiv\,& -\frac{1}{2\pi \i} \sum_{ j\neq 0} \frac{\Lambda^{|j|} \e(j x)}{j} \= -\,\frac{1}{2\pi\i} \Bigl(\text{Li}_1(\Lambda \e( x))\,-\, \text{Li}_1(\Lambda \e(- x))\Bigr)\,, \label{B1FourL}\\
B_{2\Lambda}(\{x\}) & \,\equiv\,& \frac{1}{2 \pi^2} \, \sum_{j \neq 0} \frac{\Lambda^{|j|} \e(jx)}{j^2} \,\= \,\,\,\,\,\,\,\frac{1}{2\pi^2} \Bigl(\text{Li}_2(\Lambda \e( x))\,+\, \text{Li}_2(\Lambda \e(- x))\Bigr) \,, \label{B2FourL} \\
B_{3\Lambda}(\{x\}) & \,\equiv\,& - \frac{3\,\i}{4\pi^3} \, \sum_{j \neq 0} \frac{\Lambda^{|j|} \e(jx)}{j^3}  \= -\, \frac{3\,\i}{4\pi^3} \,\Bigl(\text{Li}_3(\Lambda \e( x))\,-\, \text{Li}_3(\Lambda \e(- x))\Bigr)\,. \label{B3FourL} 
\eea
For~$x\in \mathbb{R}$ we will assume~$0\leq\Lambda<1$. In cases where the argument~$x$ happens to be complex we will assume~$0<\Lambda e^{2 \pi |\text{Im} x|}<1$\,.

\subsection*{Recalling the functions~$\log P$,~$\log Q$} \label{FunctionsPQ}

The function~$\log P$ is defined by the following double expansion
\be\label{logP}
\begin{split}
\log P(z)&\,=\,\log P(z_1,z_2;\t)\= \log P_0(z_1,z_2;\t)\,+\,{ \pi \i \t}  B_2\bigl( \{z_2\} \bigr)+\pi \i\Psi_P(z) \,\\ &\=-\frac{\i}{2\pi}  \!  
{\underset{m,n \in \IZ \atop m \neq 0}{\sum}} \; \frac{\e(n z_2 - m z_1)}{m(m\t+n)} 
\,+\,{ \pi \i \t}  B_2\bigl( \{z_2\} \bigr)+\pi \i\Psi_P(z) \,,
\end{split}
\ee
where~$\Psi_P(z)$ is an arbitrary~$\tau$ independent double periodic real function. The real part of~$\log P$ is the Kronecker-Eisenstein series. Without loss of generality,  in the main body of the manuscript we fix~$\Psi_P\=0$. This is equivalent to a trivial redefinition of the regularization ambiguity~$\Psi_0$ in~\eqref{Ambiguity}.

The function~$\log Q$ is defined by the following double expansion
\be \label{logQ}
\begin{split}
\log Q(z)&\,=\,\log Q(z_1,z_2;\t) 
 \=\log Q_0(z_1,z_2;\t)\,+\,\frac{2 \pi \i \t}{3}  B_3\bigl( \{z_2\} \bigr) \,+\, \pi\i \,\Psi_Q(z) \, \\ &\,\equiv\,  - \frac{1}{4\pi^2} \!
{\underset{m,n \in \IZ \atop m \neq 0}{\sum}} \; \frac{\e(n z_2 - m z_1)}{m(m\t+n)^{2}} 
\,+\,\frac{2 \pi \i \t}{3}  B_3\bigl( \{z_2\} \bigr) \,+\, \pi\i \,\Psi_Q(z) \,, 
\end{split}
\ee
where~$\Psi_Q(z)$ is an arbitrary~$\tau$ independent double periodic real function.  The real part of~$\log Q$ is the Bloch-Wigner elliptic dilogarithm. Without loss of generality,  in the main body of the manuscript we fix~$\Psi_Q\=0$. This is equivalent to a trivial redefinition of the regularization ambiguity~$\Psi_0$ in~\eqref{Ambiguity}.

\section{The supersymmetric partition function and the Index} \label{sec:RegularizationFlow}

In this appendix we revisit the problem of regularization of the divergent super-determinants of the underlying four-dimensional field-theory problem from a different perspective than usual.~\footnote{The gravitational counterpart of this perspective will not be studied here~\cite{Assel:2014tba,Genolini:2016ecx,Papadimitriou:2017kzw,An:2017ihs}.} This gives a first principles origin of the sequence of actions~$S_\lambda$ that was used to find a fixed point expansion of the superconformal index.

Before entering in technical discussions, let us take a couple of paragraphs to comment about regularization schemes in the context of supersymmetric partition functions. The superconformal index and the supersymmetric partition function that the gravitational side of the~$AdS/CFT$ duality sets as the appropriate observable to study~\cite{Cabo-Bizet:2018ehj}, can differ by a pre-factor.  We will show that there exist regularization schemes that, distinctly to the schemes used in~\cite{Cabo-Bizet:2018ehj,Closset:2013sxa,Assel:2014paa,Assel:2015nca}, give the superconformal index out of the divergent 4d super-determinants without the pre-factor.~In that sense the schemes below will play an analogous role to the one played by the background-subtraction method used in the gravitational side of the duality, when evaluating the onshell action~\cite{Chen:2005zj} of the dual BPS black holes~\cite{Cabo-Bizet:2018ehj,Cassani:2019mms}.

The difference in between this regularization and the ones of~\cite{Cabo-Bizet:2018ehj,Closset:2013sxa,Assel:2014paa,Assel:2015nca}~\footnote{The method used in the first three references also differs from the one used in the last one, by a finite quantity.} is a finite contribution. It would be interesting to better understand whether the counterterms accounting for such finite differences can go beyond the hypothesis of the~\emph{no-go} arguments of~\cite{Assel:2014tba}, or not; and in case the answer is yes, then how. In particular the former scheme cancels the pre-factor, meanwhile the latter two do not. As these schemes are used after supersymmetric paired contributions are cancelled, then in principle the three of them could naively be thought to be consistent with supersymmetry, and thus from this perspective, it is not clear to us why one of them is supersymmetric and the others are not. This is a puzzle to us. Perhaps an analysis along the lines of~\cite{Kuzenko:2019vvi,Bzowski:2020tue} could help to resolve it.

After this small disgresion, let us summarize. The first outcome of this appendix is to show that there exists a regularization, after supersymmetric cancellations have been already taken into consideration, that cancels the relative pre-factor between superconformal index and supersymmetric partition function.~\footnote{Note that this does not mean that the pre-factor lacks physical meaning. Indeed, the pre-factor is known to be related to 't Hooft anomaly coefficients of the corresponding theory, as it will be recalled in appendix~\ref{CasimirE}. } To claim that this scheme is consistent with supersymmetry a more detailed analysis along the lines of~\cite{Kuzenko:2019vvi,Bzowski:2020tue} needs to be pursued.~\footnote{It would be also interesting to compare with the scheme presented in~\cite{Closset:2019ucb}.} That lies beyond the scope of this paper.  The most important outcome of this section is that the freedom in choice of that regularization scheme, includes the freedom in choice of extension of the integrand (of the superconformal index) to the complex plane. This gives a first principles derivation of the extensions~$S_{\lambda}$ that we used to find the fixed point expansion of the superconformal index.

For a while we will focus on the contribution of a single chiral multiplet. In the following appendices we will reinstate the weights~$\rho\,$ and the multiplet index~$\a$ and sum over them.

\subsection{Two examples of regularization}\label{sec:Regularization}

We start by regularizing the one-loop determinants of the supersymmetric partition functions on~$S_1\times S_3$. Firstly, we illustrate two cases, that we call the elliptic and meromorphic schemes. The elliptic scheme corresponds to the double periodic extension of~\cite{Cabo-Bizet:2019eaf,Cabo-Bizet:2020nkr}. Both regularization schemes match the superconformal index at the contour of integration~$\underline{u}_2=0$.

\subsection*{General strategy to regularize}\label{subsec:RegGTR}

The contribution of a~$4d$ $\mathcal{N}=1$ chiral multiplet to the supersymmetric partition function on~$S_1\times S_3$ has been computed via supersymmetric localization.  The formal expression is~\cite{Cabo-Bizet:2018ehj}~\cite{Assel:2014paa,Closset:2013sxa}
\be\label{EffAction}
Z \=e^{-S(z)}\,\equiv\,\prod_{k\,=\,-\infty}^{\infty}\,\prod_{m,\,n\,=\,0}^{\infty} \frac{z\,+\,k\,-\,(m\,+\,\frac{1}{2})\t\,-\,(n\,+\,\frac{1}{2})\s}{z\,+\,k\,+\,(m+\frac{1}{2})\t\,+\,(n\,+\,\frac{1}{2})\s}\,
\ee
where~$S$ is the divergent one-loop effective action. The complex parameters~$\sigma=\sigma_1\,+\,\i\sigma_2$ and~$\t=\t_1\,+\,\i\,\t_2$ are in a beginning assumed to have positive imaginary part~$\sigma_2>0$ and~$\t_2>0\,$. These parameters can be thought of as a complexification of the angular velocities of rotation around the two independent rotational axis of the three-sphere. The three integer numbers~$k$,~$m$ and~$n$ label the modes associated to the non-contractible Euclidean time cycle and two contractible cycles of the three-sphere. For the moment~$z$ is just a complex variable that in due time will be identified with a linear combination of the gauge variables~$u$ and flavour and~$R$-symmetry chemical potentials. Momentarily, we will assume that~$\sigma=\t$. Later on in appendix~\ref{ExtRegul} we will show that this analysis can be extended to infinitely many other cases.

For~$\sigma=\tau$ the potential in~\eqref{EffAction} can be written as the following divergent series
\be\label{FormalLog}
S\,=\, \sum_{(k,\,m)\,\in\,\mathbb{Z}^2}\,(m\,+\,1)\,\log \Bigl( z\,+\, k\,+\, \t \,(m+1)\Bigr) \,.
\ee
The series~$S$ is formally related to 
\be
\widehat{S}\,\equiv\,\sum_{(k,\,m)\,\in\, \mathbb{Z}^2}\,\,\log\,\Bigl(z\,+\,k\,+\, \t\, (m+1)\Bigl) \, ,
\ee
by the following linear partial differential equation
\be\label{IdentityDiff}
\partial_z\,S\= \nabla_\t\,\widehat{S}\,.
\ee
The derivative
\be\label{DefConnection}
\nabla_\t \,\equiv\,\partial_{\t}\,-\,e_\t\,, \qquad \qquad e_\t\,\equiv\,\partial_\t z \, \partial_{z}\,,
\ee
is invariant under holomorphic re-parameterizations of~$z$ as a function of~$\t$. The choice of  ``frame"~$e_\t$ is part of the ambiguity in regularization. Different such choices define different regularizations. Let us choose a linear and holomorphic (in~$\tau$) re-parameterization of the complex variable~$z$
\be\label{zvsw}
z\=\chi \,+\,\t \,\xi\,,
\ee
with~$\chi$ and~$\xi$ being two auxiliary complex variables that are assumed to be independent of~$\t$. Then, the differential equation~\eqref{IdentityDiff} takes the form
\be\label{DiffEqFinal}
\partial_z\,S\= (\partial_\t\,-\, \xi\,\partial_z) \,\widehat{S}\,.
\ee
To ease the work, we restrict~\eqref{DiffEqFinal} to the domain
\be
\xi\,\in\, \mathbb{R}\,, \qquad \chi\,\in\, \mathbb{C}\,.
\ee 
We recall that one can always write a generic complex variable~$\mathfrak{X}$ such as~$z,\,\chi\,$ or~$\xi$, as a linear combination of~$1$ and~$\t$. The corresponding components are non-holomorphic functions of~$\mathfrak{X}$ and~$\t$
\be
\begin{split}
\mathfrak{X}_{1}&\=\frac{\overline{\mathfrak{X}} \,\t  \,- \, \mathfrak{X}\,\overline{\t}}{\t\,-\,\overline{\t}}\,\qquad \mathfrak{X}_{2}\=\frac{\overline{\mathfrak{X}}  \,- \, \mathfrak{X}}{\t\,-\,\overline{\t}}\,.
\end{split}
\ee
We note that~$\chi$ and~$\xi$ {are not} the real components~$z_1$ and~$z_2$. In fact~$z_1\=\chi_1$ and~$z_2\=\chi_2\,+\,\xi$, where~$\chi_1$ and $\chi_2$ are the components~$1$ and~$2$ of~$\chi$.

So far~$z$ in~\eqref{zvsw} is just a mere complex variable. We need to define its relation to the physical variables i.e~to the gauge variables~$u$ and chemical potentials~$\Delta$'s . The relation will be
\be\label{ZetaD}
u\,+\,\Delta\,\,\equiv\, z\=\chi\,+\,\t\, \xi.
\ee
In particular we can always identify the components~$1$ and~$2$ of~$u\,+\,\Delta$ with the respective ones of~$z\=\chi\,+\,\t \xi$. For short, from now on we will use the notation~$z(u)\,\equiv\, u\,+\,\Delta$.

The independent variable~$\xi\in \mathbb{R}$ labels different regularization schemes, i.e~defines different regularization conditions, each of them, associated to a differential equation~\eqref{DiffEqFinal} in the way that will be explained below.

To regularize we will follow three steps :
\begin{enumerate}
 \item Regularize the real part of~$S$ by demanding either double periodicity (in subsection~\ref{subsec:Lattice}), or meromorphy (in subsection~\ref{HolomorphicReg}) in~$u$. In both cases the regular answer turns out to be a real-harmonic function of~$\tau$ 
\be
(\partial^2_{\t_1}\,+\,\partial^2_{\t_2}) \, \text{Re}S \=0\,.
\ee
 \item  The regularized imaginary part of~$S$ is constrained by requiring meromorphy in~$\tau$. 
 
 \item There is a remaining $\t$-meromorphic ambiguity that we denote as $R$. We bound it to be, in both cases, of the form
\be\label{Rambiguity}
\begin{split}
  \log R \,\sim\, +\i \t\sim -\infty\, \quad \text{for }\quad \t \to +\,\i\,\infty\,,\qquad  \log R \,\sim\, 0\, \quad \text{for} \quad \t \to +\,\i\, 0\,.
\end{split}
\ee
\end{enumerate}
\footnote{As explained before, this condition can be relaxed. In principle, the only constraint on~$R$ is that its real part must be an arbitrary real function of~$u_2$ and~$\t$. This ambiguity has to be fixed by imposition of a physical condition. In our case such condition is matching to the superconformal index. As will be shown in the next section, the condition~\eqref{Rambiguity} is only at half-way of the constraint that we must impose. To illustrate connections to previous regularization schemes we find illustrative to leave this ambiguity free for the moment. }  We call~$R$ the~\emph{remainder function}. The function~$\log R$ determines the large-$\t$ asymptotic behaviour of the regularized version of the action~$S(z)$.

There are two technical issues that we need to highlight before moving on 
\begin{itemize}
\item The infinities to regularize, come from a zero Fourier mode~$(m,n)=0$ in the double Fourier expansions below. The Poisson summation dual of the divergent expressions will make that clear. The first step will be to excise such mode. After that, finite results are obtained, but there is still a finite ambiguity that needs to be constrained by the properties of the physical observable that one wishes to compute.
\item Some of the regular series below do not converge fast enough. This means they converge to functions which are non smooth.  To render such limit functions smooth at intermediate steps, further ``regularization" is required.~\footnote{In order to apply ABBV integration formula it is safer to use smooth functions.} To solve that issue we insert an intermediate real cut-off~$0\,\leq\,\Lambda\,<\,1$. We will implicitly assume~$\Lambda\=1^-$ at every step.~See appendix~\ref{app:P0Q0Props}.
\end{itemize}

\paragraph{An elliptic scheme} \label{subsec:Lattice}
We start by working out two particular schemes. The first one we call the~\emph{elliptic} scheme. It corresponds to the following choice of frame
\be\label{connection}
\xi\= \Delta_2\,,\qquad \Delta_2\,\in\,\mathbb{R}\,,
\ee
where~$\Delta_2$ is the second real component of the complex chemical potential~$\Delta\=\Delta_1\,+\,\Delta_2\t$.

The divergent series~$S$ and~$\widehat{S}$ are analytic in the independent complex variables~$z$ and~$\tau$. To avoid clumsiness in our presentation it is convenient to introduce the following two real variables $x$ and $y$ to denote the real and imaginary parts of $z$ i.e~$z=\,x\,+\,\i y\,$. Thus, from~\eqref{IdentityDiff} it follows that~\footnote{For a holomorphic function $h(z)$ of $z$ it follows that $\text{Re}\,\partial_z h(z) =\partial_{\text{Re}\,z} \text{Re} \,h\,.$
}
\be\label{DiffEq}
\partial_{x}\, \text{Re} S\= \,(-\,\xi \,\partial_{x}+\partial_{\t_1})\,\text{Re} \widehat{S} \, .
\ee
The series~$\widehat{S}$ is an average over the lattice $\mathbb{Z}\t+\mathbb{Z}$. Thus, at formal level we can recast it as a double Fourier expansion with the use of Poisson summation formula
\be \label{PoissonS}\nonumber
\sum_{(m,\,n)\,\in \, \mathbb{Z}^2} \, f(x+k,\,y+m) \= \, \sum_{(m,\,n)\,\in \, \mathbb{Z}^2} \widehat{f}(m,n)\, \e(x m +y n) \,.
\ee
We define the Fourier transform of $f(x,y)$ as~$\widehat{f}(m,n) \,\equiv\, \int_0^1 dx dy \, \e( -x m - y n) f(x, y)\,.$
A computation shows that
 \be
\text{Re}\widehat{S}\= - \sum_{(k,\,m)\,\in\, \mathbb{Z}^2}\, \frac{\tau_2}{2\pi}\,\frac{ \e\bigl(n z_2\, -\,m z_1\bigr)}{| m\tau\,+\,n|^2}\,.
 \ee
 
This series is divergent due to the presence of the zero mode~$(m,n)=0$. Indeed, this is the origin of the divergences in the original series $S$. From now on we excise the zero mode~$(m,n)=(0,0)$ and proceed with the new convergent expression of~$\widehat{S}$ which can be presented in the form
\be\label{ToRegular1}
\text{Re}\widehat{S}\,\equiv\, \log |P(z_1,z_2;\t)|\,,
\ee
where
\be\label{ReP}
 \log |P(z_1,z_2;\t)| \equiv   -
\frac{\imt}{2 \pi} \!  {\underset{m,n \in \IZ \atop (m,n) \neq (0,0)}{\sum}} \; \frac{\e(n z_2 - m z_1)}{|m\t+n|^{2}} \,,
\ee
is a well known real harmonic and real analytic modular invariant function of~$\t$.

If we plug~\eqref{ToRegular1} in the right-hand side of~\eqref{DiffEq} we can integrate to obtain a regular expression for Re$S$. We obtain two contributions. One contribution comes from the antiderivative with respect to~$x$ of~$\partial_{ x}\text{Re}\widehat{S}$, which equals~$-\xi$ times~$\widehat{S}$, up to an arbitrary function of~$z_2$ and~$\t $,
  \be\label{CTerm}
-\xi \int \,dx\,\partial_{x}\,  \text{Re}\widehat{S}\,=\,\,\, -\,\xi \,\log |P(z_1,z_2;\t)| \,+\,\ldots\,,
\ee   
and a second comes from the antiderivative with respect to~$x$ of the derivative of~\eqref{ToRegular1} with respect to~$\t_1$, which equals
\be\label{T1Term}
\int \,dx\,\partial_{\t_1} \text{Re}\widehat{S}\,=\,\,\,\,  \log |Q(z_1,z_2;\t)|\,+\, \ldots\,,
 \ee
where
\be
\begin{split}
\log |Q(z_1,z_2;\t)|&\equiv \frac{\i\,\tau _2}{2 \pi ^2}\,{\underset{m,n \in \IZ \atop (m,n) \neq (0,0)}{\sum}}\frac{ \left(m\, \tau _1\,+\,n\right)\, \e\bigl( n z_2 - mz_1 \bigr)}{ |m\t+n|^{4}}\,.
\end{split}
\ee
This double periodic function is closely related to the Bloch-Wigner elliptic dilogarithm~\cite{ZagierOnBloch}, as was pointed out in~\cite{Cabo-Bizet:2019eaf}, see section~$3.1$ in that reference. This function is a real harmonic function of $\tau$ but it is not modular covariant. 

Collecting partial results we obtain the regularized expression for the real part of the action after evaluating~$\xi=\Delta_2$
\be\label{EffectiveActionAlpha}
\ReSell  \= \log |Q(z_1,z_2;\t)| \,-\,\Delta_2\, \log |P(z_1,z_2;\t)|\, +\,\log |R(z_2;\t)|\,,
\ee
where~$\log |R|$ depends on~$z_2$ and~$\t$. If we assume~$\log|R|$ to be the real part of a function that is  meromorphic in~$\tau$, then we must assume it to be a real harmonic function of~$\tau\,$.

Holomorphy in~$\t$ constraints the imaginary part of the action~$\text{Im} S\,$. As~$\log |P|$ and~$\log |Q|$ are the real parts of~\eqref{logP} and~\eqref{logQ} we obtain
\be\label{SellReg}
\Sell\= \log Q(z(u)) \,-\,\Delta_2\, \log P(z(u))\, +\,\log R(z(u);\t)\,.
\ee
The asymptotic conditions~\eqref{Rambiguity} constraint~$\log R$ to be of the form
\be\label{Ambiguity0}
\log R \=  2\pi\i \t \El(z_2)\,\,+\,\pi\i \Phi(z_1,z_2)\,,
\ee
where~$\Phi$ and~$\El$ are real functions.~Note that the term proportional to~$\El$ dictates the asymptotic behaviour of~$S_{\text{ell}}$ at~$\t\,\to\, +\,\i\, \infty$. Note that when the function $E$ is a constant, we can interpret it as an insertion of a background charge (or energy). That is because~$2\pi\i \t$ is the chemical potential dual to the BPS charge~$E$.

\paragraph{A meromorphic scheme} \label{HolomorphicReg}

In this subsection we present a second scheme. It corresponds to the following choice
\be\label{connection2}
\xi\= z_2\,=\, u_2\,+\, \Delta_2 \,,  \qquad  z_2\,\in \,\mathbb{R}\,,
\ee
where~$u_2$ and~$\Delta_2$ are one of the components of~$u$ and~$\Delta$ in their linear decomposition in terms of~$1$ and~$\t$,~$u\,=\,u_1\,+\,\t u_2$ and~$\Delta\,=\,\Delta_1\,+\,\t\Delta_2\,$. 

The first two steps are identical to the ones of the previous section. Compute the anti-derivative with respect to~$x$ of the term coming from~$\xi\, \partial_{x}$ and obtain again~\eqref{CTerm}.
Then, compute the antiderivative with respect to~$x$ of the term coming from~$\partial_{\t_1}$ which is again~\eqref{T1Term}. Collecting partial results, and finally substituting~$\xi\= z_2=u_2+\Delta_2$, we obtain
\be\label{EffectiveActionAlpha2}
\begin{split}
\text{Re}S  &\=  \text{Re}S_{\text{ell}} \,-\, u_2 \log|P(z(u))|\\&\=\log |Q(z(u))| \,-\,(u_2+\Delta_2)\, \log |P(z(u))|\, +\,\log |R(z_2(u))|\,, 
\end{split}
\ee
where~$\log |R|$ is (before imposing boundary conditions~\eqref{Rambiguity}) an arbitrary function of $z_2$ and harmonic in~$\tau$. Finally, by demanding meromorphy in~$\t$ we obtain
\be\label{EffectiveActionMero2}
\begin{split}
S  &\=  S_{\text{ell}} \,-\, u_2 \log P(z(u))\\&\=\log Q(z(u)) \,-\,z_2(u)\, \log P(z(u))\, +\,\log R(z_2(u))\,, 
\end{split}
\ee
where again the asymptotic conditions~\eqref{Rambiguity} constraint~$\log R$ to be of the form
\be\label{Ambiguity}
\log R \=  2\pi\i \t \El(z_2)\,+\,\pi\i \Phi(z_1,z_2),
\ee
where~$\Phi$ and~$\El$ are real functions.  Note that this term dictates the asymptotic behaviour of $S$ at~$\t\to +\i \infty\,$. In the next section we will show that there exists a choice of~$R$ such that for~$u\in \mathbb{C}$
\be\label{SA}
\,e^{-S(z(u))}\= \Ge(z(u)+\t;\t,\t)\,.
\ee 
In these two subsections, we have assumed~$\sigma=\tau$ but that condition can be relaxed and the same regularization process applies (See appendix~\ref{ExtRegul}).

\subsection{Generalization to the case~$\s\,\neq\, \t$ }\label{ExtRegul}
In this appendix we show that the same regularization used for the cases~$\s\,=\,\t$ in section~\ref{sec:RegularizationFlow} can be adapted to more general cases with~$\s\neq\t$. This implies that everything we have said in this paper relative to the case~$\sigma=\tau$ can be straightforwardly generalized to the more general case studied in this subsection.  

Let~$\s\,=\, a \omega $ and~$\t\,=\,b \omega$ with~$a$ and~$b$ positive integers. For simplicity we will assume $a$ and $b$ to be primes, but the same result applies to the generic case. Let us define~$\widetilde{z}\= z\,+\,\frac{\t\,+\,\s}{2}$ and analyze the logarithm of~\eqref{EffAction}
\be\label{logSeries}
\sum_{k\in\mathbb{Z}}\sum_{m,n\,=\,0}^{\infty}\log \Bigl(\widetilde{z}\,+\,k\,+\, \Bigl(\frac{m}{b}+\frac{n}{a}\Bigr) ab\omega \Bigr)\,-\,\log \Bigl(\widetilde{z}\,+\,k- \Bigl(\frac{m+1}{b}+\frac{n+1}{a}\Bigr) ab \omega\Bigr)\,.
\ee
Here we have conveniently written~$m a \omega +n b \omega \= \Bigl(\frac{m}{b}+\frac{n}{a}\Bigr) a b \omega$. For positive integers~$m$ and~$n$ it is always possible to rewrite
\be
\frac{m}{b}\=\frac{r_m}{b}\,+\, \widetilde{m}\,, \qquad \frac{n}{a}\=\frac{s_n}{a}\,+\, \widetilde{n}\,,
\ee
with integers~$\widetilde{m}$,~$\widetilde{n}$,~$r_m$ and~$s_n$ ranging over
\be
\widetilde{m}\,\geq\,0\,,\quad \widetilde{n}\,\geq\,0\,,\quad  r_m\,=\,0,\,\ldots\,,b-1 \,, \quad s_n\,=\,1,\ldots,a-1\,.
\ee
It is straightforward to check that~\eqref{logSeries} takes the form
\be\label{SumStep2}
\sum_{r=0}^{a-1}\sum_{s=0}^{b-1}\sum_{k\in\mathbb{Z}}\sum_{\widetilde{m},\,\widetilde{n}\,=\,0}^{\infty}\log \Bigl(\widetilde{z}_{r,s}\,+\,k\,+\, \bigl(\widetilde{m}\,+\,\widetilde{n}\bigr) ab \omega \Bigr)\,-\,\log \Bigl(\widetilde{z}_{r,s}\,+\,k- \bigl(\widetilde{m}\,+\,\widetilde{n}\,+\,2\bigr) ab \omega\Bigr)\,,
\ee
where
\be
\widetilde{z}_{r,s}\=\widetilde{z}\,+\,r a\omega \,+\, s b \omega\,.
\ee
In reaching this expansion we have used the following redefinition in the argument of the second logarithm
\be
\widetilde{m}_{\text{after}}\=b\,-\,1\,-\,\widetilde{m}_{\text{before}}\,,\qquad  \widetilde{n}_{\text{after}}\=a\,-\,1\,-\,\widetilde{n}_{\text{before}}\,.
\ee
As the summand of~\eqref{SumStep2} only depends on the combination~$\widetilde{m}\,+\,\widetilde{n}$ we can reorganize the sum as follows, by collecting degeneracies
\be\label{Sumrs}
\sum_{r=0}^{a-1}\,\sum_{s=0}^{b-1}\,\sum_{(k,m)\in\mathbb{Z}^2}\,(m\,+\,1)\, \log \Bigl(z^{a,b}_{\,r,s}\,+\,k\,+\, (m+1)\, ab \omega \Bigr)\,.
\ee
In this expression
\be
z^{ab}_{rs}\,\equiv\,\widetilde{z}_{r,s}\,-\,a b \omega\= z\,+\,\Bigl(r+\frac{1}{2}\Bigr) a\omega \,+\, \Bigl(s+\frac{1}{2}\Bigr) b \omega\,-\, a b \omega\,.
\ee
Note that~\eqref{Sumrs} is a combination of terms like the one we regularized before in subsection~\ref{subsec:RegGTR}, see equation~\eqref{FormalLog}. Thus, assuming~\eqref{Sumrs} as starting point, one can regularize as explained in the previous subsections, but for generic values of~$\t$ and~$\sigma$, with a common complex divisor~$\omega$. 

\section{The real functions~$\El$ and~$\Phi$ } \label{sec:DoublePerMero} \label{sec:Nonabelian}

Here we fix~$R$ in such a way that the finite result coincides with the elliptic Gamma function~$\Ge(z(u)+\t;\t,\t)$ in the entire plane~$(u_1,u_2)$ with~$u\,=\,u_1\,+\,\t u_2$.~The details are given in appendix~\ref{Proofremainder}. 

We choose the same scheme as before but modifying~$z(u)$ in~\eqref{ZetaD} by substituting~$u$ by a linear combination of~$N$ complexified Cartan angles $\underline{u}$,~$\rho(u)$.~$\rho$ is a weight. This gives again~\eqref{EffectiveActionMero2} with~$u$ substituted by~$\rho(u)$. Next we must fix~$R$ by demanding that for all~$u^i\in \mathbb{C}$
\be\label{ExtensionHolPre}
\begin{split}
\Ge(z(u)+\t;\t,\t)\={R}^{-1}\,\,\frac{\,\, P_0(z_1,z_2)^{z_2}}{Q_0(z_1,z_2)} \,.\\ 
\end{split}
\ee
This equation is the exponential version of~\eqref{EffectiveActionMero2}, after assuming~\eqref{SA}, with the substitution~$u\,\mapsto\, \rho(u)$. This is the completion of the equation~(2.28) of~\cite{Cabo-Bizet:2020nkr}\cite{Zudilin}. There, this identity was defined up to a~$\tau$-independent imaginary constant. The derivation of the remainder~$R$ that fixes this pure phase ambiguity, will be postponed until appendix~\ref{Proofremainder}. Here we will just quote the result. 

We define the functions~$P_0$ and~$Q_0$ via their logarithms
\be\label{logPnewapp}
\log P_{0}\=\log P_0(z_1,z_2)\,\equiv -\frac{\i}{2\pi}  \!  
{\underset{m,n \in \IZ \atop m \neq 0}{\sum}} \; \frac{\e(n z_2 - m z_1)}{m(m\t+n)} \,,
\ee
and
\be\label{logQnewapp}
\log Q_{0}\=\log Q_0(z_1,z_2) \,\equiv\,  - \frac{1}{4\pi^2} \!
{\underset{m,n \in \IZ \atop m \neq 0}{\sum}} \; \frac{\e(n z_2 - m z_1)}{m(m\t+n)^{2}} \,.
\ee
They are related, via a redefinition of~$R$, to the functions~$\log P$ and~$\log Q$ (See appendix~\ref{FunctionsPQ}).
In the left-hand side of \eqref{ExtensionHolPre} we define
\be\label{zu}
z(u)\=\rho(u)\,+\,\Delta\,.
\ee
In the right-hand side, the variables~$z_1$ and~$z_2$ are related to the~$2$rk$(G)$ variables~$\underline{u}\,=\,\underline{u}_1\,+\,\t \underline{u}_2$ and~$\underline{v}$ as follows
\be\label{z1z2Funcuv}
\begin{split}
\,z_1(u,v)&\= z(u)\,-\,\rho(v)\,\t\,-\,\Delta_2\,\t \,, \\
\, z_2(v)&\= \rho(v) \,+\,\Delta_2\,.
\end{split}
\ee
In the right-hand side of~\eqref{ExtensionHolPre} the auxiliary variable~$v$ is fixed in terms of~$u$ i.e.
\be\label{ConstraintVU}
\underline{v}\= \underline{u}_2\,.
\ee
Note that~$z_1$ is a real function of the real variables~$\underline{u}_1$ and not of~$\underline{u}_2\,$. The~$R^{-1}$ then takes the following form
\be\label{Prefactor0}
\begin{split}
R^{-1}\,&\= R^{-1}(u,v)\,\,\equiv\, q^{-\El}\,\,\e\bigl(-\frac{\Phi}{2}\bigr)\,,
\end{split}
\ee
where~$\El$ is related to the following cubic polynomial in two variables
\be
\mathcal{A}(u,z_2(v))\=\frac{B_3(z_2(v))}{3}\,-\, \bigl(\rho(u_2)+\Delta_2\bigr)\,\frac{B_2(z_2(v))}{2}
\ee
by the following relation
\be\label{ContributionMeromorphic}
\El\,\equiv\,-\mathcal{A}(u,z_2(v))\,+\,\mathcal{A}(u,\{z_2(v)\})\,.
\ee
The polynomial~$\mathcal{A}$ is related to the contribution of the given multiplet to the cubic polynomial of anomaly coefficients. If we define
\be
\mathcal{A}(u,z_2(u))\,\=\, -\,\frac{1}{6}\,\Bigl(z_2(u)^3\, -\,\frac{z_2(u)}{2}\Bigr)\,,
\ee
then the cubic polynomial of anomaly coefficients for flavour and R-charges can be defined as
\be
\mathcal{A}_{anom}\= \sum_{\alpha}\sum_{\rho_\alpha} \mathcal{A}(0,z_{2\a}(0))\,,
\ee
where~$z_\a\=\rho_\a(u)+\Delta_\a$. More details on this can be found in section~2.4 of~\cite{Cabo-Bizet:2020nkr}.
Finally, the pure phase~$e^{\pi \i \Phi}$ takes the form
\be\label{EquPHiText}\begin{split}
\e\bigl(-\frac{\Phi}{2}\bigr)&\,\equiv\,\frac{\e\Bigl(-\,\frac{1}{2} B_1(\{z_1(u)\})\, B_2(\lfloor z_2(v)\rfloor+1)\Bigr)}{\e\Bigl(-\,\frac{1}{2} B_1(\{z_1(u)\}) \,B_2(\lfloor z_2(0^+)\rfloor+1)\Bigr)}\,, \\
&\equiv \frac{\e\Bigl(-\,\frac{1}{2} B_1(z_1(u))\, B_2(\lfloor z_2(v)\rfloor+1)\Bigr)}{\e\Bigl(-\,\frac{1}{2} B_1(z_1(u)) \,B_2(\lfloor z_2(0^+)\rfloor+1)\Bigr)}\,.
\end{split}    
\ee
In going to the second line we have used the~$2\pi\i \mathbb{Z}\,$-shift ambiguity of the exponent. In the main body of the paper we reinstated this ambiguity in the form of a piecewise integer function that we denoted with the letter~$K$ (See equation~\eqref{pragmatic}).

\subsection{The universal contribution at large-$N$ of~$R$: The susy Casimir energy}\label{CasimirE}
In a gauge anomaly-free theory,  the~$u_2$-dependence cancels out after summing over multiplets. In those theories, the non-vanishing contribution of~$\El$ to the supersymmetric partition function comes from the following two terms in~\eqref{ContributionMeromorphic}.~From the first, one obtains
\be\label{AnomaCo}
-\,\mathcal{A}_{anom}\,.
\ee
The other contribution comes from the sum over matter multiplets of the term
\be\label{SecondContributionEffAction}
\mathcal{A}(u,\{z_2(u_2)\})\= \frac{B_3(\{\rho(u_2)+\Delta_2\})}{3}\,-\, \bigl(\rho(u_2)+\Delta_2\bigr)\,\frac{B_2(\{\rho(u_2)+\Delta_2\})}{2}\,.
\ee
The contribution~\eqref{AnomaCo}, is the cubic polynomial of flavour and R-symmetry anomaly coefficients. This is a universal contribution 
to the large-$N$ effective action of complex saddles, that is linear in $\tau$~\cite{Cabo-Bizet:2020nkr}. As pointed out in~\cite{Cabo-Bizet:2020nkr}, this contribution matches the susy Casimir energy of~\cite{Assel:2015nca}.

For the complex fixed-points~$P$ that we know contribute,
\be\label{AnomalyContribu}
\mathcal{A}(u,\{z_2(u_2)\}) \Big|_{u\=P} \quad\underset{N\,\to\,\infty}{\longrightarrow}\quad0\,.
\ee
Let us explain how this happens. For~$(m,n)$ fixed points with~$m>0$, the~$\rho(u_2)$ is always of the form
\be
\frac{i-j}{K}, \qquad i,j=1,\ldots,K \in \mathbb{Z}\,,
\ee
where~$K$ is a divisor of~$N$ such that~$K\to \infty$ as~$N\to \infty$.
In that limit the sum over $i$ and $j$ can be reduced to an integral.~\eqref{AnomalyContribu} follows from the identities
\be
\int^1_0 dx B_k(x)\=0\,, \qquad \int^1_0dx\,x\, B_k(x)\=0\,.
\ee
Thus, for the meromorphic scheme
\be
\El(P) \quad\underset{N\,\to\,\infty}{\longrightarrow}\quad - \mathcal{A}_{anom}\,.
\ee
This contribution is the same for every~$P$. For the~$(1,0)$ this contribution is not present.

\section{Another choice of regularization:~$\Gamma_\lambda$}\label{app:ReminderSmooth}

As we have mentioned before. The regularization ambiguity that we have in the form of a choice of frame~$e_\tau$ and remainder~$R$, includes other schemes that are not the elliptic and meromorphic ones.~\footnote{ Note that part of the degeneracy corresponds to reparametrizations of~$S_1$ upon~$S_1$ that preserve one point. The point being a representation of the real contour of integration.} In this subsection we define a family of schemes that corresponds to the deformed action~$S_\lambda$ of section~\ref{subsec:SLambda}.

The choice of frame in this case is
\be\label{connection4}
\xi\= \rho(\{u_2\}_\lambda) \,+\,\Delta_2 \,, \qquad \rho(\{u_2\}_\lambda) \,+\,\Delta_2\in \,\mathbb{R}\,.
\ee
By~$\{\underline{u_2}\}_\lambda$ we mean the vector with components~$\{u^i_2\}_\lambda\,$; and for~$x \in\mathbb{R}$ and~$0\leq\lambda<1$ was defined in~\eqref{bracketMainText}.

Here we comment on three relevant properties of this function:
\begin{itemize}
 \item[a.]~The~$\{x\}_\lambda$ is a smooth function for~$0\leq\lambda<1$, and it is periodic~$\{x+1\}_\lambda=\{x\}_\lambda$.
\item[b.]~When~$x\to0^+$  the~$\{x^i\}_\lambda \to0\,$.
\item[c.]~When~$\lambda\to0^+$  the~$\{x\}_\lambda \to0\,$. 
\end{itemize}

Repeating the same regularization steps used in appendix~\ref{sec:Regularization}, and introducing the cut-off~$\Lambda$ in the final double series expansions we obtain 
\be\label{Glambbda}
e^{-S_{\lambda}}\,=\,{\G_\lambda}(z(u))\,\equiv\,{R}_\Lambda^{-1}\,\,\frac{\,\, P_{{0}\Lambda}(z_1,z_{2})^{z_{2}(\{u_2\}_\lambda)}}{Q_{0\Lambda}(z_1,z_{2})} \,. 
\ee  
As it will be explained in great detail in appendix~\ref{app:P0Q0Props}, to ensure continuity and smoothness of their derivatives, we will need to introduce a cut-off~$\Lambda$ in the double series representation of the logarithms of $P_0$,~$ Q_0$, and~$R^{-1}$ i.e~for~$0\leq\lambda<1\,$, we define
\be\label{logPLText}
\begin{split}
\log P_{0\Lambda}&\= -\frac{\i}{2\pi}  \!  
{\underset{m,n \in \IZ \atop m \neq 0}{\sum}} \; \Lambda^{|m|\,+\,|n|} \, \frac{\e(n z_2 - m z_1)}{m(m\t+n)}\,.
\end{split}
\ee
\be\label{logQLText}
\begin{split}
\log Q_{0\Lambda}(z;\t)&\,\equiv\, -\frac{1}{4\pi^2}  \!  
{\underset{m,n \in \IZ \atop m \neq 0}{\sum}} \; \Lambda^{|m|\,+\,|n|} \, \frac{\e(n z_2 - m z_1)}{m(m\t+n)^2}\,.
\end{split}
\ee
Moreover, we choose
\be\label{SmoothenedRm1}
\begin{split}
R^{-1}_{\Lambda}&= q^{-\El_{\Lambda}}\,\,\e{\left(- \frac{\Phi_{\Lambda}}{2}\right)}\,,
\end{split}
\ee
where
\be\label{PhaseSmooth}
\begin{split}
q^{-\El_{\Lambda}}&\,\equiv\, q^{\frac{B_3(\lfloor z_2(\{u_2\}_\Lambda) \rfloor_{\Lambda}+1)\,-\,B_3(\lfloor z_2(0^+) \rfloor+1)}{3}}\,\times\, \\ &\qquad\qquad\hfill \times\,\e\left(-\,\t\, (\{u_2\}_\Lambda+\Delta_2)\,\int^{\{u_2\}_\Lambda\,+\,\Delta_2}_{\Delta_2} d\xi \, \xi \,\,\Bigl(\sum_{m\in \mathbb{Z}} \,{\Lambda}^{|m|}\,\e(m\,\xi)\Bigr)\right)\,.\\
\e{\left(- \frac{\Phi_{\Lambda}}{2}\right)}&\,\equiv\,\e\left(-\,B_{1\Lambda}(z_1(u)))\, \,\int^{\{u_2\}_\Lambda\,+\,\Delta_2}_{\Delta_2} d\xi \, \xi \,\,\Bigl(\sum_{m\in \mathbb{Z}} \,{\Lambda}^{|m|}\,\e(m\,\xi)\Bigr)\right)\,.
\end{split}
\ee
Here by~$\lfloor\cdot\rfloor_{\Lambda}$ denotes a generic smooth version of the~$\text{floor}$ function. We note that for~$0\leq\Lambda<1\,$, the right-hand side of equation~\eqref{SmoothenedRm1} is not holomorphic in~$u$ when~$v=u_2$, but that is not a problem.
Basic algebraic manipulations upon~\eqref{PhaseSmooth} imply that
\be\label{PhaseQLambdaRelation}
\underset{\Lambda\,\to\,1}{\lim}\,\partial_{v^i} (2\, Q_\Lambda\,+\,\Phi_\Lambda)\= 0\,+\,\text{c.t.}\,.
\ee
Here~$\partial_{v^i}\= \partial_{u^i_2}\,-\,\tau\,\partial_{u^i_1}$\,. This relation was used in reaching~\eqref{EOMBA} in the main body of the paper.

\noindent Finally, properties~$a.$,~$b.$ and~$c$ imply three relevant properties of the deformed action~$S_\lambda$
\begin{itemize}
\item a.~implies that the~$S_\lambda$ is smooth in the torus~$A\times B$.
\item b.~implies that the value of~$S_{\lambda}$ at the real cycle of integration~$u^i_2=v^i=0^+$ is independent of~$\lambda\,$.
\item c.~implies that in limit~$\lambda\to0^+$ the~$S_\lambda\to S_{\text{ell}}\,$. Where~$ S_{\text{ell}}$ is essentially the elliptic deformation of~\cite{Cabo-Bizet:2019eaf,Cabo-Bizet:2020nkr}.~\footnote{Up to a term linear in~$\tau$ that does not contribute to the on-shell effective actions of~$(m,n)$ fixed points and that can be reinstated by a trivial redefinition of the remainder function~$R$.}
\end{itemize}

\section{Cauchy-Riemann conditions} \label{CRrelations}

In this section we will prove equations~\eqref{P0Theta} and~\eqref{Q0Gamma}, in that order, with the use of Cauchy-Riemann relations.
Let~$z\in \mathbb{C}$, we define its real and imaginary parts as
\be
\text{Re} z\=x\equiv z_1+\t_1 z_2\,,\quad \text{Im}(z)\=y\= \t_2\, z_2 \,.
\ee
Let~$\log G\=U+\i V$ be a holomorphic function of~$z$. Then, its real and imaginary parts~$U$ and~$V$, obey the Cauchy-Riemann relations
\be
\frac{\partial U}{\partial x}\,-\,\frac{\partial V}{\partial y}\=0\,,\quad  \frac{\partial U}{\partial y}\,+\,\frac{\partial V}{\partial x}\=0\,.
\ee
In terms of the variables~$z_1$ and~$z_2$ these relations read
\be\label{CRconditionsU}
\frac{\partial U}{\partial z_1}\,+\,\frac{\t_1}{\t_2}\frac{\partial V}{\partial z_1}\,-\,\frac{1}{\t_2}\frac{\partial V}{\partial z_2}\=0\,,\quad  \frac{1}{\t_2}\frac{\partial U}{\partial z_2}\,-\,\frac{\t_1}{\t_2}\frac{\partial U}{\partial z_2}\,+\,\frac{\partial V}{\partial z_1}\=0\,.
\ee

Let us define the real function~$U$ as
\be\label{ValueU}
\begin{split}
U\=  \text{Re}\log G\,&\equiv\,-\,\text{Re}\log Q_0(z_1,z_2;\t)\,+\,z_2 \,\text{Re}\log P_0(z_1,z_2;\t) \,.
\end{split}
\ee
For~$z_2\notin\mathbb{Z}$ the harmonic dual to $U$ can be checked to be, up to a~$\tau$ dependent constant
\be\label{ValueV}
\begin{split}
V\=& -\sum_{(m,n)\,\in\,\mathbb{Z}^2\atop m\,\neq\, 0}\frac{2 \pi  z_2 \text{Re}(m \tau +n)\,\cos \left(2 \pi  \left(m
   z_1-n z_2\right)\right)}{4 \pi ^2 m
   |\mtn|^2} \\ 
   &\,-\,\sum_{(m,n)\,\in\,\mathbb{Z}^2\atop m\,\neq\, 0}\,\frac{\Bigl(\text{Re}^2\bigl(\mtn\bigr)-\text{Im}^2\bigl(\mtn\bigr)\Bigr) \sin \left(2 \pi  \left(m z_1-n z_2\right)\right)}{4 \pi ^2 m
   |\mtn|^4} \\
   &\=-\,\text{Im}\log Q_0(z_1,z_2;\t)\,+\,z_2 \,\text{Im}\log P_0(z_1,z_2;\t)\,.
\end{split}
\ee
Thus, the function
\be\label{EqAppG}
G(z;\t)\equiv \frac{P_0 (z_1,z_2;\t)^{z_2}}{Q_0(z_1,z_2;\t)}\,
\ee
is meromorphic for~$z_2\notin \mathbb{Z}$. Moreover, in~\cite{Cabo-Bizet:2019eaf} (See below equation (3.20) in that reference) we have shown that
\be\label{EqAppG2}
\Ge(z_1+\t;\t,\t)\= \frac{1}{Q_0(z_1,0^+;\t)}\,.
\ee
Let us assume that~$0<z_2<1$. In this region we can take the limit~$z_2\to 0^+$. From equations~\eqref{EqAppG} and~\eqref{EqAppG2} it follows that in such limit 
\be
G(z_1+0^+\t;\t) \= \Ge(z_1+0^+\t+\t;\t,\t)\,.
\ee
From this initial condition and uniqueness of solutions to the Cauchy-Riemann equations it follows that for~$z\in\mathbb{C}$ and~$0<z_2<1$
\be\label{GQ}
G(z;\t) \= \Ge(z\,+\,\t;\t,\t)\,=\, \frac{P_0 (z_1,z_2;\t)^{z_2}}{Q_0(z_1,z_2;\t)}\,.
\ee
Next we note that
\be\label{InCondTheta}
\frac{\Ge(z_1+1^- \t+\t;\t,\t)}{\Ge(z_1+0^+ \t+\t;\t,\t)} \=\theta_0(z_1+ 1^-\t;\t)\= P_0 (z_1,1^-;\t)\,.
\ee
A computation shows that the real and imaginary parts of $P_0(z_1,z_2;\t)$ obey the Cauchy-Riemann relations~\eqref{CRconditionsU} for~$z_2\notin\mathbb{Z}$. Then, from the initial condition in the last equality in~\eqref{InCondTheta} and uniqueness theorem, it follows that for~$0<z_2<1$
\be\label{TP}
 \theta_0(z_1+ z_2\t;\t) \=P_0(z_1,z_2;\t)\,.
\ee
From~\eqref{GQ} and~\eqref{TP} it follows that for~$z_2\notin \mathbb{Z}$
\be\label{RelationGPQ}
 \Ge(z_1\,+\,\{z_2\}\t\,+\,\t;\t,\t)\,=\, \frac{P_0 (z_1,z_2;\t)^{\{z_2\}}}{Q_0(z_1,z_2;\t)}\,,
\ee
and
\be
 \theta_0(z_1+ \{z_2\}\t;\t)\= P_0(z_1,z_2;\t)\,.
\ee
Using quasi-periodicity properties of~$\Ge$ and~$\theta_0$ one can find other identities of these kind like for instance
\be\label{EqsComplem}
\Ge(z_1+\t (-\{-z_2\}+1);\t,\t)\= \frac{\,\,P_0(z;\t)^{-\{-z_2\}}}{Q_0(z;\t)}\,.
\ee

\subsection{Cancelling discontinuities and contact terms}\label{app:CancT}
The property~\eqref{CancellingContactTerms} is particularly useful to generate contact terms that need to be cancelled in order to relate the double periodic and meromorphic functions. From this identity it is easy to note that if one composes the function~$B_p$ with the ceiling function evaluated at real number~$x$ the new composed function is piecewise constant with jumps (from left to right) of length~$(-1)^{p+1}m^p$ at every integer value~$m$ of~$x$. This new function can be used to generate primitive functions~$B_p(x)$ of contact terms of the form
\be\label{Contact}
\sum_{m} \, (-1)^{p} \, m^p \delta(x\,+\,m) \=\frac{1}{p+1}\, \partial_x B_{p+1}(\lfloor x \rfloor\,+\,1)\,,
\ee
where the primitive is defined as
\be\label{CircB}
B^{\circ}_{p+1}(\lfloor x\rfloor\,+\,1) \,\equiv\, \left\{  \begin{array}{lllr} B_{p+1}(\lfloor x\rfloor\,+\,1)\,& \qquad \qquad &\text{for}& \qquad x\notin \mathbb{Z} \\ \frac{B_{p+1}(m\,+\,1)\,+\,B_{p+1}(m)}{2}\,& \qquad \qquad&\text{for}& \qquad x\,=\,m\in \mathbb{Z}\,\end{array} \right.\,.
\ee
The case~$x=m\in \mathbb{Z}$ will be of relevance in due time, but not for the technical manipulations that will be presented in this appendix. Thus, from now on in this appendix we will ignore the supra-index $\circ$ i.e.~we will ignore the difference between~$B^{\circ}_p\,$ and~$B_p$. Only in due time (in the main part of the manuscript) when remarking the difference between the two will turn out to be of relevance to reach partial conclusions, we will reinstate the symbol~$\circ$.

In the main body of the paper relation~\eqref{Contact} turns out to be relevant at various instances. One example being the following. Suppose we want to compute the primitive of
\be
\begin{split}
B_1(u_1) \sum_{m}\delta(v\,+\,m )&\= (B_1(u) \,-\, v\t) \sum_{m}\delta(v\,+\,m )\\
&\=B_1(u) \sum_{m}\delta(v\,+\,m ) \,+\,\t \sum_{m} \,m\, \delta(v\,+\, m)\,,
\end{split}
\ee
with respect to the variable~$v$ at fixed~$u$ and evaluate the result at the section~$u_2=  v $. Then, from property~\eqref{Contact} it follows that the answer is
\be\label{Rpold}
 B_1(u) B_1(\lfloor v\rfloor+1)\,-\,\frac{\t}{2}B_2(\lfloor v\rfloor+1)\,.
\ee
A second relevant example is the following one. Suppose we want to compute the primitive of
\be\label{Rvpold}
\begin{split}
v\,B_1(u_1) \sum_{m}\delta(v\,+\,m )&\= v\,(B_1(u) \,-\, v\t) \sum_{m}\delta(v\,+\,m )\\
&\=-\,B_1(u) \sum_{m} m\delta(v\,+\,m ) \,-\,\t \sum_{m} \,m^2\, \delta(v\,+\, m)\,,
\end{split}
\ee
with respect to the variable~$v$ at fixed~$u$  and evaluate the result on the section~$u_2= v $. From property~\eqref{Contact} it follows that the answer is
\be
 \frac{1}{2} B_1(u) B_2(\lfloor v\rfloor+1)\,-\,\frac{\t}{3}B_3(\lfloor v\rfloor+1)\,.
\ee

\paragraph{A more subtle case}
These two previous examples are closely related to another two examples that turn out to be relevant, from a technical viewpoint, to ones for our scope.
The first example is computing the primitive of
\be\label{B1perDelta}
B_1(\{u_1\}) \sum_{m}\delta(v\,+\,m )\=B_1(\{u-v\t\}) \sum_{m}\delta(v\,+\,m )\Big|_{u_2=v}\,,
\ee
with respect to the variable~$v$ at fixed~$u$ and evaluating the result at the section~$u_2=  v $. 
We first must define what we mean by 
\be\label{DefBracketComplex}
\{u-v\t\}\,,
\ee 
for~$u_2\neq v$ when the argument of the bracket is not real. In that case we define for~$x\in \mathbb{C}$
\be\label{BracketX}
\{x\}\equiv x- \lfloor P_1[x] \rfloor\,, \qquad P_1[x]\,\,\equiv\frac{\overline{\t} x -\t \overline{x}}{\overline{\t}-\t}\,\in \, \mathbb{R}\, ,
\ee
where~$P_1$ is the projector operator i.e.~$P_1=P_1^2$, that projects~$x=x_1+x_2 \t$ to~$x_1$. 

From definition~\eqref{BracketX} it follows that~$\{u-v\t\}\=\{u\}\,-\,v \t$, and the primitive of~\eqref{B1perDelta} can be written as
\be\label{Rp}
\frac{1}{2\pi\i}\log R_P(u,v) \,\equiv\, B_1(\{u\}) \bigl(B_1(\lfloor v\rfloor+1)\,+\, c_1\bigr)\,-\,\frac{\t}{2}\bigl(B_2(\lfloor v\rfloor+1)\,+\,c_2\bigr)\,.
\ee
The~$c_1=c_1(u)$ and~$c_2=c_2(u)$ are~$v$-independent quantities that we can fix at convenience. For any such choice we get
\be
\left(\partial_v \Bigl(\frac{1}{2\pi\i}\log R_P(u,v))\Bigr)\right)\Big|_{u_2=v} \=B_1(\{u_1\}) \sum_{m}\delta(v\,+\,m )\,.
\ee
Suppose we want to compute the primitive of
\be
\begin{split}
v\,B_1(\{u_1\}) \sum_{m}\delta(v\,+\,m )\,,
\end{split}
\ee
with respect to the variable~$v$ at fixed~$u$, and evaluate the result at the section~$u_2=  v $. Then, the answer is
\be\label{Rvp}
 \frac{1}{2\pi\i}\log R_{vP}(u,v) \,\equiv\, \frac{1}{2} B_1(\{u\}) \bigl(B_2(\lfloor v\rfloor+1)\,+\,c_1\bigr)\,-\,\frac{\t}{3}\bigl(B_3(\lfloor v\rfloor+1)\,+\,c_2\bigl)\,.
\ee
The~$c_1=c_1(u)$ and~$c_2=c_2(u)$ are~$v$-independent quantities that we can fix at convenience.
Namely
\be\label{EquationRvP}
\left(\partial_v \Bigl(\frac{1}{2\pi\i}\log R_{vP}(u,v))\Bigr)\right)\Big|_{u_2=v} \=v B_1(\{u_1\}) \sum_{m}\delta(v\,+\,m )\,.
\ee

\section{Double periodic and meromorphic functions} \label{app:P0Q0Props}

In this section we study the relation between the double periodic functions~$P_0$ (essentially the~$P$ that was introduced in the elliptic regularization scheme) and~$Q_0$ (essentially the~$Q$ that was introduced in the elliptic regularization scheme) and~$\theta_0$ and~$\Ge\,$.~\footnote{We shall call these \emph{double periodic} functions and sometimes (non-holomorphic)~\emph{elliptic} functions~\cite{ZagierOnBloch}.}

The precise definition of~$P_0$ and~$Q_0$ will be given below. The~$\log |P_0|$ is closely related to the real-analytic Kronecker-Eisenstein series (See equation~\eqref{logP}). The function~$\log |Q_0|$ is closely related to the Bloch-Wigner elliptic dilogarithm (See equation~\eqref{logQ}). Understanding the relations with~$\theta_0$ and~$\Ge$, is useful to understand how to construct a family of functions on the torus~$A\times B$ that flows to the meromorphic extension of the integrand of~\eqref{IndexGeneral} (truncated at a portion of a cylinder). The functions~$\log Q_0$ and~$\log P_0$ can be used as building blocks of such family of double periodic extensions. Interestingly, the relation of~$\theta_0$ and~$\Ge$ to~$P_0$ and~$Q_0$ allows us to use the latter two to understand how the former two behave in a vicinity of the real axis~$\text{Im}\t=0$ and further extend them to the lower-half plane. Existence of this extension is important in order to be able to define the contour of integration in~$\tau$-plane that could be used to extract the microcanonical index out of the single exponential blocks of the ABBV formula for the grand canonical one, see section~\ref{sec:Counting}.

\subsubsection*{The properties of~$P_0$}\label{subsec:logP0}

The~$P_0$ is a double-periodic extension of the Jacobi theta function~$\theta_0$. Equivalently, it can be defined via the following logarithmic branch
\be\label{logPnew1app}
\log P_{0}(z)\=\log P_0(z_1,z_2)\,\equiv -\frac{\i}{2\pi}  \!  
{\underset{m,n \in \IZ \atop m \neq 0}{\sum}} \; \frac{\e(n z_2 - m z_1)}{m(m\t+n)} \,.
\ee
This function is discontinuous at~$z_2\in \mathbb{Z}$. The discontinuity is clear from the following identity, which holds for~$z_2\notin \mathbb{Z}$
\be\label{P0Theta}
P_0(z) \= \theta_0(z_1+ \{z_2\}\t;\t)\,.
\ee
The curly bracket means~$\{z_2\}\equiv z_2-\lfloor z_2\rfloor$. 

The identity~\eqref{P0Theta} fails to hold at~$z_2\in \mathbb{Z}$ but it holds in the limits from the left and the right to~$z_2\in \mathbb{Z}$. However, that is not an issue for the purpose of computing integrals. That the~\eqref{P0Theta} fails to hold at~$z_2\in \mathbb{Z}$ has a simple explanation: The~$\theta_0(z_1+z_2 \t;\t)$ is a continuous function of~$z_2$. After composition with~$\{z_2\}$ the limits from the right at~$z_2\in \mathbb{Z}$ coincide with the value of the function at the point due to continuity of~$\theta_0$ and continuity from the right of the function~$\{\cdot\}$. 

On the other hand the function~$P_0$ is not continuous neither from the right nor from the left at~$z_2\in \mathbb{Z}$. The series~\eqref{logPnew1app} diverges at~$(z_1,z_2)\in \mathbb{Z}^2$. Its real part diverges to~$-\infty$ at such positions. These are the positions of the zeroes of~$\theta_0$. For any other value of~$z's$ this series converges. However, if~$z_2\in \mathbb{Z}$ and~$z_1\notin \mathbb{Z}$ this series does not converge uniformly. In that case {Dirichlet theorem\label{DirichletTheo}} states that the series converges to a function that is discontinuous at~$z_2\in \mathbb{Z}$, and {the value of the function at the discontinuity equals} the {semi-sum of the limits from both sides of the discontinuity}. This is, the function~$P_0$ is not continuous at~$z_2\in \mathbb{Z}$, neither from the right nor from the left, and its value at the discontinuity differs from that of~$\theta_0(z_1+\{z_2\};\t)$.

The discontinuity at~$z_2\in \mathbb{Z}$ from the left and the right of the double series representation~\eqref{logPnew1app} of~$\log P_0$, originates from the fact that the Fourier coefficients of these series do not decrease fast enough for large values of $m$ and $n$. We solve this with the insertion of a cut-off~$\Lambda$. We will not make it explicit at every step but we will work, unless we explicitly say otherwise, with the~$\Lambda$-smoothened~\cite{Levin,BrownLevin} double Fourier expansions
\be\label{logPL}
\begin{split}
\log P_{0\Lambda}&\= -\frac{\i}{2\pi}  \!  
{\underset{m,n \in \IZ \atop m \neq 0}{\sum}} \; \Lambda_1^{|m|}\,\Lambda_2^{|n|} \, \frac{\e(n z_2 - m z_1)}{m(m\t+n)}\,.
\end{split}
\ee
The real cut-offs~$0<\Lambda_{1}<1$ and~$0<\Lambda_{2}<1$, must be taken to~$\Lambda_1=1^-$ and~$\Lambda_2=1^-$ at the very end. The smoothened version~$\log P_{0 \Lambda}$ and its derivatives are absolutely and thus uniformly convergent for every real~$z_1$ and~$z_2$. Thus,~$\log P_{0\Lambda}$ is smooth i.e.~even divergences are also resolved by the insertion of $\Lambda$. There is a price to pay though as the piecewise meromorphy of~$\log P_0$ gets spoiled in an infinitesimal band of width of order~$1-\Lambda$ centered at~$z_2\in \mathbb{Z}$. We give an analytic description of this in terms of the cut-offs in appendix~\ref{Cutofff}. Next we will do it with distributions 

\subsection*{Recovering~$\theta_0$ from~$P_0$}

The right-hand side of~\eqref{P0Theta} can be triviallly de-periodised by using quasi-periodic properties of~$\theta_0$ (See appendix~\ref{app:QuasiP}). Here we would like to prove that, starting from the double Fourier expansion of~$P_0$ and upon the requirement of meromorphy. Quasi-periodicity will be an outcome. It is also a convenient introduction for the slightly less trivial case of elliptic Gamma functions that will be presented next.

Let us introduce the following new auxiliary complex variables~$u=u_1+\t u_2\,$, $v= u_2$ and the derivatives 
\be\label{twistedDerMBT}
\partial_u \= \partial_{u_1}\,,\qquad \partial_v\= \partial_{u_2} \, -\,\t\,\partial_{u_1}\,.
\ee
A function of~$u$ and~$v$ is a holomorphic function of~$u$ if its derivative with respect to~$v$ vanishes. 

After acting on~$\log P_0$, with~$\partial_v$, using
the Fourier expansion of~$B_1(\{z\})$ given in~\eqref{B1Four}, and the Poisson summation identity
\be
\sum_{m\in\mathbb{Z}} \e( m z) \=\sum_{m\in\mathbb{Z}} \delta(z+m)\,,
\ee
we obtain the following identity
\be \label{IdentityLogPalpba}
\begin{split}
\partial_{v} \, \log P_{0}&\=\,2\pi\i \,B_1(\{u_1\})\,\Bigl(\sum_{m\in \mathbb{Z}} \,\delta(v\,+\,m)\Bigr)\,.
\end{split}
\ee
where the derivative~$\partial_v$ is taken at fixed~$u\,\in\,\mathbb{C}$ and~$u_2\=v$.
In this expression~$u_1\,\in\, \mathbb{R}$ stands for the combination~$u\,-\,v\,\t$ with~$u\in \mathbb{C}$ and~$u_2\,=\,v$. Equation~\eqref{IdentityLogPalpba} is telling us that~$ P_0$ is meromorphic in~$u$ for every~$u_2\,\notin\,\mathbb{Z}$. It is also telling us that the lack of meromorphy corresponds to a discontinuity located at~$u_2\,\in\,\mathbb{Z}$. 

This discontinuity can be cancelled with the help of the function~$R_{P}$ that we define in~\eqref{Rp}. Namely, the combination~$P_0 \,R^{-1}_{P}$ is meromorphic in~$u$. Moreover, from the relation~\eqref{P0Theta} and uniqueness of solutions to Cauchy-Riemann relations, it follows that for complex~$u$
\be\label{Identitytheta0}
\begin{split}
\theta_0(u;\t)\= R^{-1}_{\theta_0} \,P_0(u-v\t,v) \,, \qquad v= u_2\,,\\
\end{split}
\ee
where
\be\label{Rm1theta0}
R^{-1}_{\theta_0}\,\equiv\,\frac{R^{-1}_P(u,v)}{R^{-1}_P(u,0^+)}\=\e\Bigl(\frac{1}2 \lfloor v\rfloor (1 - 2 \{u\} + (\lfloor v\rfloor+1)\tau )\Bigr)\,.
\ee
Note that~$\{u\}$ can be safely substituted by~$u$ when evaluating the exponential. Using the bracket corresponds to a~$v$-dependent choice of branch for the corresponding logarithm.

Relation~\eqref{Identitytheta0} can proven as follows: in virtue of~\eqref{P0Theta} the function 
\be\label{EqualityR}
\frac{P_0(u-v\t,v) R^{-1}_P(u,v)}{R^{-1}_P(u,0^+)}
\ee
equals the function~$\theta_0(u)$ at~$v=0^+$ and~$u\in \mathbb{R}$. Finally, Cauchy-Riemann relations in the complex extension of the variable~$u$, with the component~$u_2$ identified with~$v$, imply equality~\eqref{Identity} for all~$u\in \mathbb{C}$ with~$v=u_2$.  A much simpler way to prove it, by using quasi-periodicity properties of~$\theta_0$ can be found in appendix~\ref{app:QuasiP}.

\subsection*{Properties of~$Q_0$}

The~$P_0$ and~$Q_0$ determine a double-periodic extension of the elliptic Gamma function~$\Gamma_e$ (as will be recalled below). We can define~$Q_0$ via the following logarithmic branch 
\be\label{logQnew1}
\log Q_{0}(z)\=\log Q_0(z_1,z_2) \,\equiv\,  - \frac{1}{4\pi^2} \!
{\underset{m,n \in \IZ \atop m \neq 0}{\sum}} \; \frac{\e(n z_2 - m z_1)}{m(m\t+n)^{2}} \,.
\ee
This function is continuous, and it has no divergences for~$u_{1}$,~$u_2\,\in\, \mathbb{R}$. Its derivatives have discontinuities at~$u_2\,\in\, \mathbb{Z}$. Indeed, after acting with~$\partial_v \=\partial_{u_2}\,-\,\t \partial_{u_1}$ on the right-hand side of~\eqref{logQnew1} one obtains
\be\label{Idenitity1}
\partial_v \log Q_{0} \= \log P_{0}\,.
\ee
Then, the previous analysis for~$\log P_0$ applies. We should stress that~$P_0$ and ~$Q_0$ are always thought to be functions of~$u_1$ and~$u_2$. The meaning of~$\partial_v$ in~\eqref{Idenitity1} is that of considering the complex linear combination of~$\partial_{u_1}$ and~$\partial_{u_2}$ as stated in equation~\eqref{twistedDerMBT}.

Although we will not make it explicit at every step, we will always work, unless we indicate explicitly otherwise, with the smooth functions
\be\label{logQL}
\begin{split}
\log Q_{0\Lambda}(z;\t)&\,\equiv\, -\frac{1}{4\pi^2}  \!  
{\underset{m,n \in \IZ \atop m \neq 0}{\sum}} \; \Lambda_1^{|m|}\,\Lambda_2^{|n|} \, \frac{\e(n z_2 - m z_1)}{m(m\t+n)^2}\,.
\end{split}
\ee
The real cut-offs~$\Lambda_{1}$ and~$\Lambda_{2}$ are the same as before. The smooth version~$\log Q_{0 \Lambda}$ and its derivatives are absolutely and thus uniformly convergent for every real~$z_1$ and~$z_2$. This is, the function~$\log Q_{0\Lambda}$ is smooth.  In the very end the cut-offs must be taken to one.

\subsection*{Towards fixing the remainder function in~\eqref{EffectiveActionMero2} } 
In appendix~\ref{CRrelations} we have shown the following identity which holds for~$z_2\notin \mathbb{Z}$
\be\label{Q0Gamma}
\frac{\,\,P_0(z;\t)^{\{z_2\}}}{Q_0(z;\t)}\=\Ge(z_1+\t (\{z_2\}+1);\t,\t)\,.
\ee
Here ~$\{z_2\}\equiv z_2-\lfloor z_2\rfloor$ and
\be
P_0^{\{z_2\}}\,\equiv\,\exp(\{z_2\}\,\log P_0)\,.
\ee
The function~$Q_0$ is understood to be the exponential of~$\log Q_0$. 
This identity is saying that the combination in the left-hand side is piecewise meromorphic in~$z$. Here, we construct a combination of~$P_0$ and~$Q_0$ that is meromorphic. This is, let us de-periodise relation \eqref{Q0Gamma}.  In appendix~\ref{app:QuasiP} we do so by using quasi-periodic properties of~$\theta_0$ and~$\Ge$. Here, we would like to do it starting from a deformation of the natural ``de-periodization" of the left-hand side of~\eqref{Q0Gamma}, upon demanding meromorphy. Quasi-periodicity will be an outcome. 

We start by noticing the following identity which holds for~$v\in \mathbb{R}$
\be\label{IdenitityPQG}
\partial_v \log \Bigl(\frac{\,\,P_0^{\,v\,}}{Q_0}\Bigr) \=\,2\pi\i \,v \,B_1(\{u_1\})\,\Bigl(\sum_{m\in \mathbb{Z}} \,\delta(v\,+\,m)\Bigr)\,.
\ee
We define
\be
P_0^{\,v\,} \equiv \,\exp( v\,\log P_0)\,.
\ee
Equation~\eqref{IdenitityPQG} follows from the definitions of~$P^v_0$,~$Q_0$ and~\eqref{IdentityLogPalpba}. As before, the derivative~$\partial_v$ is taken at fixed~$u\in\mathbb{C}$ and ~$u_2=v$.
In this expression~$u_1\in \mathbb{R}$ stands for the combination~$u\,-\,v\t$ with ~$u\in \mathbb{C}$ and~$u_2=v$. Equation~\eqref{IdentityLogPalpba} is telling us that~$\frac{P_0^v}{Q_0}$ is holomorphic in~$u$ for every~$u_2\notin\mathbb{Z}\,$. It is also telling us that the lack of holomorphy corresponds to a discontinuity located at~$u_2\in\mathbb{Z}\,$. 

This discontinuity can be cancelled with the help of the function~$R_{vP}$ that we define in~\eqref{Rvp}. Namely, the combination~$\frac{P^{v}_0}{Q_0} \,R^{-1}_{vP}$ is holomorphic in~$u\,$.~\footnote{.... whenever the expression in terms of double Fourier expansions converges. If such expression diverges at some point, then that point is guarantied to be a pole.} Moreover, from the relation~\eqref{P0Theta} and uniqueness of solutions to Cauchy-Riemann relations, it follows that for~$u\in \mathbb{C}$
\be\label{Identity}
\begin{split}
\Ge(u+\t;\t,\t)\=R_{\Ge}^{-1}\,\,\frac{P_0(u-v\t,v)^v}{Q_0(u-v\t,v)} \,, \qquad v= u_2\,.
\end{split}
\ee
The remainder function~\eqref{Ambiguity} in this case is fixed to be
\be\label{remainderF}
\begin{split}
R^{-1}_{\Ge}\,\equiv\,\frac{R^{-1}_{vP}(u,v)}{R^{-1}_{vP}(u,0^+)}&\= \frac{q^{\frac{B_3(\lfloor v \rfloor+1)}{3}}\,\e\Bigl(-\,\frac{1}{2} B_1(\{u\}) B_2(\lfloor v\rfloor+1)\Bigr)}{q^{\frac{B_3(\lfloor 0^+ \rfloor+1)}{3}}\,\e\Bigl(-\,\frac{1}{2} B_1(\{u\}) B_2(\lfloor 0^+ \rfloor+1)\Bigr)}\,  \\
&\=q^{\frac{B_3(\lfloor v \rfloor+1)}{3}}\,\e\Bigl(-\,\frac{1}{2} B_1(\{u\})\bigl(B_2(\lfloor v\rfloor+1)\,-\,\frac{1}{6}\bigr)\Bigr)\, \\
&\=q^{\frac{B_3(\lfloor v \rfloor+1)}{3}}\,\e\Bigl(-\,\frac{1}{2} B_1(u)\bigl(B_2(\lfloor v\rfloor+1)\,-\,\frac{1}{6}\bigr)\Bigr)\,,
\end{split}
\ee
where~$R^{-1}_{vP}$ was defined in~\eqref{Rvp}.

Note that if~$u_1\notin\mathbb{Z}$, then~$\{u\}$ in the second line can be safely substituted by~$u$ in the third, when evaluating the exponential.~\footnote{Definition of the~$\{u\}$ when~$u$ is complex has been given in appendix~\ref{app:CancT} equation~\eqref{BracketX}.} Using the bracket corresponds to picking up a~$u$-dependent choice of branch for the corresponding logarithm. 

Relation~\eqref{Identity} is proven as follows: In virtue of~\eqref{Q0Gamma}
\be
\,\frac{R^{-1}_{vP}(u,v)}{R^{-1}_{vP}(u,0^+)}\,\frac{P_0(u-v\t,v)^v}{Q_0(u-v\t,v)}
\ee
equals the function~$\Ge(u+\t;\t,\t)$ at ~$v=0^+$ and~$u\in \mathbb{R}$. Finally, Cauchy-Riemann relations in the complex extension of the variable~$u$, with the component~$u_2$ identified with~$v$, implies equality~\eqref{Identity} for all~$u\in \mathbb{C}$ with~$v=u_2$.

\paragraph{Comments on~$R_{\Ge}$}

Note that if we change the definition of~\eqref{remainderF} by multiplying it by a phase~$\e(K)$ where we assume~$K$ to be an arbitrary piecewise-constant function taking values in the integers, then identity~\eqref{Identity} also applies to the new definition of~$R_{\Ge}$. Obviously, this is because~$\e(K)\=1$. As we have said, by using this ambiguity, which is a choice of branch, one can remove the bracket  from~$\{u\}$ in the exponent of~\eqref{remainderF}.

\subsection{Holomorphic extension of~$\theta_0$ and~$\Ge$ to the lower-half of~$\tau$-plane}\label{par:HolomorphicExtension}

We pause and comment about a subtle but important result that is implicit in the previously proven identities. Notice that the product representation~\eqref{GammaeDef} is absolutely convergent for~$\text{Im}(\t)>0$.
This does not mean that the elliptic gamma functions~$\Ge$ can not be continued to~$\text{Im}(\t)=0$. In fact~$\Ge$ is well-defined at infinitely many points in the real axis~$\text{Im}(\t)=0$. That was shown in subsection~3.5 of~\cite{Felder2000}. Specifically they showed that (at least) for irrational values of~$\t$ the function~$\Ge(z;\t+\i\epsilon)$ is continuous as the infinitesimal real~$\epsilon$ crosses~$0$ from the negatives to the positives and~\emph{viceversa}. This result is consistent with Identity~\eqref{Identity}. 

Moreover, the right-hand side of~\eqref{Identity} can be understood as the definition of~$\Ge$ in the lower-half plane~$\text{Im}(\t)<0$ i.e.~as its analytic continuation to the lower half-plane. From~\eqref{Identity} it follows that rational~$\t$'s are singularities of~$\Ge$. This means that holomorphic flows to the lower-half $\t$-plane are possible when they go across irrational values of~$\t$. That claim is consistent with the results of~\cite{Felder2000}. 

Everything we have just said for~\eqref{Identity} relative to~$\Ge$, applies also to the representation~\eqref{Identitytheta0} of~$\theta_0$.

The existence of these holomorphic flows from the upper to the lower half~$\tau$-plane is a highly non-trivial property that proves to be very hard to show existence of, in many cases.  Identities~\eqref{Identitytheta0} and~\eqref{Identity} demonstrate that such extensions exist for the quasi-elliptic functions~$\theta_0$ and $\Ge$. Their existence will be important to the analysis regarding large-$N$ counting of operators that will be presented in section~\ref{sec:Counting}. It will allow us to define the contour of integration~$C_\eta$ in section~\ref{sec:Counting}.

From these identities one can see that the index~$\mathcal{I}$ will have essential singularities at rational values of~$\tau$. Take for instance~$\tau=0$.~The asymptotic behaviour of the index at such point depends on how one approaches it. This was previously pointed out in~\cite{Cabo-Bizet:2019osg}. More generally, for rational values~$\tau=-\frac{n}{m}$ an analogous conclusion holds, as pointed out in~\cite{Cabo-Bizet:2020nkr}. We encountered this issue in section~\ref{sec:Counting}. There it was further argued that these essential singularities are of paramount importance for operator counting.

Moreover, the relation of~$\theta_0$ and~$\Ge$ to~$P_0$ and~$Q_0$ allows us to use the latter two to understand how the former two behave in a vicinity of the real axis~$\text{Im}(\t)=0$ and further extend them to the lower-half plane.

\subsection*{Relation between~$Q_0$ and~$\Ge$ from quasi-periodicity}\label{app:QuasiP}

In this appendix we prove~\eqref{Identity} with the use of quasi-periodicity properties of~$\Ge$. We start from the logarithm of relation \eqref{RelationGPQ}
\be \label{logtP}\begin{split}
\log G(z_1+\t \{z_2\};\t)&\,=\,\log \Ge(z_1+\t z_2+\t;\t,\t)\=- \log Q_0(z;\t)+\{z_2\} \log P_0(z;\t) \, \\
&\=-\log Q_0(z;\t)+\{z_2\} \log \theta_0(z_1+\t \{z_2\};\t)\,.
\end{split}
\ee
Next we prove a couple of identities that will help in ``de-periodising"~\eqref{logtP}. Assume without loss of generality that~$\lfloor z_2\rfloor>0$ then
\be\label{thetaIdApp}\begin{split}
\log \theta_0(z_1+\t \{z_2\};\t)&\= \log\theta_0(z;\t)+2\pi\i \sum_{i=1}^{\lfloor z_2\rfloor} (z-i\t-\frac{n}{2}) \\
&\= \log\theta_0(z;\t)+2\pi\i \lfloor z_2\rfloor (z-\frac{n}{2}) -2\pi\i \t \frac{\lfloor z_2\rfloor(\lfloor z_2\rfloor+1)}{2}\,.
\end{split}
\ee
The final answer for the case~$\lfloor z_2 \rfloor<0$ is the same. In this equation~$n\in\mathbb{Z}$ is an arbitrary quantity that corresponds to a choice of branch. Note that $n$ can be a discontinuous function of~$z$. Note that identity~\eqref{thetaIdApp}
is a version of~\eqref{Rm1theta0} in terms of logarithms after identifying~$z$ and~$u$ and~$z_2$ with~$v$.
The second identity is
\be\label{GIdApp}\begin{split}
\log G(z_1+\t \{z_2\};\t)&\=\log G(z;\t)- \sum_{i=0}^{\lfloor z_2 \rfloor-1} \log \theta_0(z-i\t;\t)
\\
&\=\log G(z;\t)- \sum_{i=0}^{\lfloor z_2 \rfloor-1} \bigl(\log \theta_0(z;\t)+2\pi\i\sum_{j=1}^{i}(z- j\t-\frac{n}{2})\bigr) \\
&\=\log G(z;\t)- \lfloor z_2 \rfloor \log \theta_0(z;\t)-2\pi\i \frac{\lfloor z_2\rfloor(\lfloor z_2\rfloor-1)}{2} (z-\frac{n}{2})  \\
& \qquad\,+\frac{2\pi\i\t}{6} \lfloor z_2\rfloor (\lfloor z_2\rfloor^2-1) \,.
\end{split}
\ee
Finally, from~\eqref{logtP},~\eqref{thetaIdApp} and~\eqref{GIdApp} it follows that
\be
\log \Ge (z_1+\t z_2+\t;\t,\t)\= -\log Q_0(z_1,z_2;\t) \,+\,z_2\log P_0(z_1,z_2;\t)\,+\,\log R_{\Ge}^{-1}\,,
\ee
where  
\be\label{EqRApp}
\log R_{\Ge}^{-1}\=\,\frac{2\pi\i \t}{3} B_3(\lfloor z_2\rfloor+1 ) \,-\,\pi \i\, \bigl(z\,-\,\frac{n}{2}\bigr)\Bigl(B_2(\lfloor z_2\rfloor + 1)\,-\,\frac{1}{6}\Bigr) .
\ee
After the following identifications~$z\=u \= u_1+\t v$, one can check that the exponential of~\eqref{EqRApp} maps to~\eqref{remainderF}. We note that the two equations were computed in different ways. Equation~\eqref{remainderF} was computed via ``differential" methods, meanwhile~\eqref{EqRApp} was computed via quasi-periodicity properties. They both represent the same object, which as~\emph{a priori} expected, can be recast as a remainder function of the regularization procedure introduced in section~\ref{sec:Regularization}.

\subsection{Placing the discontinuities at~$u=0$ for every~$\rho$} \label{Proofremainder}
This corresponds to a proper choice of integration constants~$c_1(u)$ and~$c_2(u)$, piece-wise continuous functions of $u$, to be more precise, in the solution to equation~\eqref{EquationRvP} given in~\eqref{Rvp}.
Let us start by assuming~{$-1<\Delta_2<1\,$}.  A choice of remainder function~$R$ that solves equation~\eqref{ExtensionHolPre} is
\be\label{Prefactor}
\begin{split}
R^{-1}\,&\= R^{-1}(u,v)\,\equiv\, \, \frac{R^{-1}_{vP}(z(u),z_2(v))}{R^{-1}_{vP}(z(u),z_2(0^+))}\\\
&\= \frac{q^{\frac{B_3(\lfloor z_2(v) \rfloor+1)}{3}}\,\e\Bigl(-\,\frac{1}{2} B_1(\{z(u)\}) B_2(\lfloor z_2(v)\rfloor+1)\Bigr)}{q^{\frac{B_3(\lfloor z_2(0^+) \rfloor+1)}{3}}\,\e\Bigl(-\,\frac{1}{2} B_1(\{z(u)\}) B_2(\lfloor z_2(0^+)\rfloor+1)\Bigr)}\,, \\
&\=  \frac{q^{\frac{B_3(\lfloor z_2(v) \rfloor+1)}{3}}\,\e\Bigl(-\,\frac{1}{2} B_1(z(u)) B_2(\lfloor z_2(v)\rfloor+1)\Bigr)}{q^{\frac{B_3(\lfloor z_2(0^+) \rfloor+1)}{3}}\,\e\Bigl(-\,\frac{1}{2} B_1(z(u)) B_2(\lfloor z_2(0^+)\rfloor+1)\Bigr)}\,,
\end{split}
\ee
where~$R_{vP}^{-1}$ was defined in~\eqref{Rvp}. In going from the second to the third line we have used the~$2\pi\i \mathbb{Z}\,$-shift ambiguity of the exponent.
Moreover, as explained in appendix~\ref{app:CancT} below~\eqref{CircB}, when the argument of the function $floor$ hits integer values, we must use the definitions of~$B^\circ_2$ and~$B^\circ_3$ that were given in~\eqref{CircB}, instead of the Bernoulli polynomials~$B_2$ and~$B_3\,$. This is what Dirichlet theorem indicates to do.
The bracket function of the complex variable~$z(u)$ is defined by replacing the real component~$z_1$ by~$\{z_1\}$ and leaving~$z_2$ unchanged (See definition \eqref{BracketX}). The term in the denominator of~\eqref{Prefactor} is independent of~$v$. It corresponds to a choice of integration constant that guarranties~$R^{-1}\to 1$ to vanish as~$v^i\to 0^+$.

For~$-1<\Delta_2<1$ this choice of~$R^{-1}$ corresponds to the choice~$c_1=-\frac{1}{6}$ and~$c_2\=0$ in~\eqref{Rvp}.

To prove that the choice~\eqref{Prefactor} solves~\eqref{ExtensionHolPre} we proceed as follows. Start by noticing that for~$v\in \mathbb{R}$
\be\label{IdenitityPQG2}
\partial_{v^i} \log\Bigl(\frac{\,\,P_0^{\,z_2(v)\,}}{Q_0}\Bigr) \=\,2\pi\i\,\rho^i \, z_2(v) \,B_1(\{z(u)\})\,\Bigl(\sum_{m\in \mathbb{Z}} \,\delta(z_2\,+\,m)\Bigr)\,.
\ee
This derivative is taken at fixed~$u^i\,$. The Dirac-delta term comes from a discontinuity of the function upon which~$\partial_{v^i}$ is acting. Such discontinuity is cancelled by adding a multiplying pre-factor~$R^{-1}$, the one defined in~\eqref{Prefactor}, to the discontinuous function~${\,\,P_0^{\,z_2(v)\,}}/{Q_0}$. This was explained in appendix~\ref{app:CancT}.

Finally, in virtue of identity~\eqref{Q0Gamma}, it follows that upon substitution of~\eqref{Prefactor} in~\eqref{ExtensionHolPre}, the latter holds at least for~$v=0^+$ for~$0\leq\Delta_2<1$. This together with uniqueness of solutions to Cauchy-Riemann relations, imply that~\eqref{ExtensionHolPre} holds for any~$u\in \mathbb{C}$ and~$v=u_2$. More precisely, in virtue of~\eqref{Q0Gamma} and for~$0\leq\Delta_2<1$ the function
\be\label{ExpreR}
\,\frac{R^{-1}_{vP}(z(u),z_2(v))}{R^{-1}_{vP}(z(u),z_2(0^+))}\,\frac{P_0(z_1(u,v),z_2(v))^{z_2(v)}}{Q_0(z_1(u,v),z_2(v))}
\ee
equals~$\Ge(\rho(u)+\Delta+\t;\t,\t)$ at~$v=0^+$ and~$u\in \mathbb{R}$. At last, Cauchy-Riemann relations in the complex extension of the variable~$u$, with the component~$u_2$ identified with~$v$, imply equality~\eqref{ExtensionHolPre} to hold upon using the definition of $R$ given in~\eqref{Prefactor}, for all~$u\in \mathbb{C}$ and~$v=u_2\,$. 

\begin{figure}\centering
\includegraphics[width=6.8cm]{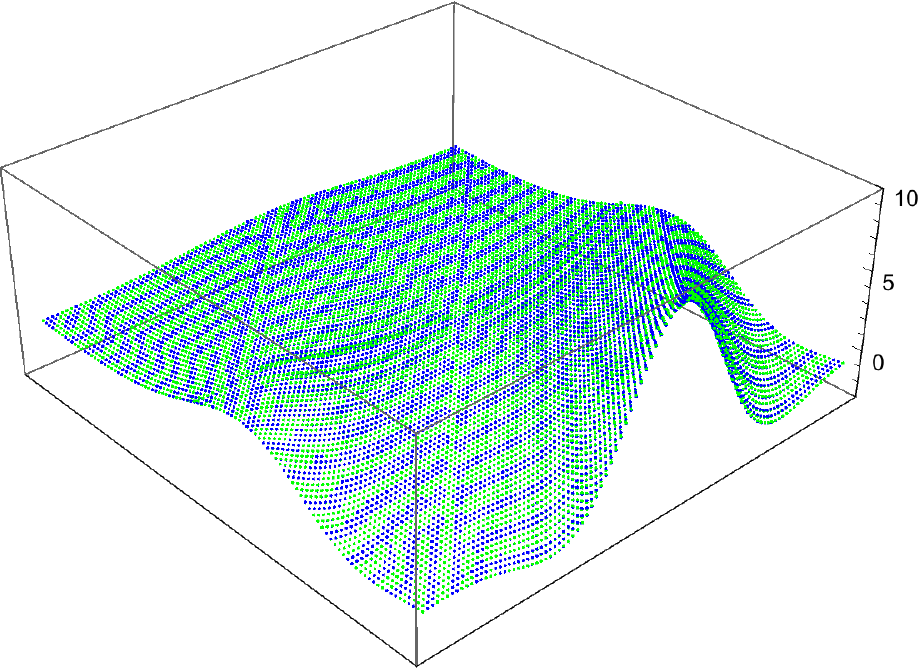} 
\qquad 
\includegraphics[width=6.8cm]{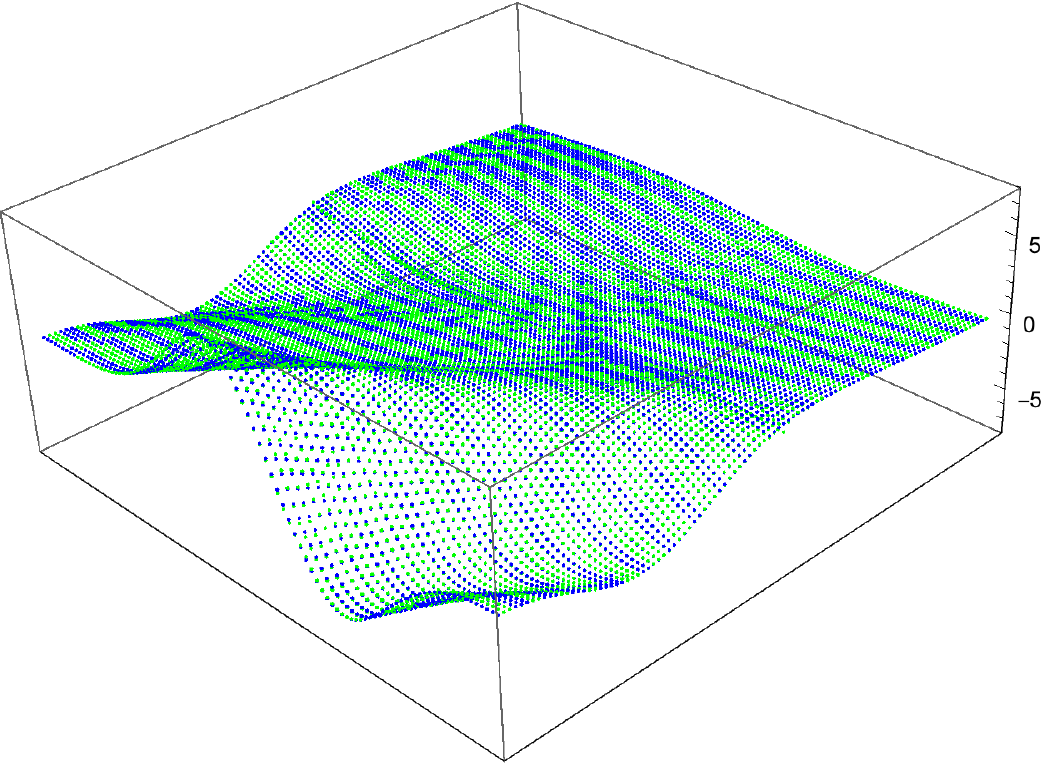}
 \caption{ The plots of the real (left) and imaginary (right) parts of the left- and right-hand sides of~\eqref{ExtensionHolPre} for gauge group~$U(1)$ and~$\rho(u)=u$, as a function of~$0<u_1<1$ and~$0<u_2<1$, and for~$\tau =1 + \i/3$,~$\Delta_1=0.4325$,~$\Delta_2= 0.54731$. The blue (resp. green) points are obtained out of the left (resp. right)-hand side of~\eqref{ExtensionHolPre}. We have used the product representation of~$\Ge(z;\t,\t)$ for the left hand side. Both product and double Fourier representations were truncated at level~$20$. More precisely, the product representation has been truncated at the term~$q^{20}$.}
  \label{fig:GammaComparison}
\end{figure}

If~$\Delta_2$ is not in between~$0$ and~$1$ we can not identify the limit~$v=0^+$ of~\eqref{ExpreR} with the left-hand side of~\eqref{Q0Gamma} (for~$u\in \mathbb{R}$). In that case the proof of~\eqref{ExtensionHolPre} needs to be slightly modified by using an alternative to~\eqref{Q0Gamma} that we wrote in~\eqref{EqsComplem}. After identifying the limit~$v=0^+$ of~\eqref{ExpreR} with the left-hand side of that equation (for~$u\in \mathbb{R}$) it follows that~\eqref{ExtensionHolPre} holds also when~$\Delta_2$ ranges in between~$-1$ and~$0$ i.e.~$-1<\Delta_2\leq0\,$.

\paragraph{A non Abelian formula for~$\theta_0$ in terms of~$P_0$}
From completeness we present a non Abelian formula that is analogous to~\eqref{ExtensionHolPre}.

This one relates the~$\theta_0$ and~$P_0$ when evaluated at non Abelian arguments~$z(u)=\rho(u)+\Delta$, for any value of~$u^{i}$'s and~$0\leq\Delta_2<1$.  The formula is
\be\label{IdentityNonabelianTheta0}
\begin{split}
\theta_0(z(u);\t)\= {R}_{\theta_0}^{-1} \,P_0(z_1,z_2) \,, \qquad v= u_2\,,\\
\end{split}
\ee
where~$z_1$ and~$z_2$ are the ones defined in~\eqref{z1z2Funcuv} and
\be
\begin{split}
{R}_{\theta_0}^{-1}&\= {R}^{-1}_{\theta_0}(u,v) \,\equiv\,\frac{R^{-1}_P(z(u),z_2(v))}{R^{-1}_P(z(u),z(0^+))}\, \\
&\=\frac{q^{\frac{B_2(\lfloor z_2(v) \rfloor\,+\,1)}{2}}\,\e{\Bigl(-B_1(\{z(u)\}) B_1(\lfloor z_2(v)\rfloor+1)\Bigr)}}{q^{\frac{B_2(\lfloor z_2(0^+)\rfloor\,+\,1)}{2}}\,\e{\Bigl(-B_1(\{z(u)\}) B_1(\lfloor z_2(0^+)\rfloor+1)\Bigr)}} \,\\
&\=\frac{q^{\frac{B_2(\lfloor z_2(v) \rfloor\,+\,1)}{2}}\,\e{\Bigl(-B_1(z(u)) B_1(\lfloor z_2(v)\rfloor+1)\Bigr)}}{q^{\frac{B_2(\lfloor z_2(0^+)\rfloor\,+\,1)}{2}}\,\e{\Bigl(-B_1(z(u)) B_1(\lfloor z_2(0^+)\rfloor+1)\Bigr)}}\,.
\end{split}
\ee
In going from the second to the third line we have used the~$2\pi\i \mathbb{Z}\,$-shift ambiguity of the exponent.
This formula can be proven by following analogous steps to the ones followed in reaching~\eqref{ExtensionHolPre}. It can also be checked to follow from~\eqref{P0Theta} and the use of quasi-periodic properties of~$\theta_0\,$.

\subsection{Analysis in the presence of cut-off} \label{Cutofff}

In this subsection we arrive to the relations~\eqref{IdentityLogPalpba} and~\eqref{Idenitity1} with the use of cut-offs~$0<\Lambda_{1,2}<1$.~\footnote{We are abusing notation in the sense that now we consider two independent cut-offs~$\Lambda_1$ and~$\Lambda_2$ instead of one. The single cut-off mentioned in the main body of this paper corresponds to~$\Lambda_1=\Lambda_2=\Lambda$.}

The second of these relations follows trivially from~\eqref{logPL} and \eqref{logQL}
\be
\begin{split}
\partial_v \log Q_{0\Lambda}\=\log P_{0\Lambda} \,.
\end{split}
\ee
The first one follows from
\be\label{DeltaIllustration}
\begin{split}
\,\partial_v  \log P_{0\Lambda}&\=  \!  
{\underset{m,n \in \IZ \atop m \neq 0}{\sum}} \; \Lambda_1^{|m|}\,\Lambda_2^{|n|} \, \frac{\e(n u_2 - m u_1)}{m} \,\\&\= 2 \pi\i \,B_{1\Lambda}(\{u_1\}) \sum_{n} \Lambda^{|n|}_2 \e( n u_2)\, \\
&\= 2 \pi\i\, \,B_{1\Lambda}(\{u_1\})\,\frac{1\, -\, \Lambda_2^2}{1 \,+\, \Lambda_2^2 \,-\, 
 2\, \Lambda_2 \cos(2\pi u_2)}\,.  \\
\end{split}
\ee

Note that the right-hand side is a pure phase, and it is~$\tau$-independent at fixed~$u_1$ and~$u_2$. Note also that upon taking the limit~$\Lambda_2\to 1$ \eqref{DeltaIllustration} vanishes everywhere except for at the points $z_2\in \mathbb{Z}$ where the result goes to infinity. Moreover, a computation shows that if one integrates the pre-factor in the right-hand side of~\eqref{DeltaIllustration} along the segment~$-\epsilon<v<\epsilon$, with $\epsilon$ being an infinitesimal positive real number, and extracts the leading contribution to such result after expanding it around~$\Lambda=1$, the final result turns out to be independent of~$\epsilon$ and it matches the analogous integral of the right-hand side of \eqref{IdentityLogPalpba}. These facts together with periodicity under the shifts~$u_2\mapsto u_2+1$, imply that in the limit~$\Lambda\to 1$ the right-hand side of~\eqref{DeltaIllustration} approaches the Dirac-delta term in the right-hand side of~\eqref{IdentityLogPalpba}.


\begin{thebibliography}{10}

\bibitem{Maldacena:1997re2}
J.~M. Maldacena, {\it {The Large N limit of superconformal field theories and
  supergravity}},  {\em Int. J. Theor. Phys.} {\bf 38} (1999) 1113--1133,
  [\href{http://arxiv.org/abs/hep-th/9711200}{{\tt hep-th/9711200}}].

\bibitem{Strominger:1996sh}
A.~Strominger and C.~Vafa, {\it {Microscopic origin of the Bekenstein-Hawking
  entropy}},  {\em Phys. Lett.} {\bf B379} (1996) 99--104,
  [\href{http://arxiv.org/abs/hep-th/9601029}{{\tt hep-th/9601029}}].

\bibitem{Benini:2015eyy}
F.~Benini, K.~Hristov, and A.~Zaffaroni, {\it {Black hole microstates in
  AdS$_{4}$ from supersymmetric localization}},  {\em JHEP} {\bf 05} (2016)
  054, [\href{http://arxiv.org/abs/1511.04085}{{\tt arXiv:1511.04085}}].

\bibitem{Cabo-Bizet:2018ehj}
A.~Cabo-Bizet, D.~Cassani, D.~Martelli, and S.~Murthy, {\it {Microscopic origin
  of the Bekenstein-Hawking entropy of supersymmetric AdS$_{5}$ black holes}},
  {\em JHEP} {\bf 10} (2019) 062, [\href{http://arxiv.org/abs/1810.11442}{{\tt
  arXiv:1810.11442}}].

\bibitem{Choi:2018hmj}
S.~Choi, J.~Kim, S.~Kim, and J.~Nahmgoong, {\it {Large AdS black holes from
  QFT}},  \href{http://arxiv.org/abs/1810.12067}{{\tt arXiv:1810.12067}}.

\bibitem{Benini:2018ywd}
F.~Benini and P.~Milan, {\it {Black Holes in 4D $\mathcal{N}$=4
  Super-Yang-Mills Field Theory}},  {\em Phys. Rev. X} {\bf 10} (2020), no.~2
  021037, [\href{http://arxiv.org/abs/1812.09613}{{\tt arXiv:1812.09613}}].

\bibitem{Cabo-Bizet:2019eaf}
A.~Cabo-Bizet and S.~Murthy, {\it {Supersymmetric phases of 4d $ \mathcal{N} $
  = 4 SYM at large $N$}},  {\em JHEP} {\bf 09} (2020) 184,
  [\href{http://arxiv.org/abs/1909.09597}{{\tt arXiv:1909.09597}}].

\bibitem{Cabo-Bizet:2020nkr}
A.~Cabo-Bizet, D.~Cassani, D.~Martelli, and S.~Murthy, {\it {The large-$N$
  limit of the 4d $ \mathcal{N} $ = 1 superconformal index}},  {\em JHEP} {\bf
  11} (2020) 150, [\href{http://arxiv.org/abs/2005.10654}{{\tt
  arXiv:2005.10654}}].


\bibitem{Benini:2018mlo}
F.~Benini and P.~Milan, {\it {A Bethe Ansatz type formula for the
  superconformal index}},  {\em Commun. Math. Phys.} {\bf 376} (2020), no.~2
  1413--1440, [\href{http://arxiv.org/abs/1811.04107}{{\tt arXiv:1811.04107}}].

\bibitem{Gutowski:2004yv}
J.~B. Gutowski and H.~S. Reall, {\it {General supersymmetric AdS(5) black
  holes}},  {\em JHEP} {\bf 04} (2004) 048,
  [\href{http://arxiv.org/abs/hep-th/0401129}{{\tt hep-th/0401129}}].

\bibitem{Cvetic:2004hs}
M.~Cvetic, H.~Lu, and C.~Pope, {\it {Charged Kerr-de Sitter black holes in five
  dimensions}},  {\em Phys. Lett. B} {\bf 598} (2004) 273--278,
  [\href{http://arxiv.org/abs/hep-th/0406196}{{\tt hep-th/0406196}}].

\bibitem{Cvetic:2004ny}
M.~Cvetic, H.~Lu, and C.~Pope, {\it {Charged rotating black holes in five
  dimensional U(1)**3 gauged N=2 supergravity}},  {\em Phys. Rev. D} {\bf 70}
  (2004) 081502, [\href{http://arxiv.org/abs/hep-th/0407058}{{\tt
  hep-th/0407058}}].

\bibitem{Kunduri:2005zg}
H.~K. Kunduri and J.~Lucietti, {\it {Notes on non-extremal, charged, rotating
  black holes in minimal D=5 gauged supergravity}},  {\em Nucl. Phys.} {\bf
  B724} (2005) 343--356, [\href{http://arxiv.org/abs/hep-th/0504158}{{\tt
  hep-th/0504158}}].

\bibitem{Kunduri:2006ek}
H.~K. Kunduri, J.~Lucietti, and H.~S. Reall, {\it {Supersymmetric multi-charge
  AdS$_5$ black holes}},  {\em JHEP} {\bf 04} (2006) 036,
  [\href{http://arxiv.org/abs/hep-th/0601156}{{\tt hep-th/0601156}}].

\bibitem{Silva:2006xv}
P.~J. Silva, {\it {Thermodynamics at the BPS bound for Black Holes in AdS}},
  {\em JHEP} {\bf 10} (2006) 022,
  [\href{http://arxiv.org/abs/hep-th/0607056}{{\tt hep-th/0607056}}].

\bibitem{Hosseini:2017mds}
S.~M. Hosseini, K.~Hristov, and A.~Zaffaroni, {\it {An extremization principle
  for the entropy of rotating BPS black holes in AdS$_{5}$}},  {\em JHEP} {\bf
  07} (2017) 106, [\href{http://arxiv.org/abs/1705.05383}{{\tt
  arXiv:1705.05383}}].

\bibitem{Closset:2017bse}
C.~Closset, H.~Kim, and B.~Willett, {\it {$ \mathcal{N} $ = 1 supersymmetric
  indices and the four-dimensional A-model}},  {\em JHEP} {\bf 08} (2017) 090,
  [\href{http://arxiv.org/abs/1707.05774}{{\tt arXiv:1707.05774}}].

\bibitem{Goldstein:2020yvj}
K.~Goldstein, V.~Jejjala, Y.~Lei, S.~van Leuven, and W.~Li, {\it {Residues,
  modularity, and the Cardy limit of the 4d $\mathcal{N}=4$ superconformal
  index}},  \href{http://arxiv.org/abs/2011.06605}{{\tt arXiv:2011.06605}}.

\bibitem{Cassani:2019mms}
D.~Cassani and L.~Papini, {\it {The BPS limit of rotating AdS black hole
  thermodynamics}},  {\em JHEP} {\bf 09} (2019) 079,
  [\href{http://arxiv.org/abs/1906.10148}{{\tt arXiv:1906.10148}}].

\bibitem{Kantor:2019lfo}
G.~K\'antor, C.~Papageorgakis, and P.~Richmond, {\it {AdS$_7$ Black-Hole
  Entropy and 5D $\mathcal{N}=2$ Yang-Mills}},
  \href{http://arxiv.org/abs/1907.02923}{{\tt arXiv:1907.02923}}.

\bibitem{Suh:2018qyv}
M.~Suh, {\it {On-shell action and the Bekenstein-Hawking entropy of
  supersymmetric black holes in $AdS_6$}},
  \href{http://arxiv.org/abs/1812.10491}{{\tt arXiv:1812.10491}}.

\bibitem{Bobev:2019zmz}
N.~Bobev and P.~M. Crichigno, {\it {Universal spinning black holes and theories
  of class $ \mathcal{R} $}},  {\em JHEP} {\bf 12} (2019) 054,
  [\href{http://arxiv.org/abs/1909.05873}{{\tt arXiv:1909.05873}}].

\bibitem{Lezcano:2019pae}
A.~Gonz\'alez~Lezcano and L.~A. Pando~Zayas, {\it {Microstate counting via
  Bethe Ans\"atze in the 4d $ \mathcal{N} $ = 1 superconformal index}},  {\em
  JHEP} {\bf 03} (2020) 088, [\href{http://arxiv.org/abs/1907.12841}{{\tt
  arXiv:1907.12841}}].

\bibitem{Benini:2020gjh}
F.~Benini, E.~Colombo, S.~Soltani, A.~Zaffaroni, and Z.~Zhang, {\it
  {Superconformal indices at large $N$ and the entropy of AdS$_5$ $\times$
  SE$_5$ black holes}},  {\em Class. Quant. Grav.} {\bf 37} (2020), no.~21
  215021, [\href{http://arxiv.org/abs/2005.12308}{{\tt arXiv:2005.12308}}].


\bibitem{Lanir:2019abx}
A.~Lanir, A.~Nedelin, and O.~Sela, {\it {Black hole entropy function for toric
  theories via Bethe Ansatz}},  {\em JHEP} {\bf 04} (2020) 091,
  [\href{http://arxiv.org/abs/1908.01737}{{\tt arXiv:1908.01737}}].

\bibitem{ArabiArdehali:2019orz}
A.~Arabi~Ardehali, J.~Hong, and J.~T. Liu, {\it {Asymptotic growth of the 4d $
  \mathcal{N} $ = 4 index and partially deconfined phases}},  {\em JHEP} {\bf
  07} (2020) 073, [\href{http://arxiv.org/abs/1912.04169}{{\tt
  arXiv:1912.04169}}].

\bibitem{Murthy:2020rbd}
S.~Murthy, {\it {The growth of the $\frac{1}{16}$-BPS index in 4d
  $\mathcal{N}=4$ SYM}},  \href{http://arxiv.org/abs/2005.10843}{{\tt
  arXiv:2005.10843}}.

\bibitem{Agarwal:2020zwm}
P.~Agarwal, S.~Choi, J.~Kim, S.~Kim, and J.~Nahmgoong, {\it {AdS black holes
  and finite N indices}},  \href{http://arxiv.org/abs/2005.11240}{{\tt
  arXiv:2005.11240}}.

\bibitem{Choi:2018vbz}
S.~Choi, J.~Kim, S.~Kim, and J.~Nahmgoong, {\it {Comments on deconfinement in
  AdS/CFT}},  \href{http://arxiv.org/abs/1811.08646}{{\tt arXiv:1811.08646}}.

\bibitem{Honda:2019cio}
M.~Honda, {\it {Quantum Black Hole Entropy from 4d Supersymmetric Cardy
  formula}},  {\em Phys. Rev.} {\bf D100} (2019), no.~2 026008,
  [\href{http://arxiv.org/abs/1901.08091}{{\tt arXiv:1901.08091}}].

\bibitem{ArabiArdehali:2019tdm}
A.~Arabi~Ardehali, {\it {Cardy-like asymptotics of the 4d $ \mathcal{N}=4 $
  index and AdS$_{5}$ blackholes}},  {\em JHEP} {\bf 06} (2019) 134,
  [\href{http://arxiv.org/abs/1902.06619}{{\tt arXiv:1902.06619}}].

\bibitem{Kim:2019yrz}
J.~Kim, S.~Kim, and J.~Song, {\it {A 4d $N=1$ Cardy Formula}},
  \href{http://arxiv.org/abs/1904.03455}{{\tt arXiv:1904.03455}}.

\bibitem{Cabo-Bizet:2019osg}
A.~Cabo-Bizet, D.~Cassani, D.~Martelli, and S.~Murthy, {\it {The asymptotic
  growth of states of the 4d $ \mathcal{N}=1 $ superconformal index}},  {\em
  JHEP} {\bf 08} (2019) 120, [\href{http://arxiv.org/abs/1904.05865}{{\tt
  arXiv:1904.05865}}].

\bibitem{Amariti:2019mgp}
A.~Amariti, I.~Garozzo, and G.~Lo~Monaco, {\it {Entropy function from toric
  geometry}},  \href{http://arxiv.org/abs/1904.10009}{{\tt arXiv:1904.10009}}.

\bibitem{Crichigno:2020ouj}
P.~M. Crichigno and D.~Jain, {\it {The 5d Superconformal Index at Large $N$ and
  Black Holes}},  {\em JHEP} {\bf 09} (2020) 124,
  [\href{http://arxiv.org/abs/2005.00550}{{\tt arXiv:2005.00550}}].

\bibitem{Goldstein:2019gpz}
K.~Goldstein, V.~Jejjala, Y.~Lei, S.~van Leuven, and W.~Li, {\it {Probing the
  EVH limit of supersymmetric AdS black holes}},  {\em JHEP} {\bf 02} (2020)
  154, [\href{http://arxiv.org/abs/1910.14293}{{\tt arXiv:1910.14293}}].

\bibitem{Copetti:2020dil}
C.~Copetti, A.~Grassi, Z.~Komargodski, and L.~Tizzano, {\it {Delayed
  Deconfinement and the Hawking-Page Transition}},
  \href{http://arxiv.org/abs/2008.04950}{{\tt arXiv:2008.04950}}.

\bibitem{Agarwal:2020pol}
P.~Agarwal, K.-H. Lee, and J.~Song, {\it {Classification of large N
  superconformal gauge theories with a dense spectrum}},
  \href{http://arxiv.org/abs/2007.16165}{{\tt arXiv:2007.16165}}.

\bibitem{Agarwal:2019crm}
P.~Agarwal and J.~Song, {\it {Large N Gauge Theories with a Dense Spectrum and
  the Weak Gravity Conjecture}},  \href{http://arxiv.org/abs/1912.12881}{{\tt
  arXiv:1912.12881}}.

\bibitem{Larsen:2019oll}
F.~Larsen, J.~Nian, and Y.~Zeng, {\it {AdS$_{5}$ black hole entropy near the
  BPS limit}},  {\em JHEP} {\bf 06} (2020) 001,
  [\href{http://arxiv.org/abs/1907.02505}{{\tt arXiv:1907.02505}}].

\bibitem{Nian:2020qsk}
J.~Nian and L.~A. Pando~Zayas, {\it {Toward an effective CFT$_{2}$ from $
  \mathcal{N} $ = 4 super Yang-Mills and aspects of Hawking radiation}},  {\em
  JHEP} {\bf 07} (2020) 120, [\href{http://arxiv.org/abs/2003.02770}{{\tt
  arXiv:2003.02770}}].

\bibitem{David:2020ems}
M.~David, J.~Nian, and L.~A. Pando~Zayas, {\it {Gravitational Cardy Limit and
  AdS Black Hole Entropy}},  {\em JHEP} {\bf 11} (2020) 041,
  [\href{http://arxiv.org/abs/2005.10251}{{\tt arXiv:2005.10251}}].

\bibitem{Melo:2020amq}
J.~a.~F. Melo and J.~E. Santos, {\it {Stringy corrections to the entropy of
  electrically charged supersymmetric black holes with $\mathrm{AdS}_5\times
  S^5$ asymptotics}},  \href{http://arxiv.org/abs/2007.06582}{{\tt
  arXiv:2007.06582}}.

\bibitem{David:2020jhp}
M.~David and J.~Nian, {\it {Universal Entropy and Hawking Radiation of
  Near-Extremal AdS$_4$ Black Holes}},
  \href{http://arxiv.org/abs/2009.12370}{{\tt arXiv:2009.12370}}.

\bibitem{Larsen:2020lhg}
F.~Larsen and S.~Paranjape, {\it {Thermodynamics of Near BPS Black Holes in
  AdS$_4$ and AdS$_7$}},  \href{http://arxiv.org/abs/2010.04359}{{\tt
  arXiv:2010.04359}}.

\bibitem{Zaffaroni:2019dhb}
A.~Zaffaroni, {\it {Lectures on AdS Black Holes, Holography and Localization}},
   2019.
\newblock \href{http://arxiv.org/abs/1902.07176}{{\tt arXiv:1902.07176}}.

\bibitem{Duistermaat:1982vw}
J.~Duistermaat and G.~Heckman, {\it {On the Variation in the cohomology of the
  symplectic form of the reduced phase space}},  {\em Invent. Math.} {\bf 69}
  (1982) 259--268.

\bibitem{1983InMat..72..153D}
J.~J. {Duistermaat} and G.~J. {Heckman}, {\it {Addendum to : On the variation in
  the cohomology of the symplectic form of the reduced phase space.}},  {\em
  Inventiones Mathematicae} {\bf 72} (Feb., 1983) 153--158.

\bibitem{Witten:1982im}
E.~Witten, {\it {Supersymmetry and Morse theory}},  {\em J. Diff. Geom.} {\bf
  17} (1982), no.~4 661--692.

\bibitem{berline1983}
N.~Berline and M.~Vergne, {\it Zeros d'un champ de vecteurs et classes
  caracteristiques equivariantes},  {\em Duke Math. J.} {\bf 50} (06, 1983)
  539--549.

\bibitem{Atiyah:1984px}
M.~Atiyah and R.~Bott, {\it {The Moment map and equivariant cohomology}},  {\em
  Topology} {\bf 23} (1984) 1--28.

\bibitem{audin2012torus}
M.~Audin, {\em Torus Actions on Symplectic Manifolds}.
\newblock Progress in Mathematics. Birkh{\"a}user Basel, 2012.

\bibitem{Cordes:1994fc}
S.~Cordes, G.~W. Moore, and S.~Ramgoolam, {\it {Lectures on 2-d Yang-Mills
  theory, equivariant cohomology and topological field theories}},  {\em Nucl.
  Phys. B Proc. Suppl.} {\bf 41} (1995) 184--244,
  [\href{http://arxiv.org/abs/hep-th/9411210}{{\tt hep-th/9411210}}].

\bibitem{Niemi:1994ej}
A.~J. Niemi and K.~Palo, {\it {Equivariant Morse theory and quantum
  integrability}},  \href{http://arxiv.org/abs/hep-th/9406068}{{\tt
  hep-th/9406068}}.

\bibitem{Blau:1995rs}
M.~Blau and G.~Thompson, {\it {Localization and diagonalization: A review of
  functional integral techniques for low dimensional gauge theories and
  topological field theories}},  {\em J. Math. Phys.} {\bf 36} (1995)
  2192--2236, [\href{http://arxiv.org/abs/hep-th/9501075}{{\tt
  hep-th/9501075}}].

\bibitem{Pestun:2016qko}
V.~Pestun, {\it {Review of localization in geometry}},  {\em J. Phys. A} {\bf
  50} (2017), no.~44 443002, [\href{http://arxiv.org/abs/1608.02954}{{\tt
  arXiv:1608.02954}}].

\bibitem{JeffreyLectures}
L.~Jeffrey. \url{http://www.math.toronto.edu/~jeffrey/mat1312/lec10.eqcoh.pdf}.

\bibitem{Alekseev:2000fe}
A.~Alekseev, {\it {Notes on equivariant localization}},  {\em Lect. Notes
  Phys.} {\bf 543} (2000) 1--24.

\bibitem{Cremonesi:2014dva}
S.~Cremonesi, {\it {An Introduction to Localisation and Supersymmetry in Curved
  Space}},  {\em PoS} {\bf Modave2013} (2013) 002.

\bibitem{Romelsberger:2005eg}
C.~Romelsberger, {\it {Counting chiral primaries in N = 1, d=4 superconformal
  field theories}},  {\em Nucl. Phys.} {\bf B747} (2006) 329--353,
  [\href{http://arxiv.org/abs/hep-th/0510060}{{\tt hep-th/0510060}}].

\bibitem{Kinney:2005ej}
J.~Kinney, J.~M. Maldacena, S.~Minwalla, and S.~Raju, {\it {An Index for 4
  dimensional super conformal theories}},  {\em Commun. Math. Phys.} {\bf 275}
  (2007) 209--254, [\href{http://arxiv.org/abs/hep-th/0510251}{{\tt
  hep-th/0510251}}].

\bibitem{Hong:2018viz}
J.~Hong and J.~T. Liu, {\it {The topologically twisted index of $ \mathcal{N} $
  = 4 super-Yang-Mills on T$^{2} \times S^{2}$ and the elliptic genus}},  {\em
  JHEP} {\bf 07} (2018) 018, [\href{http://arxiv.org/abs/1804.04592}{{\tt
  arXiv:1804.04592}}].

\bibitem{Janik:2007pm}
R.~A. Janik and M.~Trzetrzelewski, {\it {Supergravitons from one loop
  perturbative N=4 SYM}},  {\em Phys. Rev.} {\bf D77} (2008) 085024,
  [\href{http://arxiv.org/abs/0712.2714}{{\tt arXiv:0712.2714}}].

\bibitem{Janik2008}
R.~A. Janik and M.~Trzetrzelewski, {\it {Supergravitons from one loop
  perturbative N=4 super Yang-Mills theory}},  {\em Physical Review D -
  Particles, Fields, Gravitation and Cosmology} {\bf 77} (2008), no.~8 1--22,
  [\href{http://arxiv.org/abs/0712.2714}{{\tt arXiv:0712.2714}}].

\bibitem{Berkooz:2006wc}
M.~Berkooz, D.~Reichmann, and J.~Simon, {\it {A Fermi Surface Model for Large
  Supersymmetric AdS(5) Black Holes}},  {\em JHEP} {\bf 01} (2007) 048,
  [\href{http://arxiv.org/abs/hep-th/0604023}{{\tt hep-th/0604023}}].
  
\bibitem{Chang:2013fba}
C.-M. Chang and X.~Yin, {\it {1/16 BPS states in $\mathcal N=$ 4
  super-Yang-Mills theory}},  {\em Phys. Rev.} {\bf D88} (2013), no.~10 106005,
  [\href{http://arxiv.org/abs/1305.6314}{{\tt arXiv:1305.6314}}].

\bibitem{Dolan:2008qi}
F.~Dolan and H.~Osborn, {\it {Applications of the Superconformal Index for
  Protected Operators and q-Hypergeometric Identities to N=1 Dual Theories}},
  {\em Nucl. Phys. B} {\bf 818} (2009) 137--178,
  [\href{http://arxiv.org/abs/0801.4947}{{\tt arXiv:0801.4947}}].

\bibitem{Spiridonov:2010em}
V.~P. Spiridonov, {\it {Elliptic beta integrals and solvable models of
  statistical mechanics}},  {\em Contemp. Math.} {\bf 563} (2012) 181--211,
  [\href{http://arxiv.org/abs/1011.3798}{{\tt arXiv:1011.3798}}].

\bibitem{Baxter1996}
L.~A. Baxter and R.~G. Bartle, {\it {The Elements of Integration and Lebesgue
  Measure.}},  {\em The Statistician} {\bf 45} (1996), no.~4 528.

\bibitem{BrownLevin}
F.~C.~S. {Brown} and A.~{Levin}, {\it {Multiple Elliptic Polylogarithms}},
  {\em arXiv e-prints} (Oct, 2011) arXiv:1110.6917,
  [\href{http://arxiv.org/abs/1110.6917}{{\tt arXiv:1110.6917}}].

\bibitem{GonzalezLezcano:2020yeb}
A.~Gonz\'{a}lez~Lezcano, J.~Hong, J.~T. Liu, and L.~A. Pando~Zayas, {\it
  {Sub-leading Structures in Superconformal Indices: Subdominant Saddles and
  Logarithmic Contributions}},  \href{http://arxiv.org/abs/2007.12604}{{\tt
  arXiv:2007.12604}}.

\bibitem{Nekrasov:2009uh}
N.~A. Nekrasov and S.~L. Shatashvili, {\it {Supersymmetric vacua and Bethe
  ansatz}},  {\em Nucl. Phys. B Proc. Suppl.} {\bf 192-193} (2009) 91--112,
  [\href{http://arxiv.org/abs/0901.4744}{{\tt arXiv:0901.4744}}].

\bibitem{Nekrasov:2009ui}
N.~A. Nekrasov and S.~L. Shatashvili, {\it {Quantum integrability and
  supersymmetric vacua}},  {\em Prog. Theor. Phys. Suppl.} {\bf 177} (2009)
  105--119, [\href{http://arxiv.org/abs/0901.4748}{{\tt arXiv:0901.4748}}].

\bibitem{Gutowski:2004ez}
J.~B. Gutowski and H.~S. Reall, {\it {Supersymmetric AdS$_5$ black holes}},
  {\em JHEP} {\bf 02} (2004) 006,
  [\href{http://arxiv.org/abs/hep-th/0401042}{{\tt hep-th/0401042}}].

\bibitem{Brezin:1977sv}
E.~Brezin, C.~Itzykson, G.~Parisi, and J.~Zuber, {\it {Planar Diagrams}},  {\em
  Commun. Math. Phys.} {\bf 59} (1978) 35.

\bibitem{Marino:2015yie}
M.~Mari\~{n}o, {\em {Instantons and Large N}: {An Introduction to Non-Perturbative
  Methods in Quantum Field Theory}}.
\newblock Cambridge University Press, 9, 2015.

\bibitem{Anninos:2020ccj}
D.~Anninos and B.~M\"uhlmann, {\it {Notes on matrix models (matrix musings)}},
  {\em J. Stat. Mech.} {\bf 2008} (2020) 083109,
  [\href{http://arxiv.org/abs/2004.01171}{{\tt arXiv:2004.01171}}].

\bibitem{Dolan:2007rq}
F.~Dolan, {\it {Counting BPS operators in N=4 SYM}},  {\em Nucl. Phys. B} {\bf
  790} (2008) 432--464, [\href{http://arxiv.org/abs/0704.1038}{{\tt
  arXiv:0704.1038}}].

\bibitem{Witten:2010cx}
E.~Witten, {\it {Analytic Continuation Of Chern-Simons Theory}},  {\em AMS/IP
  Stud. Adv. Math.} {\bf 50} (2011) 347--446,
  [\href{http://arxiv.org/abs/1001.2933}{{\tt arXiv:1001.2933}}].

\bibitem{Felder2000}
G.~Felder and A.~Varchenko {\em Advances in Mathematics} {\bf 156} (2000),
  no.~1 44--76, [\href{http://arxiv.org/abs/9907061}{{\tt 9907061}}].

\bibitem{Assel:2014tba}
B.~Assel, D.~Cassani, and D.~Martelli, {\it {Supersymmetric counterterms from
  new minimal supergravity}},  {\em JHEP} {\bf 11} (2014) 135,
  [\href{http://arxiv.org/abs/1410.6487}{{\tt arXiv:1410.6487}}].

\bibitem{Genolini:2016ecx}
P.~Benetti~Genolini, D.~Cassani, D.~Martelli, and J.~Sparks, {\it {Holographic
  renormalization and supersymmetry}},  {\em JHEP} {\bf 02} (2017) 132,
  [\href{http://arxiv.org/abs/1612.06761}{{\tt arXiv:1612.06761}}].

\bibitem{Papadimitriou:2017kzw}
I.~Papadimitriou, {\it {Supercurrent anomalies in 4d SCFTs}},  {\em JHEP} {\bf
  07} (2017) 038, [\href{http://arxiv.org/abs/1703.04299}{{\tt
  arXiv:1703.04299}}].

\bibitem{An:2017ihs}
O.~S. An, {\it {Anomaly-corrected supersymmetry algebra and supersymmetric
  holographic renormalization}},  {\em JHEP} {\bf 12} (2017) 107,
  [\href{http://arxiv.org/abs/1703.09607}{{\tt arXiv:1703.09607}}].

\bibitem{Closset:2013sxa}
C.~Closset and I.~Shamir, {\it {The $\mathcal{N}=1$ Chiral Multiplet on
  $T^2\times S^2$ and Supersymmetric Localization}},  {\em JHEP} {\bf 03}
  (2014) 040, [\href{http://arxiv.org/abs/1311.2430}{{\tt arXiv:1311.2430}}].

\bibitem{Assel:2014paa}
B.~Assel, D.~Cassani, and D.~Martelli, {\it {Localization on Hopf surfaces}},
  {\em JHEP} {\bf 08} (2014) 123, [\href{http://arxiv.org/abs/1405.5144}{{\tt
  arXiv:1405.5144}}].

\bibitem{Assel:2015nca}
B.~Assel, D.~Cassani, L.~Di~Pietro, Z.~Komargodski, J.~Lorenzen, and
  D.~Martelli, {\it {The Casimir Energy in Curved Space and its Supersymmetric
  Counterpart}},  {\em JHEP} {\bf 07} (2015) 043,
  [\href{http://arxiv.org/abs/1503.05537}{{\tt arXiv:1503.05537}}].

\bibitem{Chen:2005zj}
W.~Chen, H.~Lu, and C.~N. Pope, {\it {Mass of rotating black holes in gauged
  supergravities}},  {\em Phys. Rev.} {\bf D73} (2006) 104036,
  [\href{http://arxiv.org/abs/hep-th/0510081}{{\tt hep-th/0510081}}].
  

\bibitem{Kuzenko:2019vvi}
S.~M. Kuzenko, A.~Schwimmer, and S.~Theisen, {\it {Comments on Anomalies in
  Supersymmetric Theories}},  {\em J. Phys. A} {\bf 53} (2020), no.~6 064003,
  [\href{http://arxiv.org/abs/1909.07084}{{\tt arXiv:1909.07084}}].

\bibitem{Bzowski:2020tue}
A.~Bzowski, G.~Festuccia, and V.~Proch\'azka, {\it {Consistency of
  supersymmetric 't Hooft anomalies}},
  \href{http://arxiv.org/abs/2011.09978}{{\tt arXiv:2011.09978}}.
   

\bibitem{Closset:2019ucb}
C.~Closset, L.~Di~Pietro, and H.~Kim, {\it {'t Hooft anomalies and the
  holomorphy of supersymmetric partition functions}},  {\em JHEP} {\bf 08}
  (2019) 035, [\href{http://arxiv.org/abs/1905.05722}{{\tt arXiv:1905.05722}}].

\bibitem{ZagierOnBloch}
D.~Zagier, {\it {The Bloch-Wigner-Ramakrishnan polylogarithm function}},  {\em
  Math. Annalen} {\bf 286} (1990) 613--624.

\bibitem{Zudilin}
V.~{Pa{\textcommabelow s}ol} and W.~{Zudilin}, {\it {A study of elliptic gamma
  function and allies}},  {\em arXiv e-prints} (Dec., 2017) arXiv:1801.00210,
  [\href{http://arxiv.org/abs/1801.00210}{{\tt arXiv:1801.00210}}].

\bibitem{Levin}
A.~Levin, {\it {Elliptic polylogarithms: An analytic theory}},  {\em Compositio
  Mathematica} {\bf 12} (1997) 106: 267.

\end{thebibliography}
\bibliographystyle{JHEP}

\providecommand{\href}[2]{#2}\begingroup\raggedright\endgroup

\end{document}